\documentclass[twocolumn,trackchanges,resetfootnote]{aastex7}

\usepackage{comment}
\usepackage{multirow}
\usepackage[flushleft]{threeparttable}
\usepackage{graphicx}
\usepackage{amsmath}
\usepackage{breqn}

\shortauthors{Asada et al.}
\accepted{October 28, 2025}
\submitjournal{ApJ}
\graphicspath{{./}{figures/}}

\begin{document}

\title{Earliest Galaxy Evolution in the CANUCS+Technicolor fields:\\
Galaxy Properties at $z\sim10-16$ seen with the Full NIRCam Medium and Broad Band Filters}

\author[orcid=0000-0003-3983-5438,sname=Asada,gname=Yoshihisa]{Yoshihisa Asada}
\altaffiliation{Dunlap Fellow}
\affiliation{Waseda Research Institute for Science and Engineering, Faculty of Science and Engineering, Waseda University,\\ 3-4-1 Okubo, Shinjuku, Tokyo 169-8555, Japan}
\affiliation{Dunlap Institute for Astronomy and Astrophysics, 50 St. George Street, Toronto, Ontario M5S 3H4, Canada}
\email[show]{yoshi.asada@utoronto.ca}

\author[orcid=0000-0002-4201-7367,gname=Chris, sname=Willott]{Chris J. Willott}
\affiliation{NRC Herzberg, 5071 West Saanich Rd, Victoria, BC V9E 2E7, Canada}
\email{chris.willott@nrc.ca}  

\author[orcid=0000-0002-9330-9108,gname=Adam, sname=Muzzin]{Adam Muzzin}
\affiliation{Department of Physics and Astronomy, York University, 4700 Keele St., Toronto, Ontario M3J 1P3, Canada}
\email{muzzin@yorku.ca}

\author[orcid=0000-0001-5984-0395]{Maru{\v s}a Brada{\v c}}
\affiliation{Faculty of Mathematics and
Physics, University of Ljubljana, Jadranska ulica 19, SI-1000 Ljubljana, Slovenia}
\affiliation{Department of Physics and Astronomy, University of California Davis, 1 Shields Avenue, Davis, CA 95616, USA}
\email{marusa.bradac@fmf.uni-lj.si}  

\author[orcid=0000-0003-2680-005X]{Gabriel Brammer}
\affiliation{Cosmic Dawn Center (DAWN), Denmark}
\affiliation{Niels Bohr Institute, University of Copenhagen, Jagtvej 128, DK-2200 Copenhagen N, Denmark}
\email{gabriel.brammer@nbi.ku.dk}

\author[orcid=0000-0001-8325-1742]{Guillaume Desprez}
\affiliation{Kapteyn Astronomical Institute, University of Groningen, P.O. Box 800, 9700AV Groningen, The Netherlands}
\email{guillaume.desprez@protonmail.com}

\author[orcid=0000-0001-9298-3523]{Kartheik G. Iyer}
\affiliation{Columbia Astrophysics Laboratory, Columbia University, 550 West 120th Street, New York, NY 10027, USA}
\email{kgi2103@columbia.edu}

\author[0000-0001-9002-3502]{Danilo Marchesini}
\affiliation{Department of Physics \& Astronomy, Tufts University, 574 Boston Avenue, Suite 304, Medford, MA 02155, USA}
\email{Danilo.Marchesini@tufts.edu}

\author[orcid=0000-0003-3243-9969]{Nicholas S. Martis}
\affiliation{Faculty of Mathematics and Physics, University of Ljubljana, Jadranska ulica 19, SI-1000 Ljubljana, Slovenia}
\email{nicholas.martis@fmf.uni-lj.si}


\author{Ga\"{e}l Noirot}
\affiliation{Space Telescope Science Institute, 3700 San Martin Drive, Baltimore, Maryland 21218, USA}
\email{gnoirot@stsci.edu}

\author[orcid=0000-0001-8830-2166]{Ghassan T. E. Sarrouh}
\affiliation{Department of Physics and Astronomy, York University, 4700 Keele St., Toronto, Ontario M3J 1P3, Canada}
\email{gsarrouh@yorku.ca}

\author[orcid=0000-0002-7712-7857]{Marcin Sawicki}
\affiliation{Department of Astronomy and Physics and Institute for Computational Astrophysics, Saint Mary's University, 923 Robie Street, Halifax, Nova Scotia B3H 3C3, Canada}
\email{marcin.sawicki@smu.ca}

\author[orcid=0009-0000-8716-7695]{Sunna Withers}
\affiliation{Department of Physics and Astronomy, York University, 4700 Keele St., Toronto, Ontario M3J 1P3, Canada}
\email{sunnaw@my.yorku.ca}



\author[orcid=0000-0001-7201-5066]{Seiji Fujimoto}
\affiliation{Dunlap Institute for Astronomy and Astrophysics, 50 St. George Street, Toronto, Ontario M5S 3H4, Canada}
\affiliation{David A. Dunlap Department of Astronomy and Astrophysics, University of Toronto,\\ 50 St. George Street, Toronto, Ontario M5S 3H4, Canada}
\email{seiji.fujimoto@utoronto.ca}

\author[orcid=0009-0001-0778-9038]{Giordano Felicioni}
\affiliation{Faculty of Mathematics and
Physics, University of Ljubljana, Jadranska ulica 19, SI-1000 Ljubljana, Slovenia}
\email{giordano.felicioni@fmf.uni-lj.si}

\author[orcid=0009-0007-8470-5946]{Ilias Goovaerts}
\affiliation{Space Telescope Science Institute, 3700 San Martin Drive, Baltimore, Maryland 21218, USA}
\email{igoovaerts@stsci.edu}

\author[orcid=0009-0000-2101-1938]{Jon Jude\v{z}}
\affiliation{Faculty of Mathematics and
Physics, University of Ljubljana, Jadranska ulica 19, SI-1000 Ljubljana, Slovenia}
\email{jon.judez@fmf.uni-lj.si}

\author[orcid=0009-0009-9848-3074]{Naadiyah Jagga}
\affiliation{Department of Physics and Astronomy, York University, 4700 Keele St., Toronto, Ontario M3J 1P3, Canada}
\email{jagga@yorku.ca}

\author[]{Maya Merchant}
\affiliation{Niels Bohr Institute, University of Copenhagen, Jagtvej 128, DK-2200 Copenhagen N, Denmark}
\email{jzr171@alumni.ku.dk}

\author[orcid=0000-0001-8115-5845]{Rosa M. M\'erida}
\affiliation{Department of Astronomy and Physics and Institute for Computational Astrophysics, Saint Mary's University, 923 Robie Street, Halifax, Nova Scotia B3H 3C3, Canada}
\email{rosa.meridagonzalez@smu.ca}

\author[orcid=0000-0002-6265-2675]{Luke Robbins}
\affiliation{Department of Physics \& Astronomy, Tufts University, 574 Boston Avenue, Suite 304, Medford, MA 02155, USA}
\email{andrew.robbins@tufts.edu}


\begin{abstract}

We present a sample of $z_{\rm phot}\sim10-16$ galaxies by exploiting one of the richest JWST NIRCam imaging data, taken in the CANUCS survey in Cycle 1 and the Technicolor (TEC) survey in Cycle 2.
The combination of the CANUCS+TEC provides multi-epoch, deep NIRCam images in all medium bands (MBs) and broad bands (BBs) onboard NIRCam (22 filters in total), over $\sim23\ {\rm arcmin}^2$ in three independent lines of sight.
We select high-$z$ galaxy candidates based on photometric redshifts, and obtain eight candidates at $z\sim10-16$, including a very robust candidate at $z\sim15.4$.
The ultraviolet (UV) luminosity function (LF) from our sample is consistent with 
previous JWST studies showing a scatter of $\sim0.6$ dex across the literature, marking the significance of the field-to-field variance in interpreting galaxy abundance measurements at $z>10$.
We find that the UV LF moderately evolves at $z>10$, and the LF normalization and the luminosity density decline by a factor of $\sim7$ from $z\sim11$ to $z\sim15$, indicating less steep evolution than $z<11$. 
We highlight the importance of MB filters, not only to minimize the contamination by low-$z$ interlopers but also to maximize the completeness.
In particular, faint and less blue galaxies could be missed when the sample is built solely on BB data. 
The contamination and incompleteness of BB-only selected samples can bias our views of earliest galaxy evolution at $z>10$, including the UV LF by $\sim0.6$ dex, the size-magnitude relation by $\sim0.6$ dex, and the UV slope-magnitude relation by $\Delta\beta_{\rm UV}\sim-0.3$.

\end{abstract}



\section{Introduction}
Finding and characterizing the earliest galaxies is one of the major quests in modern astronomy.
The James Webb Space Telescope (JWST) has made a breakthrough in this field with its capability of finding $z\gtrsim10$ galaxies.
Numerous $z>10$ galaxy candidates have been identified with NIRCam imaging \citep[e.g.,][]{Naidu2022ApJ,Bouwens2023MNRAS,Donnan2023,Donnan2023MNRAS,Harikane2023,Perez_Gonzalez2023,Adams2024ApJ,Hainline2024,Willott2024ApJ,Castellano2025arXiv,Perez_Gonzalez2025}.
The statistical search for high-$z$ galaxy candidates with NIRCam imaging is in particular essential to probe the earliest galaxy evolution. 
Tracking the early evolution of galaxy number density, e.g., the UV luminosity function (LF), directly unveils how first galaxies have emerged from the initial matter perturbation and how they have grown with baryon accretion in dark matter halos \citep[e.g.,][]{Bouwens2021,Harikane2022}.

Various NIRCam imaging surveys have been carried out in the first three years of JWST observation, most of which employed the standard Broad-band (BB) filter configurations.
This has been a good strategy as BB filters are wide in wavelength coverage, offering good spectral energy distribution (SED) coverage but also facilitating deep photometry with efficient integration times. Given that the Lyman break feature is strong and does not require high spectral resolution, the BB approach in the initial years of JWST has been sensible and highly efficient at discovering $z > 10$ galaxy candidates.
Thus far, many $z_{\rm phot}\gtrsim10$ galaxy samples are built relying on the BB-only photometry data, including galaxy candidates even up to $z_{\rm phot}\sim25$ \citep[e.g.,][]{Castellano2025arXiv,Perez_Gonzalez2025}.
However, the selection of high-$z$ galaxies with BB-only data needs to be treated with caution, because the BB photometry spectral energy distributions (SEDs) of interlopers at lower-$z$ with strong emission lines combined with a Balmer break or a red, dusty continuum can mimic the high-$z$ galaxy SED and contaminate the high-$z$ galaxy sample.
Perhaps the most prominent example of this is the galaxy
CEERS-9331, which was originally identified as a bright $z\sim16$ galaxy candidate by several independent authors based on the seven BB filters + F410M photometric data \citep[e.g.,][]{Donnan2023,Harikane2023}, but subsequent NIRSpec observation confirmed it as a dusty extreme emission line galaxy at $z=4.9$ \citep{Arrabal_Haro2023}.

Particularly at $z>9.5$, all strong rest-frame optical emission lines are redshifted beyond NIRCam filter coverage, and the Lyman-$\alpha$ break at $\lambda_{\rm rest}=1216\ {\rm \AA}$ is the only signature to identify $z>9.5$ galaxy candidates from photometry. 
The combination of NIRCam Medium-band (MB) and BB filter observations is one way to minimize contamination when building robust $z>9.5$ galaxy samples \citep[e.g.,][]{Adams2025}.
NIRCam MB observations have successfully shown that they can identify strong emission lines from the flux excess in MB filters \citep[e.g.,][]{Withers2023,Sarrouh2024ApJ,Suess2024ApJ,Martis2025arXiv} and therefore they are extremely efficient at identifying low-$z$ strong line emitters that contaminate high-$z$ candidates such as in \citet{Arrabal_Haro2023}.
MB+BB observations also allow us to precisely locate the sharp break in the continuum at the Lyman break.
Thus MB+BB filter observations should unambiguously distinguish low-$z$ interlopers with strong emission lines vs real high-$z$ galaxies with sharp continuum drop-outs.

Not only can MBs efficiently rule out interlopers, they can also help properly classify real candidates previously considered more likely to be low-$z$ contaminants with BB-only data.
One good showcase of this is GS-z14-0 at $z_{\rm spec}=14.18$ \citep{Carniani2025AAS}.
GS-z14-0 was originally identified from NIRCam BB imaging observations in Cycle 1, but it was thought to be at $z\sim3.5$ with the BB-only data \citep[][]{Williams2024ApJ}. After obtaining MB observations in Cycle 2, GS-z14-0 was identified as a secure $z\sim14$ galaxy candidate \citep[][]{Robertson2024ApJ}, and the subsequent spectroscopic follow-up observation proved it to be at $z_{\rm spec}=14.18$ \citep{Carniani2024,Carniani2025AAS}.

\begin{figure}[t]
\includegraphics[width=0.95\linewidth]{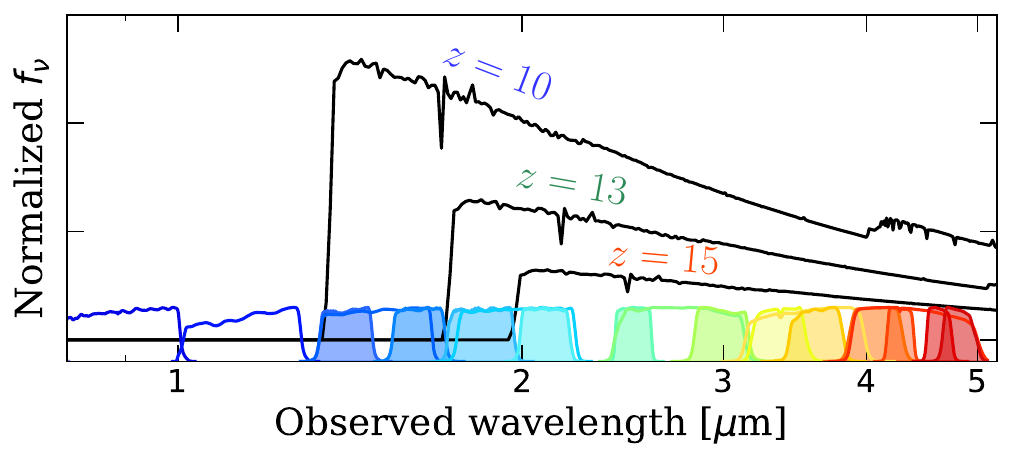}
\caption{Illustration showing the NIRCam filter transmission curves and the Lyman-$\alpha$ break feature in $z>10$ galaxies. The MB filters are shown by filled curves.
}
\label{fig:LyBreak_filters}
\end{figure}

Specifically, the combination of F140M, F150W, and F162M is powerful in selecting $z\sim10-13$ galaxies, and that of F182M, F200W, and F210M in selecting $z\sim13-16$ galaxies, as they simultaneously cover the Lyman break wavelength range with two MBs and one BB overlapping each other, and they can locate the sharp drop-out from photometry (or strong emission line excess for low-$z$ interlopers; Figure \ref{fig:LyBreak_filters}).
\citet{Robertson2024ApJ} exploited a deep MB+BB NIRCam survey program, the JADES Origins Field \citep[JOF;][]{Eisenstein2023arXiv}, to derive the galaxy UV LF at $z>9.5$ based on 14 NIRCam filter images (seven MBs and seven BBs) in a very deep, single NIRCam pointing.
As pointed out by \citet{Robertson2024ApJ}, galaxy abundance measurements in such a small area are affected by cosmic variance, and it is necessary to enlarge the survey volume for $z>10$ galaxies with a comprehensive set of MB+BB filters.

In this paper, we present UV LF measurements at $z>9.5$ using data from the CAnadian NIRISS Unbiased Cluster Survey \citep[CANUCS; GTO-1208, PI: Willott;][]{Willott2022PASP} in Cycle 1 and the JWST in Technicolor program (TEC; GO-3362, PI: Muzzin) in Cycle 2.
Combining CANUCS+TEC provides the full set of NIRCam BB+MB filter observations (+ two narrow-band filters) in three independent NIRCam pointings.
Observing three independent sight lines with the rich MB+BB filters adds an invaluable constraint on the early UV LF evolution, by building a new, very clean $z>10$ galaxy sample in addition to the previous work in the JOF.
We also explore the potential bias due to low-$z$ interloper contamination in BB-only data selection of $z\gtrsim10$ galaxies, by comparing our main $z\gtrsim10$ candidates built from MB+BB observations and those from a degraded BB-only photometry catalog.

The paper is structured as follows.
Sec.~\ref{sec:data} describes the imaging data and data processing we use in this work.
Sec.~\ref{sec:selection} presents the sample selection of high-$z$ galaxies and the completeness simulation of our selection.
Sec.~\ref{sec:result} gives the main result of this paper, including the UV LF at $z>9.5$ and the redshift evolution of the UV luminosity density.
Sec.~\ref{sec:discussion} is the discussion focusing on what can affect measurements of high-$z$ galaxy abundance in JWST surveys, including cosmic variance and low-$z$ interlopers in BB-only data selections.
The section also discusses the possible incompleteness of BB-only selection sample for faint and less blue galaxies at high-$z$, and quantitatively evaluate the potential bias by the contamination and incompleteness with BB-only data in various aspects of galaxy properties at $z>10$ and future prospects from scheduled JWST Cycle 4 observations.
Sec.~\ref{sec:summary} finally provides the summary of the paper.
Throughout the paper, we assume a flat $\Lambda$CDM cosmology with $\Omega_{m}=0.3$, $\Omega_{\Lambda}=0.7$, and $H_0=70\ {\rm km\ s^{-1}\ Mpc^{-1}}$, and all magnitudes are quoted in the AB system \citep{Oke1983ApJ}.

\section{Data}\label{sec:data}
\subsection{Imaging Data and Photometry}\label{subsec:image_photometry}
We use JWST/NIRCam imaging data from CANUCS in Cycle 1 and TEC in Cycle 2 of the three NIRCam flanking fields (NCFs). 
The overall survey description and full details of the imaging reduction are presented in \citet[SA25 hereafter]{Sarrouh2025arXiv}.
Combining the multi-epoch two surveys, a total of three NIRCam pointings are observed with all NIRCam BB and MB filters\footnote{except for the MACS~1149 flanking field: F162M and F250M are missing due to a program definition error}, in addition to two narrow-band filters of F164N and F187N.
Each NIRCam pointing is an associated flanking field of the strong lensing cluster field (MACS~0416, MACS~1149, and A~370), and we refer to these flanking fields as NCF fields.
As we only use the flanking fields, the effect of gravitational lensing is negligibly small in this work.
Table \ref{tab:summary_depth} presents the limiting magnitudes of all available NIRCam filters in each field.
The table also includes the information about the observing cycle (Cycle 1 by CANUCS program or Cycle 2 by TEC program).

\begin{deluxetable}{lccc}
    \label{tab:summary_depth}
    \tablecaption{Summary of JWST/NIRCam and HST/ACS observations in CANUCS+Technicolor fields.
  	}
    \tablewidth{0pt}
    \tablehead{
    \colhead{Filter (Obs.~Cycle)} & \colhead{MACS~0416} & \colhead{MACS~1149} & \colhead{A~370}
    }
    \startdata
    F070W (Cy.~2) & 29.74 & 29.69 & 29.65 \\
    F090W (Cy.~1) & 30.20 & 30.16 & 30.09 \\
    F115W (Cy.~1) & 30.23 & 30.17 & 30.05 \\
    F140M (Cy.~1) & 29.51 & 29.34 & 29.28 \\
    F150W (Cy.~1) & 30.35 & 30.41 & 30.18 \\
    F162M (Cy.~1) & 29.53 &  ...  & 29.43 \\
    F164N (Cy.~2) & 27.38 & 27.40 & 27.34 \\
    F182M (Cy.~1) & 30.02 & 30.01 & 29.90 \\
    F187N (Cy.~2) & 27.37 & 27.47 & 27.44 \\
    F200W (Cy.~2) & 30.02 & 29.99 & 29.90 \\
    F210M (Cy.~1) & 29.90 & 29.84 & 29.79 \\
    F250M (Cy.~1) & 29.30 &  ...  & 29.11 \\
    F277W (Cy.~1) & 30.60 & 30.65 & 30.49 \\
    F300M (Cy.~1) & 29.71 & 29.58 & 29.52 \\
    F335M (Cy.~1) & 29.97 & 29.93 & 29.84 \\
    F356W (Cy.~2) & 30.13 & 30.06 & 30.03 \\
    F360M (Cy.~1) & 29.85 & 29.85 & 29.76 \\
    F410M (Cy.~1) & 29.67 & 29.68 & 29.57 \\
    F430M (Cy.~2) & 28.91 & 28.90 & 28.91 \\
    F444W (Cy.~1) & 29.91 & 29.82 & 29.75 \\
    F460M (Cy.~2) & 28.55 & 28.48 & 28.50 \\
    F480M (Cy.~2) & 28.53 & 28.47 & 28.51 \\
    \hline
    F435W & 30.09 & 30.01 & 30.05 \\
    F606W & 29.64 & 29.59 & 29.80 \\
    F814W & 29.73 & 29.19 & 29.63 \\
    \hline
    Area (arcmin$^2$)\tablenotemark{a} & 7.57 & 7.88 & 7.87 \\
    \enddata
    \tablenotetext{}{\textbf{Notes.} For each filter, the 3-sigma limiting magnitudes (in ABmag) are quoted. The limiting magnitudes are measured in $0.\!\!^{\prime\prime}16$-diameter aperture, by randomly distributing empty apertures in the background sky region of each image (without PSF-convolution) and performing aperture correction based on the PSF EEs.
    }
    \tablenotetext{a}{The unmasked sky area from completeness simulations is quoted.
    }
\end{deluxetable}

We also utilize HST optical imaging observations in F435W, F606W, and F814W filters on Advanced Camera for Surveys (ACS), taken in previous observations by the Hubble Frontier Fields Parallels program \citep{Lotz2017ApJ}. 
The HST/ACS footprints partially overlap with the NIRCam area (typically $\sim50\ \%$ of NIRCam field-of-views), so we use HST/ACS photometry when available for photometric redshift computation (Sec.~\ref{subsec:photoz}), but do not adopt S/N cuts in HST/ACS filter images in the sample selection.
There are also several HST Wide Field Camera 3 infra-red filter images available in these fields, but we do not use them in this work, given the wavelength overlap with NIRCam and the superior quality of NIRCam images in resolution and sensitivity.

We exploit the background-subtracted images and the source detection catalogs given by SA25, but we perform custom photometry to maximize the signal-to-noise ratio (S/N) for faint, small sources barely resolved with NIRCam imaging.
In SA25, the source detection is performed on the $\chi_{\rm mean}$ image following \citet{Drlica-Wagner2018ApJS}, which is built by co-adding all available JWST and HST-optical filters.
At the position of each detected source, we perform fixed aperture photometry with a $0.\!\!^{\prime\prime}16$-diameter aperture on the background-subtracted images that have not been convolved to a common PSF. We correct the aperture photometry in each filter based on the PSF encircled energies (EEs) of the filter, since the galaxies are small.
The photometry is then corrected for Galactic extinction with the \citet{Fitzpatrick1999PASP} extinction curve assuming $R_V=3.1$, to obtain the color (i.e., SEDs) of each detected source.
The total flux of each source is estimated by scaling the SED with the Kron-to-aperture-corrected flux ratios in the F277W image.
The photometric errors are computed by measuring 2000 empty aperture fluxes in the background sky region and applying the same correction factors as the fixed aperture fluxes (i.e., color correction, Galactic dust attenuation correction, and aperture correction).


\subsection{Photometric Redshift}\label{subsec:photoz}
We select high-$z$ galaxies using photometric redshifts (photo-$z$) in this work.
The photo-$z$ is estimated using the template-fitting code \texttt{EAzY-py} \citep{Brammer2008ApJ} under almost the same configuration as in SA25.
A total of 15 galaxy template spectra are used in the fit, 12 of which are from FSPS \citep{Conroy2010ascl} and the other three are custom templates based on \citet[see SA25 for details of the modification]{Larson2023ApJL}.
No redshift prior is assumed, while systematic uncertainties of $5\ \%$ of observed fluxes are added to the flux error budget in quadrature.
The galaxy template spectra are corrected for the inter-galactic medium (IGM) absorption following the prescription by \citet{Asada2025ApJL}, which takes into account the effect of increasing Ly$\alpha$ damping-wing absorption at $z>7$, on top of the canonical IGM transmission curve \citep{Inoue2014MNRAS}.
We stress that the use of the IGM modeling by \citet{Asada2025ApJL} is expected to resolve the photo-$z$ overestimation at $z\gtrsim7$, which has been commonly seen in previous works just using classical IGM prescriptions, and this reduces a systematic uncertainty with regard to the redshift and absolute magnitude estimation of high-$z$ galaxies.
Under this configuration, we compute the photo-$z$ probability function (PDZ) of each source from $z=0$ to $20$, and obtain the best photo-$z$ estimation ($z_{\rm ml}$) based on the peak of the PDZ.


\section{Sample Selection}\label{sec:selection}

\begin{deluxetable*}{lccccccc}
    \label{tab:highz_cand}
    \tablecaption{High-redshift galaxy candidates in CANUCS+Technicolor fields.
  	}
    \tablewidth{0pt}
    \tablehead{
    \colhead{Name} & \colhead{R.A.} & \colhead{Decl.} & \colhead{$z_{\rm ml}$} & \colhead{$M_{\rm UV}$} & \colhead{$\beta_{\rm UV}$} & \colhead{$r_{\rm eff}$ [pc]} & \colhead{$P(z<7)$} \\
    \colhead{(1)} & \colhead{(2)}& \colhead{(3)} & \colhead{(4)} & \colhead{(5)} & \colhead{(6)} & \colhead{(7)} & \colhead{(8)}
    }
    \startdata
    CANUCS-5223684 & 177.414290 & 22.32748 & $9.65^{+0.32}_{-0.35}$ & $-18.37$ & $-2.14 \pm 0.22$ & $17.1 \pm 2.3$ & 4.96e-04 \\
    CANUCS-3209488 & 64.169709 & -24.10793 & $9.69^{+0.24}_{-0.36}$ & $-19.23$ & $-1.88 \pm 0.27$ & $487.9 \pm 27.3$\tablenotemark{a} & 2.76e-09 \\
    CANUCS-2215881 & 40.056129 & -1.61568 & $9.78^{+0.13}_{-0.43}$ & $-19.09$ & $-2.20 \pm 0.14$ & $109.8 \pm 83.3$ & 8.00e-11 \\
    CANUCS-5221204 & 177.386651 & 22.30266 & $10.15^{+0.02}_{-1.42}$ & $-18.89$ & $-2.67 \pm 0.37$ & $207.6 \pm 91.1$ & 2.17e-03 \\
    CANUCS-2219868 & 40.032579 & -1.60055 & $10.32^{+0.15}_{-0.26}$ & $-19.52$ & $-2.33 \pm 0.16$ & $112.4 \pm 32.6$\tablenotemark{b} & 1.66e-04 \\
    CANUCS-5208401 & 177.404008 & 22.32253 & $10.69^{+0.33}_{-0.38}$ & $-18.75$ & $-1.93 \pm 0.30$ & $15.9 \pm 2.1$ & 6.52e-02 \\
    CANUCS-5203757 & 177.408240 & 22.28612 & $11.23^{+0.14}_{-0.10}$ & $-19.77$ & $-2.18 \pm 0.07$ & $15.3 \pm 2.0$ & 9.92e-19 \\
    CANUCS-3214552 & 64.133557 & -24.07987 & $15.43^{+0.26}_{-0.57}$ & $-18.98$ & $-1.80 \pm 0.34$ & $29.2 \pm 1.3$ & 5.05e-03 \\
    \enddata
    \tablenotetext{}{(1) Source IDs in the CANUCS catalog by SA25. (2) Right Ascension in J2000. (3) Declination in J2000. (4) Best photo-$z$ estimations. Uncertainties are from the 16th- to 84th-percentile range of $P(z)$. (5) Absolute rest UV magnitudes. (6) UV slopes measured from NIRCam photometry. (7) The intrinsic, PSF-deconvolved effective radii $r_{\rm eff}$ in the NIRCam F277W image. (8) Low-$z$ solution probabilities.
    }
    \tablenotetext{a}{The source appears to have two components in the NIRCam image, but fails to fit with two S\'{e}rsic components and results of fitting a single component are quoted.}
    \tablenotetext{b}{Fitted with two S\'{e}rsic components, and the size of the main clump is quoted.}
\end{deluxetable*}

We present the high-$z$ galaxy sample as defined at $z>9.5$ in this work. 
To minimize the low-$z$ interlopers into the sample, we adopt the following selection criteria:
\begin{equation}
\label{eqn:selection}
    \begin{split}
        S/N_{\rm Blue} < 3,\\
        S/N_{\rm F277W} > 9,\\
        \int_{0}^{7} P(z) dz < 0.1, \\
        9.5 < z_{\rm ml} < 16, \\
    \end{split}
\end{equation}
where $S/N_{\rm Blue}<3$ means the signal-to-noise ratio (S/N) in bluer filters (F070W, F090W, and F115W) are all less than 3.
Requiring S/Ns smaller than 3 in three independent filters would remove $\sim0.3\ \%$ of real sources by the random errors.
If we required S/Ns smaller than 2 instead, we would miss $\sim6\ \%$ of real candidates due to the random noise.
The limit of low-$z$ solution probability is applied to select only high-$z$ sources with high confidence level and remove possible low-$z$ interlopers that have double-peaked $P(z)$.
We also adopt the redshift upper limit of $16$ to remove any spurious NIRCam long wave (LW)-only sources: 
this is because the relatively large wavelength coverage gap between NIRCam short-wave (SW) filters and LW filters, 
unmasked bad pixels in the LW detectors, broader PSF size at longer wavelength, or the combination of them can easily produce fake high-$z$ candidates that are detected only in NIRCam LW filters. 
The motivation of this work is to provide the cleanest sample of $z>9.5$ galaxies possible, relying on the rich NIRCam MB+BB observations that are capable to robustly capture the Lyman break, thus we limit the redshift up to $z<16$ where the dropout can still be detected with F200W and F210M on the SW detectors.

A total of eight galaxies are selected by the criteria (equation \ref{eqn:selection}), and we use them for the subsequent luminosity function analysis.
Three of them (CANUCS ID: 2215881, 2219868, and 5203757) were reported in our previous work \citep{Willott2024ApJ} in the same fields, but the rest are new.
We note that the previous work only used CANUCS Cycle 1 observations and utilized smaller number of NIRCam filters.
The five new sources were not included in the previous work mostly because of their non-negligible low-$z$ solution probability, which becomes negligible with the new NIRCam observations from the TEC servey in Cycle 2.
On the other hand, two sources included in the previous work are not selected in this work, and they are both due to low S/N in F277W.
For each high-$z$ candidate, we also measure the rest-UV beta slope $\beta_{\rm UV}$ and the effective radius $r_{\rm eff}$ in the NIRCam F277W image.
For $\beta_{\rm UV}$ measurements, we fit a pure power law $f_\lambda\propto\lambda^\beta$ to the NIRCam photometry that corresponds to $\lambda_{\rm rest}<3000\ $ \AA\ and covers only the filters redward of the Ly$\alpha$ transition.
Thanks to the extensive NIRCam observations, the $\beta_{\rm UV}$ measurements are based on 5 to 9 NIRCam data points.
For $r_{\rm eff}$ measurements, we perform forward modeling assuming the S\'{e}rsic light profile with \texttt{Galfit} \citep{Peng2010AJ}.
We assume a single S\'{e}rsic component by default, but we adopt two-component fitting when the high-resolution NIRCam images show two clumps (CANUCS-2219868).
Table~\ref{tab:highz_cand} lists the eight candidates in this work. 
The eight galaxies are found at $z\sim9.5$ to $z\sim15$, and they are typically faint, moderately blue, and compact galaxies.
We note that the gravitational lens magnifications of the eight galaxies are negligibly small (median $\langle\mu\rangle=1.02$, maximum $\mu_{\rm max}=1.08$).

\subsection{A remarkable $z>15$ galaxy candidate: CANUCS-3214552}\label{subsec:CANUCS_z15_0}
\begin{figure*}[t]
\plotone{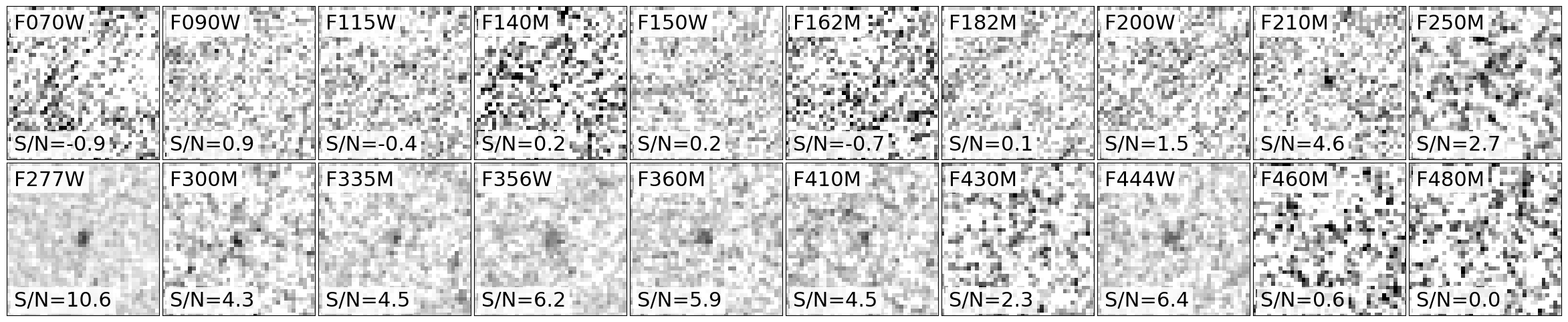}
\plotone{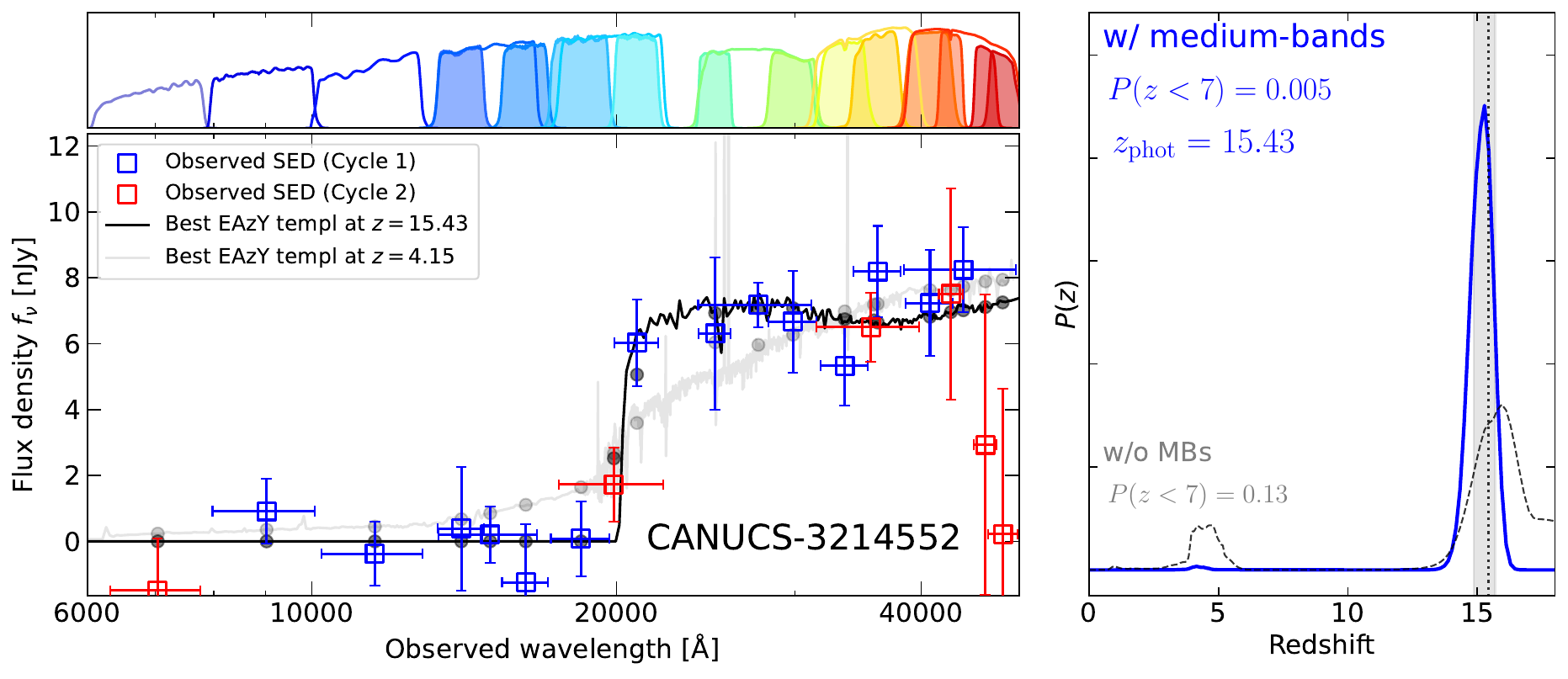}
\caption{The highest redshift candidate of CANUCS-3214552 at $z_{\rm phot}\sim15.4$. \textit{Top}: cutouts of the source in all NIRCam filter images, except for the two narrow band filters. Cutouts are $1.\!\!^{\prime\prime}6$ on the side. Image scalings are homogeneous.
\textit{Bottom}: the SED and the PDZ of CANUCS-3214552. The SED shows a clear dropout between F210M and F200W, and the sharp drop between this narrow wavelength range strongly supports the high-$z$ solution template (black solid curve in the bottom left panel) and results in negligible low-$z$ probability (blue curve in the bottom right panel).
There is no variability between Cycle 1 and Cycle 2 observation with 1 year separation (blue and red squares in bottom left), which rules out the possibility of $z\sim4$ SN contamination.
If MB data were not available, the SED could also be fit by a low-$z$ solution (gray curve in bottom left) and the PDZ would have a non-negligible secondary peak at $z\sim4$ (gray dashed line in bottom right).
Only multiple-epoch rich MB+BB observations can secure this galaxy as a very robust $z\sim15$ galaxy candidate.
}
\label{fig:3214552}
\end{figure*}

The highest redshift candidate among the sample is CANUCS-3214552, whose best photo-$z$ estimate is $z_{\rm ml}=15.43^{+0.26}_{-0.57}$ with a negligible low-$z$ possibility ($P(z<7)=5.1\times10^{-3}$; Figure \ref{fig:3214552}).
The galaxy is faint and relatively red in UV ($M_{\rm UV}=-19.0$ mag and $\beta_{\rm UV}=-1.8$), and the sharp dropout is captured between the F182M+F200W and F210M filters. 
Together with the flat SED at longer wavelengths, the sharp break in the narrow wavelength range between F182M and F210M and the dropout in F200W strongly support the $z=15.4$ solution over the low-$z$ ($z=4$) solution (the blue solid curve in the bottom right panel of Figure \ref{fig:3214552}).
Indeed, this source would have a non-negligible secondary low-$z$ peak in the PDZ if only the standard set of NIRCam eight filters (seven BBs + F410M) were available (the dashed curve in the panel) and would not be selected as a secure high-$z$ candidate.
We note that the lack of Ly$\alpha$ line and the rounded shape of the Ly$\alpha$-break in the best-fit template spectrum is due to the increased neutral Ly$\alpha$ damping wing absorption in the IGM modeling \citep{Asada2025ApJL}.

Moreover, the multi-epoch observations confirmed there is no evidence for variability.
Several studies have shown that some types of supernovae (SNe) at $2<z<5$ have strong breaks around rest 4000 \AA\ wavelength, which can mimic the color of highest-$z$ LBGs \citep[e.g.,][]{Yan2023ApJS,DeCoursey2025ApJ}.
\cite{DeCoursey2025ApJ} suggested that a high fraction of point-like, $z\sim16$ candidates could be SNe at $z\sim4$.
However in the case of CANUCS-3214552, the galaxy was observed in Cycle 1 and Cycle 2 with one year separation and the NIRCam photometry agrees well between the two epochs (e.g., F335M and F360M taken in Cycle 1 vs. F356W taken in Cycle 2).

CANUCS-3214552 is thus one of the most reliable high-$z$ galaxy candidates at $z>15$, which could have been secured only with the multi-epoch, full NIRCam MB+BB observations.
The galaxy shows a relatively red rest-UV color, in contrast to recent reports of very blue colors for the most of faint highest-$z$ galaxies \citep[e.g.,][]{Cullen2024MNRAS}, and it could suggest the galaxy represents a galaxy population that has been missed in previous surveys using the standard NIRCam BB configuration.
We provide a detailed discussion on this point later (Sec.~\ref{subsubsec:UV_slope}).

\begin{figure}[t]
\centering
\includegraphics[width=0.9\linewidth]{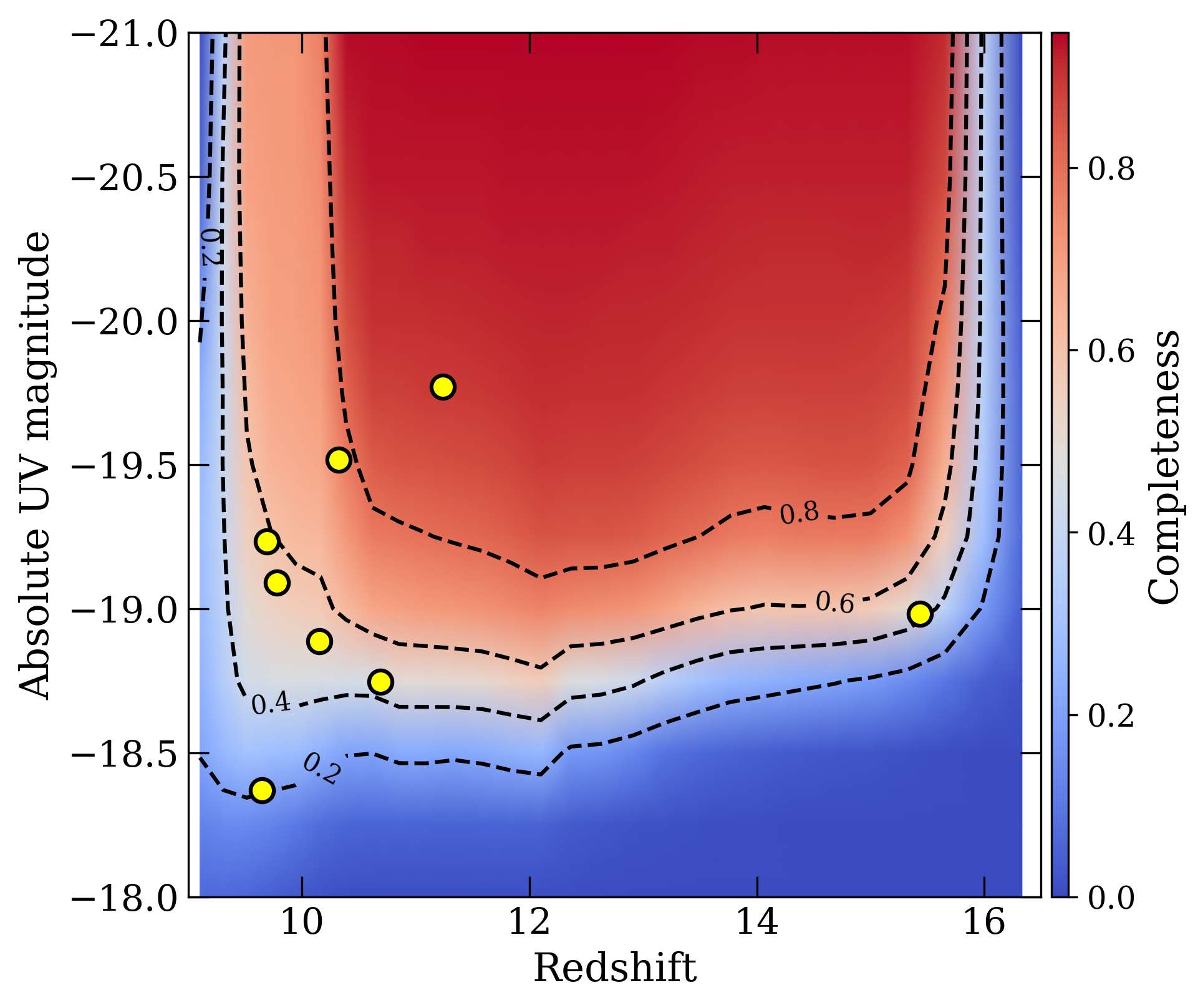}
\caption{High-$z$ galaxy selection completeness in the absolute UV magnitude vs redshift plane. Yellow points present our main $z>9.5$ galaxy sample selected by criteria (\ref{eqn:selection}). Our selection is 50 \% complete down to $M_{\rm UV}\sim-18.7$ mag over the full redshift range of $10\lesssim z\lesssim16$. Though, only one galaxy is found at $z>12$ whereas seven are found at $z<12$, which indicates redshift evolution of the UV LF across this redshift range.}
\label{fig:Completeness}
\end{figure}

\subsection{Completeness simulation}\label{subsec:completeness}
A luminosity function estimation requires computation of the effective volume that has been searched, taking account of the effect of detection and selection completeness. Our selection criteria (equation \ref{eqn:selection}) are fully automated and we do not apply additional cuts based on visual inspection. Thus, we can estimate the completeness by inserting simulated sources into our data and repeating the detection and selection procedure to measure the recovered fraction. We follow the completeness simulation prescription of \citet{Willott2024ApJ} with a few modifications.

For the simulated galaxy size distribution, \citet{Willott2024ApJ} used a redshift-dependent size-$M_{\rm UV}$ relationship derived from a photometric-redshift-selected sample of high-$z$ galaxies from \citet{Morishita2024}. As we show in Sec.~\ref{subsubsec:size-muv}, the $z>9.5$ galaxies from the current study plus those from the JOF that also includes MB photometry \citep{Robertson2024ApJ} have sizes considerably smaller (median offset 0.6 dex) than that of the \citet{Morishita2024} relationship. Therefore, we offset the \citet{Morishita2024} relationship by 0.6 dex for the size distribution of the simulated galaxies.

The SEDs of the simulated galaxies incorporate the updated Larson templates and IGM absorption used in deriving photometric redshifts (Sec.~\ref{subsec:photoz}), as described in detail in SA25. We adopt a set of 4 template spectra with a range of stellar age, dust attenuation and emission line strength that gives a F200W-F277W color distribution comparable to that of our sample.

A grid of $M_{\rm UV}$ and redshift covering the range of this study was set up. At each point in this plane, simulated galaxies were randomly inserted into all the NIRCam images of the three fields. In total, 850,000 galaxies were inserted, but not more than 500 per NIRCam module at a time to prevent crowding of simulated sources. These images were then analyzed using the same detection and segmentation process, followed by the sample selection criteria of equation \ref{eqn:selection}. As in \citet{Willott2024ApJ}, it is found that $\sim10\%$ of bright objects are not recovered due to overlap with existing galaxies in the images.

Figure \ref{fig:Completeness} shows the completeness of our high-$z$ galaxy selection in the  $M_{\rm UV}$ vs.~redshift plane.
Our selection is $\sim50\ \%$ complete down to $M_{\rm UV}\sim-18.7$ at $z\gtrsim10$.
The observed $M_{\rm UV}$ and redshift values of our high-$z$ sample are also plotted in the figure (yellow points). The distribution of observed galaxies agrees well with the predicted completeness from the simulation (e.g., only a few galaxies in the $<50\ \%$ complete area), which supports the validity of our simulation configuration.
In particular, if we were to use the original size-$M_{\rm UV}$ relation of \citet{Morishita2024} in the simulation, five of the eight galaxies in our sample would be located in the $<50\ \%$ completeness region in the $M_{\rm UV}$-$z$ plane.

\section{Results}\label{sec:result}
\subsection{Evolving UV Luminosity Functions}\label{subsec:LFparams}
Knowing the completeness of our survey and high-$z$ galaxy selection, we can compute the UV luminosity function and its redshift evolution at $z\gtrsim10$.
Although our selection is nearly complete down to $M_{\rm UV}\sim-19$ mag beyond $z=12$, there is only one galaxy found in the total $\sim23\ {\rm arcmin}^2$ unmasked area at $z>12$, which suggests a significant evolution of the galaxy UV luminosity function at these redshifts.
We parameterize the UV LF in an analytical form with redshift-dependent parameters, and directly fit to data without binning in redshift or magnitude to avoid any effect by binning, considering the paucity of $z\gtrsim11$ galaxies.
We adopt the Schechter form \citep{Schechter1976ApJ} for the LF,
\begin{dmath}\label{Eqn:schechter}
    \phi_{\rm UV}(M_{\rm UV},z) = \frac{\ln(10)}{2.5}\phi_\star(z)\ 10^{0.4(\alpha+1)(M_{\rm UV}^\star-M_{\rm UV})} \exp\left[-10^{0.4(M_{\rm UV}^\star-M_{\rm UV})}\right],
\end{dmath}
where $\alpha$ and $M_{\rm UV}^\star$ are the faint-end slope and the characteristic UV absolute magnitude, respectively, and $\phi_\star(z)$ is the redshift-dependent normalization parameter.
We assume a simple redshift dependency, as done by \citet{Robertson2024ApJ},
\begin{equation}
    \log_{10}\phi_\star(z) = \log_{10}\phi_{\star,0} + \eta\ (z-z_0),
\end{equation}
to reduce the free parameters in the fit.
The reference redshift $z_0$ is fixed to $z_0=10$ throughout the paper unless otherwise specified.
We also fix $\alpha$ and $M_{\rm UV}^\star$ to $\alpha=-2.1$ and $M_{\rm UV}^\star=-21.0$ mag referring to previous studies of UV LFs at $z>10$ \citep[e.g.,][]{Adams2024ApJ,Donnan2024,Robertson2024ApJ,Willott2024ApJ}.
Thus the model LF has two free parameters of $\phi_{\star,0}$ and $\eta$.
Although there are some evidence preferring the double-power law function over the Schechter function in high-$z$ UV LFs \citep[e.g.,][]{Harikane2022ApJS,Donnan2024}, we assume the Schechter function in this work because we do not have a constraint on the bright-end of LFs to distinguish the Schechter function and double-power law and the Schechter function has one fewer parameter.
We confirmed that all results in this paper do not change even if we adopt the double-power law function instead of the Schechter form.

We follow \citet{Marshall1983} for the LF parameter fitting, so that we can also take into account the non-detections in the relatively bright (brighter than $M_{\rm UV}\sim-19$ mag) galaxies at $z\gtrsim12$.
We derive the best estimations of $\phi_{\star,0}$ and $\eta$ to minimize the function $S=-2\ln \mathcal{L}$, where $\mathcal{L}$ is the likelihood and
\begin{dmath}
    \ln \mathcal{L} = \sum_{i}^{N} \left[ \phi_{\rm UV}(M_{{\rm UV}\ i}, z_i) p(M_{{\rm UV}\ i}, z_i) \right]  - \iint \phi_{\rm UV}(M_{\rm UV},z)p(M_{\rm UV}, z) \frac{dV}{dz}\ dz\ dM_{\rm UV}.
\end{dmath}
The first term is summed over all observed high-$z$ sample galaxies $i$, and the second term is the integral over all possible range of $M_{\rm UV}$ and $z$.
$p(M_{\rm UV},z)$ is the completeness at the point of $(M_{\rm UV},z)$, as estimated in Sec.~\ref{subsec:completeness}, and $dV/dz$ is the differential volume element at $z$. The combination of $p(M_{\rm UV},z)$ and $dV/dz$ can thus consider the effective sky area  of all three fields at $(M_{\rm UV},z)$, taking account for the completeness.
We use a python implementation of the MCMC \citep[\texttt{emcee;}][]{emcee} to obtain the posterior distributions of $\phi_{\star,0}$ and $\eta$, assuming flat priors between $-8<\log_{10}\phi_{\star,0} < -2$ and $-3<\eta<3$.
Table \ref{tab:LF_params} present the fitting results.
The best estimations are from the median of the marginalized posterior distribution, and the uncertainties are from 16th- and 84th-percentiles of the posterior.
We here stress again that the fit is based on all redshift ranges where the completeness has non-negligible value and also accounts for the non-detections of galaxies in the $M_{\rm UV}$-$z$ plane.

\begin{deluxetable}{lcc}
    \label{tab:LF_params}
    \tablecaption{Parameters of redshift-dependent Schechter UV LFs at $z>9.5$ inferred from CANUCS+TEC data.
  	}
    \tablewidth{\textwidth}
    \tablehead{
    \colhead{Parameter} & \colhead{Prior} & \colhead{Posterior}
    }
    \startdata
    $\log_{10}(\phi_{\star,0}/{\rm Mpc^{-3}\ mag^{-1}})$ & $\mathcal{U}(-8,-2)$ & $-4.89^{+0.17}_{-0.20}$ \\
    $\eta$ & $\mathcal{U}(-3,3)$ & $-0.21^{+0.11}_{-0.12}$ \\
    $\alpha$ & ... & $-2.1^*$\\
    $M_{\rm UV}^\star/{\rm mag}$ & ... & $-21.0^*$ \\
    \enddata
    \tablenotetext{}{\textbf{Notes.} The normalization parameter $\phi_\star$ is assumed to evolve depending on redshifts as $\log_{10}\phi_\star (z) = \log_{10}(\phi_{\star,0}) + \eta(z-10)$, and the fitting results for the two parameters $\eta$ and $\log_{10}(\phi_{\star,0})$ are given. Values with an asterisk are fixed in the fit.
    }
\end{deluxetable}

The log-linear rate of the UV LF redshift evolution $\eta$ obtained in this work ($\eta=-0.21^{+0.11}_{-0.12}$) is consistent with previous work in the JOF that used a similarly rich NIRCam MB+BB filter set \citep{Robertson2024ApJ}.
This means that the logarithmic galaxy abundance declines at a rate of $-0.21$ per unit redshift at $z\gtrsim10$; for example, the number density should decrease by a factor of $\sim7$ from $z\sim11$ to $z\sim15$.
On the other hand, the normalization parameter at the reference redshift $\phi_{\star,0}$ in this work is smaller than \cite{Robertson2024ApJ} by a factor of $\sim0.6$ dex (after correcting the difference in the reference redshift $z_0$ and degeneracy between $\phi_{\star,0}$ and $M_{\rm UV}^\star$).
This implies that cosmic variance (CV) can significantly affect galaxy number density measurement by at least $0.6$ dex.

\begin{figure*}[t]
\centering
\begin{minipage}{\columnwidth}
    \centering
    \includegraphics[width=\columnwidth]{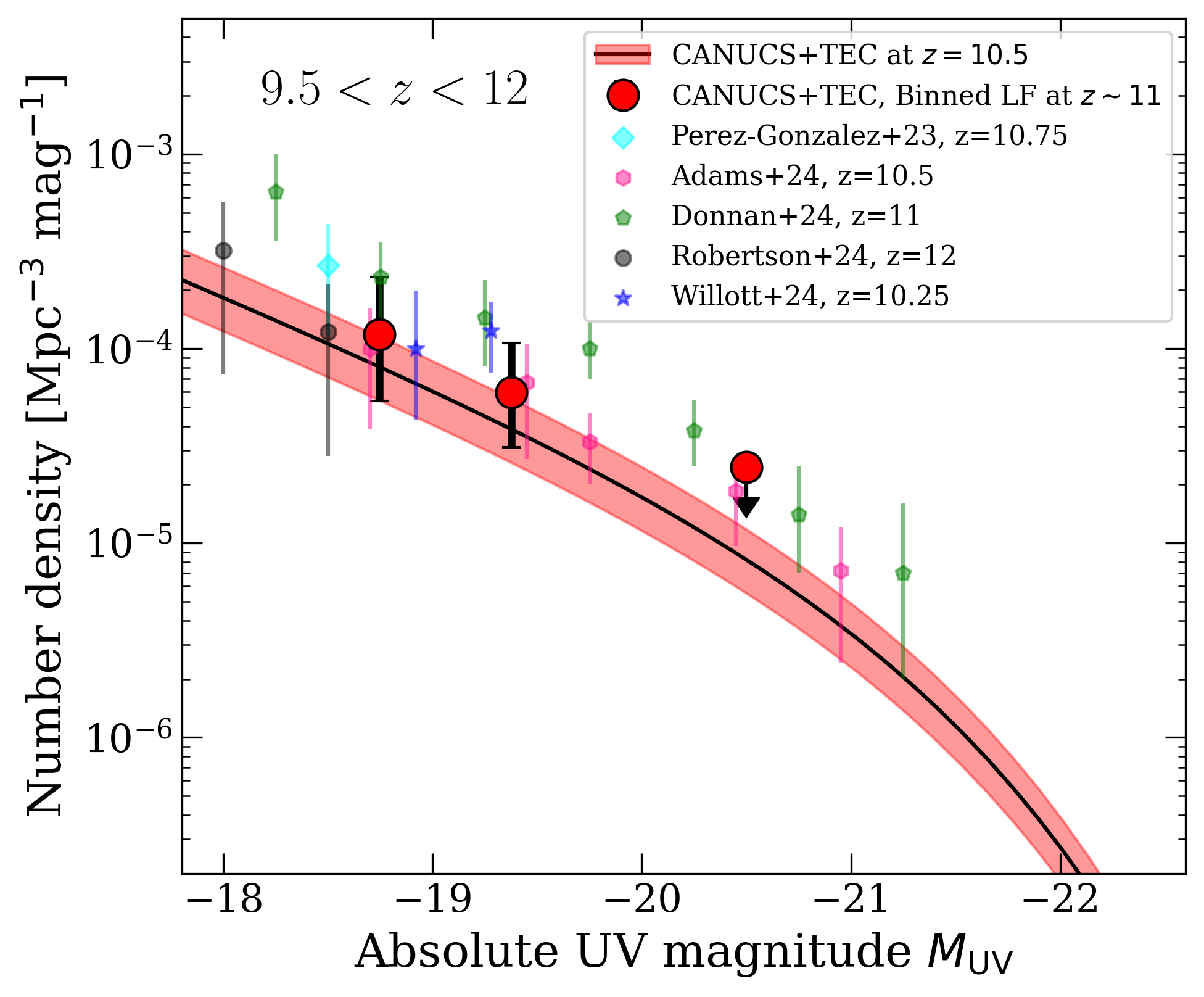}
\end{minipage}
\hspace{0.5cm}
\begin{minipage}{\columnwidth}
    \centering
    \includegraphics[width=\columnwidth]{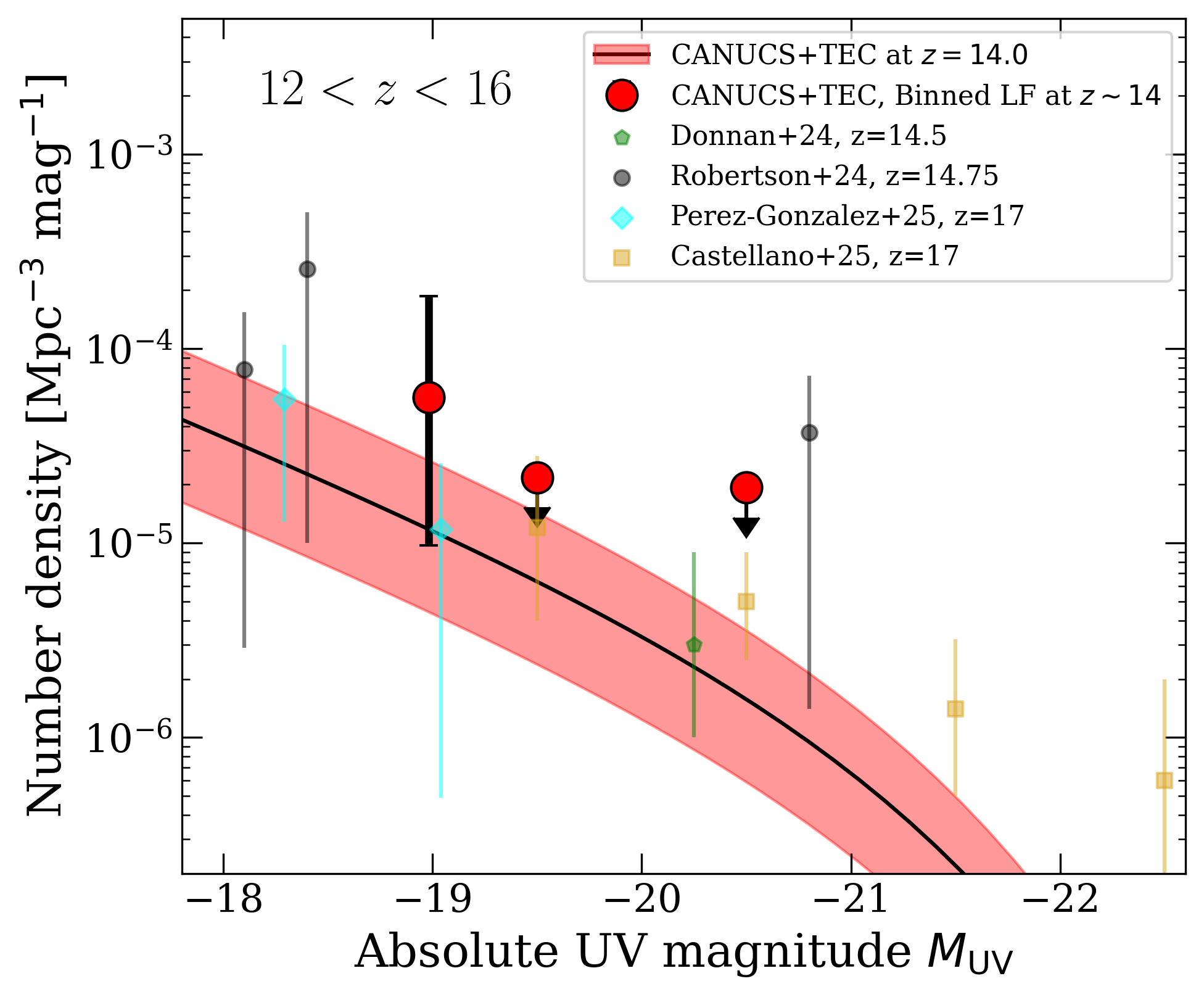}
\end{minipage}
\caption{The UV LFs at $z\sim11$ (left) and $z\sim14$ (right) from CANUCS+TEC data. Black solid curve is the best estimation of the redshift-evolving UV LF obtained in Sec.~\ref{subsec:LFparams} at $z=10.5$ (left) and $z=14$ (right), with the red-shaded area showing 16th- to 84th-percentile.
Red filled circles with black error bars present the binned UV LFs in the $9.5<z<12$ bin (left) and $12<z<16$ bin (right) measured in Sec.~\ref{subsec:binLF}.
UV LF measurements in literature at similar redshifts are also shown for comparison \citep[][]{Perez_Gonzalez2023, Adams2024ApJ, Donnan2024, Robertson2024ApJ, Willott2024ApJ, Perez_Gonzalez2025, Castellano2025arXiv}.
}
\label{fig:UV_LFs}
\end{figure*}


\subsection{Binned UV LFs}\label{subsec:binLF}
Although we estimate the evolving UV LFs without binning the sample galaxies in the previous section, deriving the binned LFs is still useful to compare with LF measurements in literature.
We use the binned $1/V_a$ method of \citet{Avni1980} and obtain the LFs at $9.5<z<12$ and $12<z<16$ separately.
The available effective volume for a galaxy $i$ in a redshift/magnitude bin with the size of $\Delta z$ and $\Delta M_{\rm UV}$ is
\begin{equation}
    V_a^i = \iint p(M_{\rm UV}, z) \frac{dV}{dz}\ dz\ dM_{\rm UV},
\end{equation}
considering the effect of selection completeness.
The LF in the bin is then computed as
\begin{equation}
    \Phi = \sum_i^{N} \frac{1}{V_a^i} \frac{1}{\Delta M_{\rm UV}}\ {\rm Mpc^{-3}\ mag^{-1}}.
\end{equation}
Uncertainties on the number density is based on Poisson statistics with the correction for small numbers by \citet{Gehrels1986}.

\begin{deluxetable}{cccc}
    \label{tab:binLF}
    \tablecaption{Binned UV LFs
  	}
    \tablewidth{240pt}
    \tablehead{
    \multicolumn{2}{c}{$M_{\rm UV}$ bin} & \colhead{$\langle M_{\rm UV} \rangle$} &
    \colhead{$\Phi$} \\
    \colhead{lower} & \colhead{upper} & \colhead{} & \colhead{$10^{-5}$ Mpc$^{-3}$ mag$^{-1}$}
    }
    \startdata
    \multicolumn{4}{c}{Redshift bin: $9.5<z<12$}\\
    \hline
    -21.0 & -20.0 & ... & $<2.46$ \\
    -20.0 & -19.0 & -19.38 & $5.97^{+4.72}_{-2.86}$ \\
    -19.0 & -18.0 & -18.75 & $11.8^{+11.5}_{-6.4}$ \\
    \hline
    \multicolumn{4}{c}{Redshift bin: $12<z<16$}\\
    \hline
    -21.0 & -20.0 & ... & $<1.93$ \\
    -20.0 & -19.0 & ... & $<2.18$ \\
    -19.0 & -18.0 & -18.98 & $5.63^{+12.96}_{-4.66}$ \\
    \enddata
    \tablenotetext{}{\textbf{Notes.} Upper limits are $1\sigma$.
    }
\end{deluxetable}

Table \ref{tab:binLF} and Figure \ref{fig:UV_LFs} presents the result.
We quote the $1\sigma$ upper limits when no galaxy is detected in a bin.
In the figure, we also show the marginalized LFs at $z=10.5$ and $z=14$ obtained from the MCMC sampling in Sec.~\ref{subsec:LFparams} (black solid curves with red shaded regions).
In the lower redshift bin at $z\sim11$ (left panel in Figure \ref{fig:UV_LFs}), our LF measurements are consistent with or at the lower envelope of other recent JWST studies \citep[][]{Perez_Gonzalez2023,Adams2024ApJ,Donnan2024,Robertson2024ApJ,Willott2024ApJ,Whitler2025arXiv}.
The spread of LF measurements across the literature is roughly $\sim0.6$ dex, which can be understood as the effect of the CV.
In the higher redshift bin at $z\sim14$ (right panel in the figure), although we have only one detection in the faintest magnitude bin, the non-detection of relatively bright ($M_{\rm UV}<-19$ mag) galaxies in the fairly large survey volume puts a tight constraint on the UV LFs, which results in a decline of galaxy abundance at $z>12$ ($\eta=-0.2$ as obtained in Sec.\ref{subsec:LFparams}).
We note that our LF measurement at $z\sim14$ is comparable to or even lower than recent reports of UV LFs at $z\sim17$ \citep{Castellano2025arXiv,Perez_Gonzalez2025}, which mostly used only NIRCam BB data to select high-$z$ galaxy candidates.
This is possibly due to the low-$z$ interloper contaminants when selecting high-$z$ candidates only based on BB-only catalogs, rather than the effect of CV or very shallow UV LF evolution at $z>15$ (see Sec.~\ref{subsubsec:BB_only_LFs}).

\subsection{Cosmic UV Luminosity Density}\label{subsec:UV_lum_dens}
\begin{figure*}[t]
\centering
\includegraphics[width=0.7\textwidth]{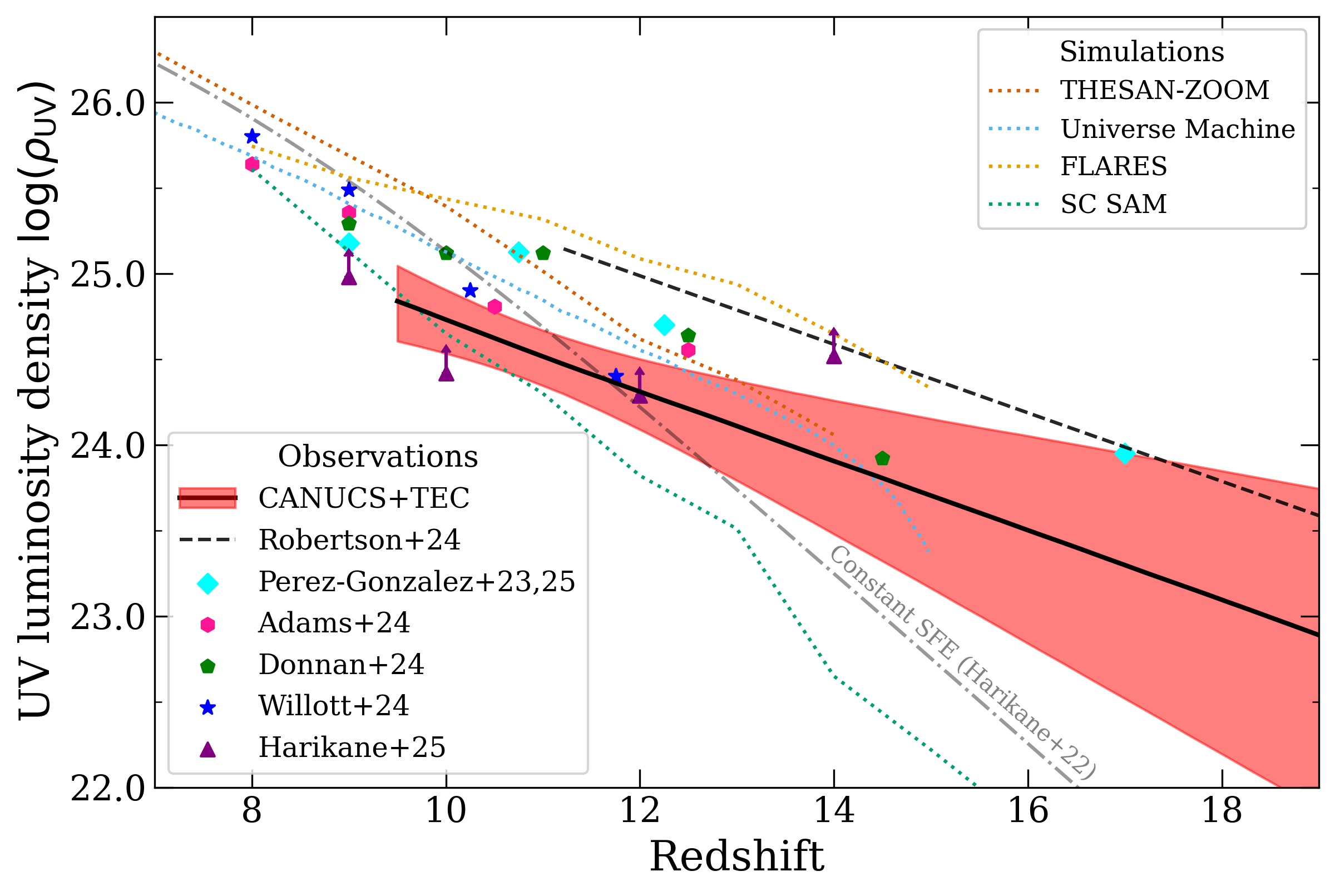}
\caption{The redshift evolution of cosmic UV luminosity density.
Black solid line is the median estimation from CANUCS+TEC data in this work, based on the redshift-evolving UV LF measurement in Sec.~\ref{subsec:LFparams}, and red shaded area shows the 16th- to 84th-percentile range.
Black dashed line is the $\rho_{\rm UV}$ measurement from JOF observation \citep{Robertson2024ApJ}, which uses similarly rich NIRCam MB+BB filters in high-$z$ galaxy selection.
For comparison, literature $\rho_{\rm UV}$ measurements from other JWST observations are shown filled small symbols \citep[][]{Perez_Gonzalez2023,Adams2024ApJ,Donnan2024,Willott2024ApJ,Harikane2025ApJ}.
Gray dash-dot curve marks the prediction of $\rho_{\rm UV}$ evolution assuming a constant star-formation efficiency at $z>6$ by \citet{Harikane2022}.
Dotted curves present predictions by several theoretical simulations (THESAN-ZOOM, \citealt{Kannan2025}; Universe Machine, \citealt{Behroozi2020}; FLARES, \citealt{Vijayan2021,Wilkins2023}; SC SAM, \citealt{Yung2024}).
The $\rho_{\rm UV}$ evolution slope in our work at $z\gtrsim10$ (black solid line) is somewhat less steep than previous measurements at $z\sim8$ and in a great agreement with that from JOF data, and starts to deviate from the constant SFE prediction above $z\sim11$.
}
\label{fig:rho_UVs}
\end{figure*}

Given the parameterized UV LF are measured as a function of redshift directly from the data in Sec.~\ref{subsec:LFparams}, we can directly compute the redshift evolution of UV luminosity density ($\rho_{\rm UV}(z)$) from the UV LF measurement.
We integrate the LF in the Schechter functional form (Eqn.~\ref{Eqn:schechter}) down to $M_{\rm UV}=-17$ mag, and obtain the marginalized $\rho_{\rm UV}(z)$ estimation from the MCMC sampling in Sec.~\ref{subsec:LFparams}.
Figure \ref{fig:rho_UVs} shows the median estimation (black solid) with the 16th-84th percentile (red shaded) from our results, comparing with the literature \citep[][]{Perez_Gonzalez2023,Perez_Gonzalez2025, Adams2024ApJ,Donnan2024,Robertson2024ApJ,Willott2024ApJ,Harikane2025ApJ}.
The overall $\rho_{\rm UV}$ value in this work is somewhat lower than most JWST studies, and in line with \citet{Adams2024ApJ} and \citet{Willott2024ApJ}.
Some literature claims that significant modifications of dust, IMF, or star formation efficiencies (SFEs) in model assumptions are required based on the overabundance of UV bright galaxies at $z>10$ \citep[e.g.,][]{Finkelstein2024ApJ,Harikane2025ApJ,Perez_Gonzalez2025}.
However, our result together with \citet{Adams2024ApJ} and \citet{Willott2024ApJ} should suggest that such significant modifications are not necessarily required.

The log-linear rate of the UV LF evolution $\eta$ is directly connected to the $\rho_{\rm UV}(z)$ redshift evolution, and our result indicates $\rho_{\rm UV}(z)$ decline as $d\log(\rho_{\rm UV})/dz=-0.21^{+0.11}_{-0.12}$ at $10<z<16$.
This is less steep than measurements at $z\sim8-11$ by \citet{Adams2024ApJ, Willott2024ApJ} and consistent with \citet{Robertson2024ApJ} at $z=11-16$.
Importantly, this is also less steep than the prediction by the simple constant extrapolation of the SFE from lower redshift \citep[e.g.,][the gray dash-dot line in Figure \ref{fig:rho_UVs}]{Harikane2022}.
These results suggest $\rho_{\rm UV}$ evolution gets shallower above $z\sim11$, and one of the possible interpretations is the SFE may start to deviate at $z>11$ from the simple constant extrapolation from lower redshift.

Figure \ref{fig:rho_UVs} also compares observations to predictions by several simulations: THESAN-ZOOM \citep{Kannan2025}, Universe Machine \citep{Behroozi2020}, FLARES \citep{Vijayan2021,Wilkins2023}, and SC SAM \citep{Yung2024}.
Simulations are well converged up to $z\sim8$ but their predictions spread at $z>8$. However, observational constraints are also scattered at $z>10$, and it is thus still difficult to decisively conclude one physical model is favored over others.
The large scatter at $z>10$ in observational data points seems not only due to the Poisson error on small sample size but also to the higher CVs and/or low-$z$ interloper contamination, which will be discussed further in the following section.


\section{What can bias our views of earliest galaxy evolution}\label{sec:discussion}
\subsection{Cosmic variance}\label{subsec:CV}
A component of observational uncertainty in measuring the high-$z$ galaxy abundance is the CV \citep[e.g.,][]{Steinhardt2021, Desprez2024, Willott2024ApJ, Jespersen2025ApJ}.
As shown in Sec.~\ref{subsec:LFparams} and \ref{subsec:UV_lum_dens}, UV LFs and $\rho_{\rm UV}$ measurements in our CANUCS+TEC fields are $\sim$0.6 dex lower than in the JOF by \citet{Robertson2024ApJ}, which can be interpreted as the effect of Poisson + CV uncertainties.

The effect of CV appears larger than nominal predictions based on semi-analytical simulations .
There are nine galaxies found at $11.5<z<16$ in the JOF, and the predicted uncertainty due to CV on the JOF number count is 23 \% (1-sigma), slightly smaller than the Poisson uncertainty \citep{Trenti2008}.
The combined fractional error would be 41 \%, which leads to $\sim0.2$ dex uncertainty in the number density estimation including UV LFs, though, our observation shows galaxy abundance measurements can actually be affected by as much as 0.6 dex at this high redshift.
Galaxy number statistics at high redshift can thus be considerably affected by field-to-field variance, more severely than the nominal semi-analytical models' predictions, which is presumably due to bursty star-formation \citep[e.g.,][]{Sun2023ApJ} or non-universal IMFs \citep[e.g.,][]{Yung2024} in the early universe that can enhance the field-to-field variance.
Therefore, building a larger sample of robust high-$z$ galaxies over wide areas in multiple lines of sight is essential to ascertain the earliest galaxy evolution at $z>10$.

\subsection{Low-$z$ interlopers when selecting $z\gtrsim10$ galaxies only with Broad Bands}\label{subsubsec:BB_only_LFs}
Another potential bias is due to low-$z$ interlopers, particularly when selecting $z\gtrsim10$ galaxies based on BB-only data.
To explore this effect, we make a degraded photometry catalog with all MB filters removed except for F410M (hereafter, the BB-only catalog). This catalog contains the HST/ACS F435W, F606W, F814W, and JWST/NIRCam F070W, F090W, F115W, F150W, F200W, F277W, F356W, F410M and F444W filters, and we compare our original results with this catalog.
We run \texttt{EAzY} on this BB-only catalog with exactly the same configuration as in Sec.~\ref{subsec:photoz}, and select high-$z$ galaxies from the same criteria (\ref{eqn:selection}).

The BB-only catalog selection high-$z$ sample consists of 10 galaxies, four of them are common with our main high-$z$ sample from the full MB+BB photometry (Sec.~\ref{sec:selection}), while six of them are not.
The six galaxies are selected as high-$z$ when only BB filter photometry is available, but turn out to be lower-$z$ with the full MB+BB data ("False Positives").
The False Positives, except for CANUCS-2225353, appear as good-looking dropouts with negligible low-$z$ solution probability (less than $\sim$5 \%) when the BB-only catalog is used, so it seems inevitable to select them as high-$z$ from BB-only selection (see Appendix \ref{apx:low-z} for details of the False Positives).
CANUCS-2225353 does pass all selection criteria using the BB-only catalog and is selected as a $z\sim16$ candidate with F200W-dropout, but the source is also detected in F150W (which is not included as a "blue" filter in the criteria). In the following computation of the BB-only UV LF, we therefore exclude CANUCS-2225353 and use the remaining nine galaxies as the \textit{high-$z$ sample} from the BB-only catalog.
The comparison of the BB-only selection high-$z$ sample with our main high-$z$ sample indicates that as much as half of $z\gtrsim10$ galaxy candidates could actually be low-$z$ interlopers when only BB photometry is available.
Of course this fraction largely depends on the selection criteria, data quality (e.g. depth), or observation line of sights (not only clustering of $z>10$ galaxies but also lower-$z$ interloper clustering can affect this fraction), but this fraction is roughly consistent with simulations using mock galaxy catalogs by \citet{Adams2025}.
Thus the low-$z$ interloper contamination is not negligible, particularly at $z\gtrsim10$ when only BB photometry data is available.

The main population of low-$z$ interlopers in the BB-only selection high-$z$ sample is (very) dusty extreme emission line galaxies (EELGs) at $z>2$.
They typically show large emission line excesses due to [O{\sc iii}]4959/5007+H$\beta$ and H$\alpha$ lines on a red continuum at rest optical wavelengths, while being completely undetected in the rest UV at $\lambda_{\rm rest}\lesssim3000\ {\rm \AA}$, which makes their BB SEDs look as if they are $z>10$ galaxies (see; e.g., Figure \ref{fig:2217931}, \ref{fig:2225353}, \ref{fig:5217969}, and \ref{fig:5219487} in Appendix \ref{apx:low-z}).
Their rest UV light is tremendously attenuated by dust, while the nebular emission lines are not and reaching the equivalent width of ${\rm EW_{0}\sim2000}\ {\rm \AA}$.
Some of these extremely reddened EELGs are spectroscopically confirmed in their rest-optical emission lines \citep[e.g.,][]{Withers2023,Bisigello2025AA}, but their rest UV properties are almost unknown and there is no good template to replicate the extreme population.
The lack of good templates results in poor fits with low-$z$ solutions but rather the SED is well-fitted by blue, high-$z$ solutions, when only BB data are available.
However with MB observations, the emission lines are clearly identified by the MB flux excesses which strongly supports the low-$z$ solution, and the observed SED is best-fitted with a low-$z$ template although the rest UV non-detection is poorly fitted.

\begin{figure}[t]
\centering
\includegraphics[width=0.95\linewidth]{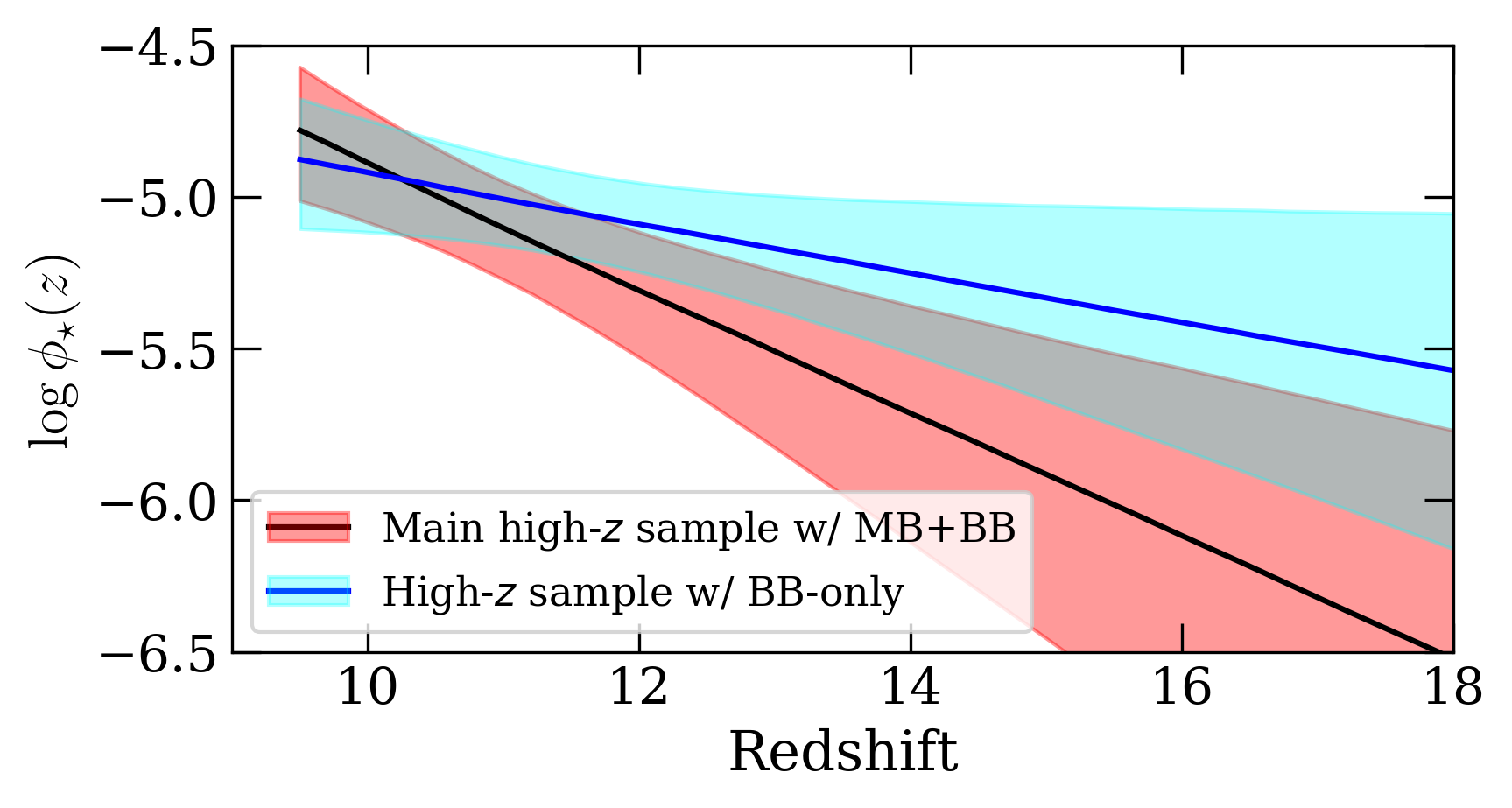}
\caption{The effect of low-$z$ interlopers on UV LF measurement. Compared is the redshift evolution of $\phi_\star(z)$, obtained from the main high-$z$ sample selected with the full MB+BB (red) and from high-$z$ candidates based on BB-only degraded catalog (blue). When building high-$z$ galaxy sample from BB-only catalog, the low-$z$ interlopers can result in $\sim0.6$ dex overestimation of galaxy abundance at $z\gtrsim10$.}
\label{fig:phistar_comp}
\end{figure}

We then investigate the effect of the low-$z$ interloper contaminants on the UV LF measurements.
We follow the same procedure as Sec.~\ref{sec:result} on the BB-only catalog selection high-$z$ sample and derive the best estimate UV LF redshift evolution from it.
In this BB-only UV LF measurement, we also perform the simulation as we do in Sec.~\ref{subsec:completeness} with removing the MBs to obtain the completeness of the BB-only catalog selection of $9.5<z<16$ galaxies.
The completeness gets lower in general with the BB-only catalog selection as compared to the standard high-$z$ selection with MB+BBs.

Figure \ref{fig:phistar_comp} compares the estimated $\phi_\star(z)$ evolution from the main high-$z$ galaxy sample (Sec.~\ref{subsec:LFparams}) and that from the BB-only UV LF measurement.
The overestimation of the UV LF normalization parameter is severe at higher redshift and it reaches up to $\sim0.6$ dex at $z\sim15$. The overestimation of $\phi_\star(z)$ similarly affects the $\rho_{\rm UV}(z)$ measurement.
The effect of low-$z$ interlopers is particularly critical at $z\gtrsim15$, because the galaxy abundance at $z\gtrsim15$ becomes very low while the corresponding lower-$z$ range of the interlopers enters intermediate-$z$ range of $z\sim4$, where extreme line emitters start to commonly populate \citep[e.g.,][]{Withers2023, Eisenstein2023arXiv}.
Indeed, the two brightest galaxy candidates in the BB-only catalog selection are both False Positive misidentified as $z\sim16$ galaxies brighter than $M_{\rm UV}=-20$ mag, similar to CEERS-9331 \citep{Arrabal_Haro2023}.
Low-$z$ interlopers can thus bias the UV LF and UV luminosity density measurements by $\sim0.6$ dex in BB-only studies.
The brighter magnitude bins at higher-$z$ are more likely to be contaminated, and thus LF bright-end studies at $z\gtrsim10$ need particular caution.

\subsection{Effects of interlopers on other galaxy statistics}
The considerable contamination by low-$z$ interlopers in the $z\gtrsim10$ galaxy sample from the BB-only catalog raises caution not only in UV LFs but also in other statistical studies of earliest galaxy properties.
In the following, we compare the False Positives and the robust high-$z$ galaxies identified with MB+BB observations in several aspects to reveal the potential effects of low-$z$ interloper contamination in high-$z$ galaxy studies at $z\gtrsim10$.
We here also utilize the JOF $z>10$ galaxy sample \citep{Robertson2024ApJ} as an additional robust high-$z$ galaxy sample, since they select the sample based on similarly rich NIRCam MB+BB filters. The median redshift of CANUCS+TEC+JOF sample is $\langle z \rangle=11.5$.

\subsubsection{Size-$M_{\rm UV}$ relation}
\label{subsubsec:size-muv}

\begin{figure}[t]
\centering
\includegraphics[width=0.95\linewidth]{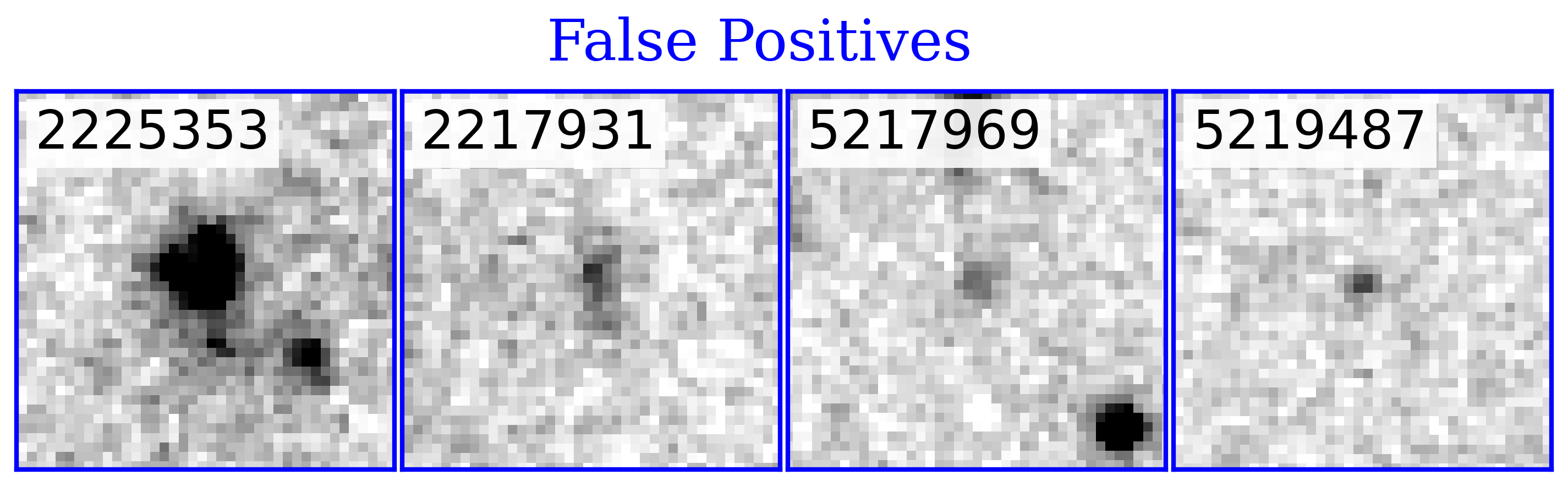}
\includegraphics[width=0.95\linewidth]{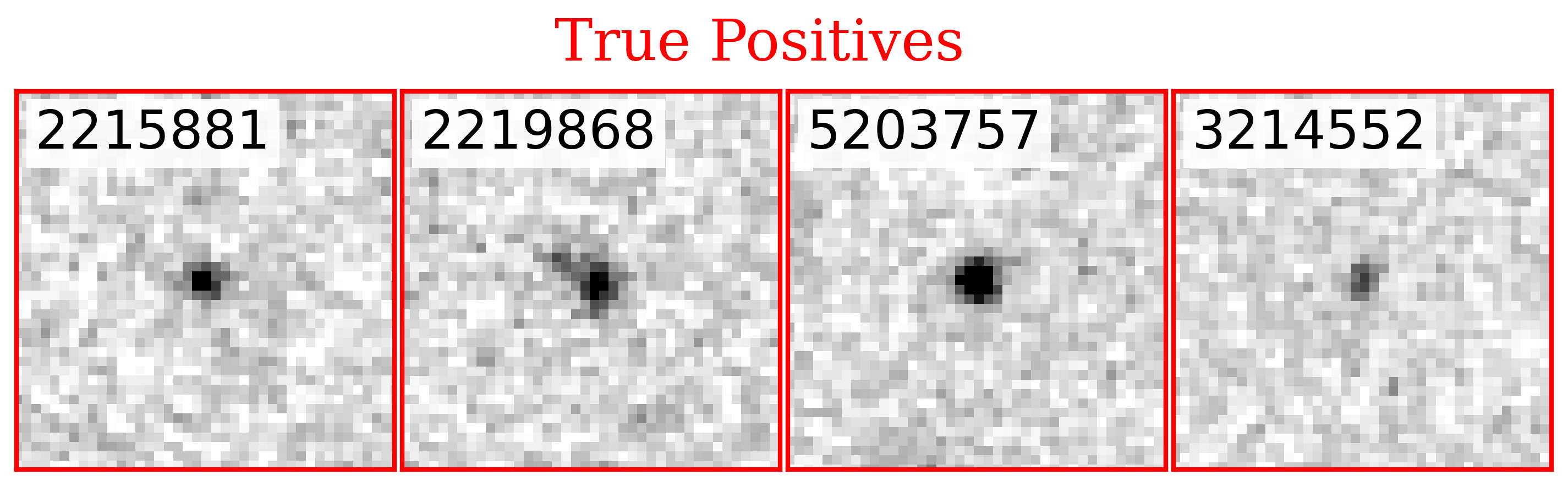}
\includegraphics[width=0.95\linewidth]{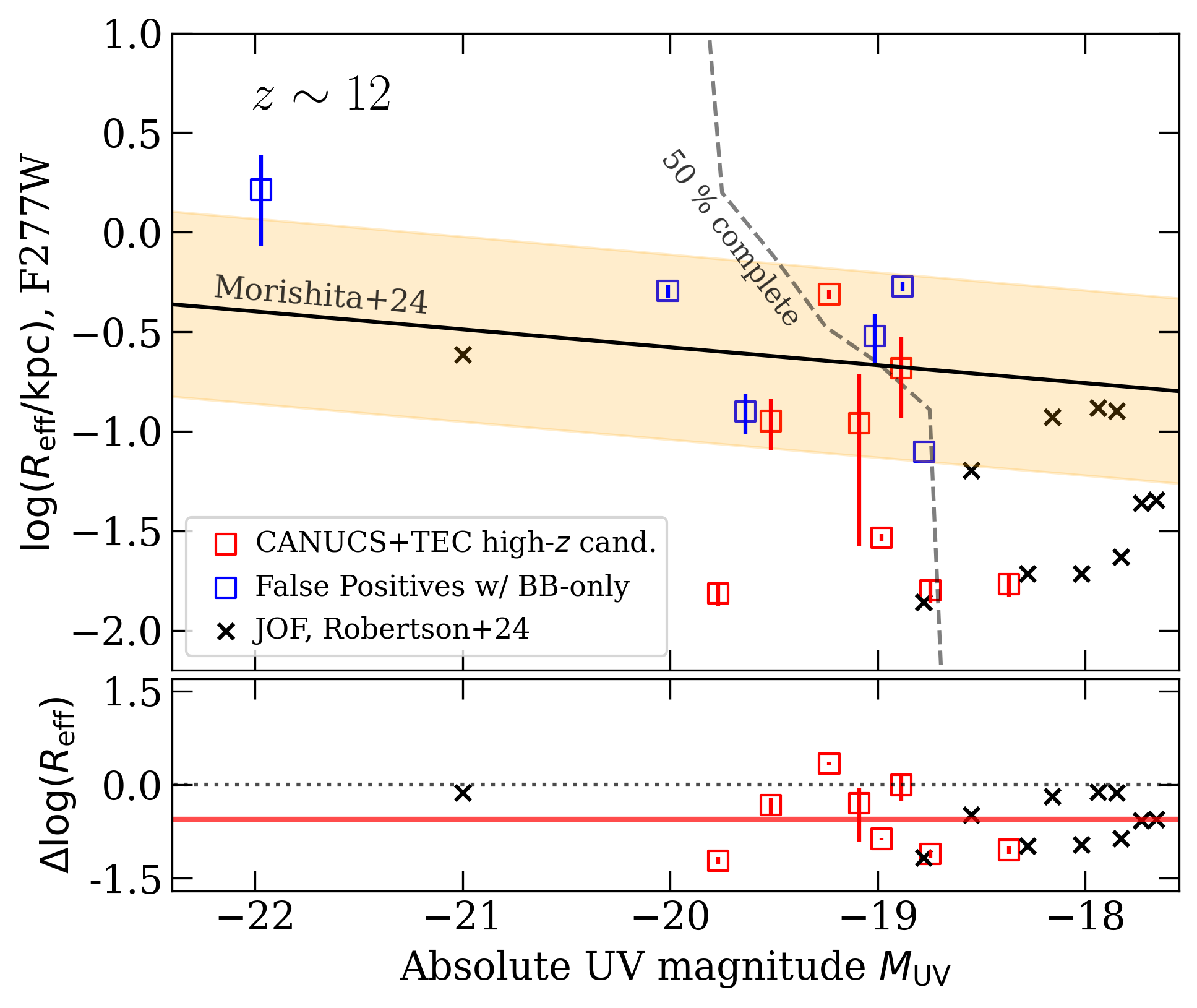}
\caption{\textit{Top}: cutouts of four examples of False Positives in the F277W filter. They are $1.\!\!^{\prime\prime}6$ on the side.
\textit{Middle}: cutouts of four True Positives in the F277W filter.
\textit{Bottom}: distribution of high-$z$ galaxy candidates in the size-$M_{\rm UV}$ plane.
High-$z$ candidates selected with extensive MB+BB filters (red squares, this work; black crosses, JOF) are systematically smaller than a previous calibration of the size-$M_{\rm UV}$ relation at $z\sim12$ by \citet[black solid line]{Morishita2024}.
On the other hand, the False Positives (blue squares) appear systematically larger than the main high-$z$ and JOF candidates and would align well with the size-$M_{\rm UV}$ relation by \citet{Morishita2024}.
The bottom sub-panel shows the systematic offset in $\log(R_{\rm eff})$ of the main high-$z$ candidates from the predicted size by \citet{Morishita2024} relation.
The median offset is $-0.6$ dex (red line).
}
\label{fig:size_mag}
\end{figure}

The size-$M_{\rm UV}$ relation is one of the key scaling relations of galaxies and it is an \textit{a priori} assumption in deriving the completeness for the galaxy luminosity function.
Figure \ref{fig:size_mag} shows the distribution of the robust high-$z$ galaxy sample from MB+BB data in the size-$M_{\rm UV}$ plane (red squares and black crosses; main sample in this work and from \citealt{Robertson2024ApJ}, respectively).
The robust high-$z$ galaxy sample distribution is systematically smaller than the size-$M_{\rm UV}$ relation at $11<z<13$ previously calibrated by \citet[black line in the figure; the shaded area presents the 1.5-sigma scatter]{Morishita2024}.
This systematic offset cannot be fully explained solely by incompleteness: our sample is $50 \%$ complete up to $\log(R_{\rm eff}/{\rm kpc})\sim-0.5$ at $M_{\rm UV}=-19.0$ mag (gray dashed line in the figure), which corresponds to the size-$M_{\rm UV}$ relation median of \citet{Morishita2024}, while almost all the robust high-$z$ galaxy sample lies well below the \citet{Morishita2024} relation.
The systematic offset is $\Delta \log(R_{\rm eff})=-0.6$ dex (red line in the bottom panel).
On the other hand, if we measure the size and absolute UV magnitude of the False Positives based on the BB-only photometry catalog and photo-$z$ estimation from the BB-only data, they distribute very well along the previously calibrated relation (blue squares).
Figure \ref{fig:size_mag} thus demonstrates that the False Positives typically appear larger and brighter than the real high-$z$ galaxies, and the size-$M_{\rm UV}$ relation can be biased towards larger sizes when selecting high-$z$ candidates based on BB-only data.
In the completeness simulation (Sec.~\ref{subsec:completeness}), we thus assume a modified size-$M_{\rm UV}$ relation of \citet{Morishita2024} offset 0.6 dex smaller. The good agreement between the estimated completeness with this modification and the observed high-$z$ galaxies in the $M_{\rm UV}$-$z$ plane validates this offset.

\subsubsection{Rest UV slopes}\label{subsubsec:UV_slope}
Previous JWST studies have found rest-UV slopes, $\beta_{\rm UV}$, are blue at the highest redshifts \citep[e.g.,][]{Cullen2024MNRAS,Topping2024}, and some of them suggest galaxies at $z>10$ are remarkably blue and could be completely dust-free \citep{Cullen2024MNRAS}.
On the other hand, galaxies selected from our CANUCS+TEC data and from JOF \citep{Robertson2024ApJ} are typically less blue ($\langle \beta_{\rm UV}\rangle=-2.2$ in CANUCS+TEC and $\langle \beta_{\rm UV}\rangle=-2.5$ in JOF; see also \citealt{Martis2025arXiv} for the high-$z$ red galaxy population in CANUCS+TEC fields).
This could indicate that rest-UV slope studies can also be biased in some cases with BB-only data due to bluer galaxies being easier to distinguish from low-$z$ interlopers and to fewer filters to compute $\beta_{\rm UV}$.

\begin{figure}[t]
\centering
\includegraphics[width=0.95\linewidth]{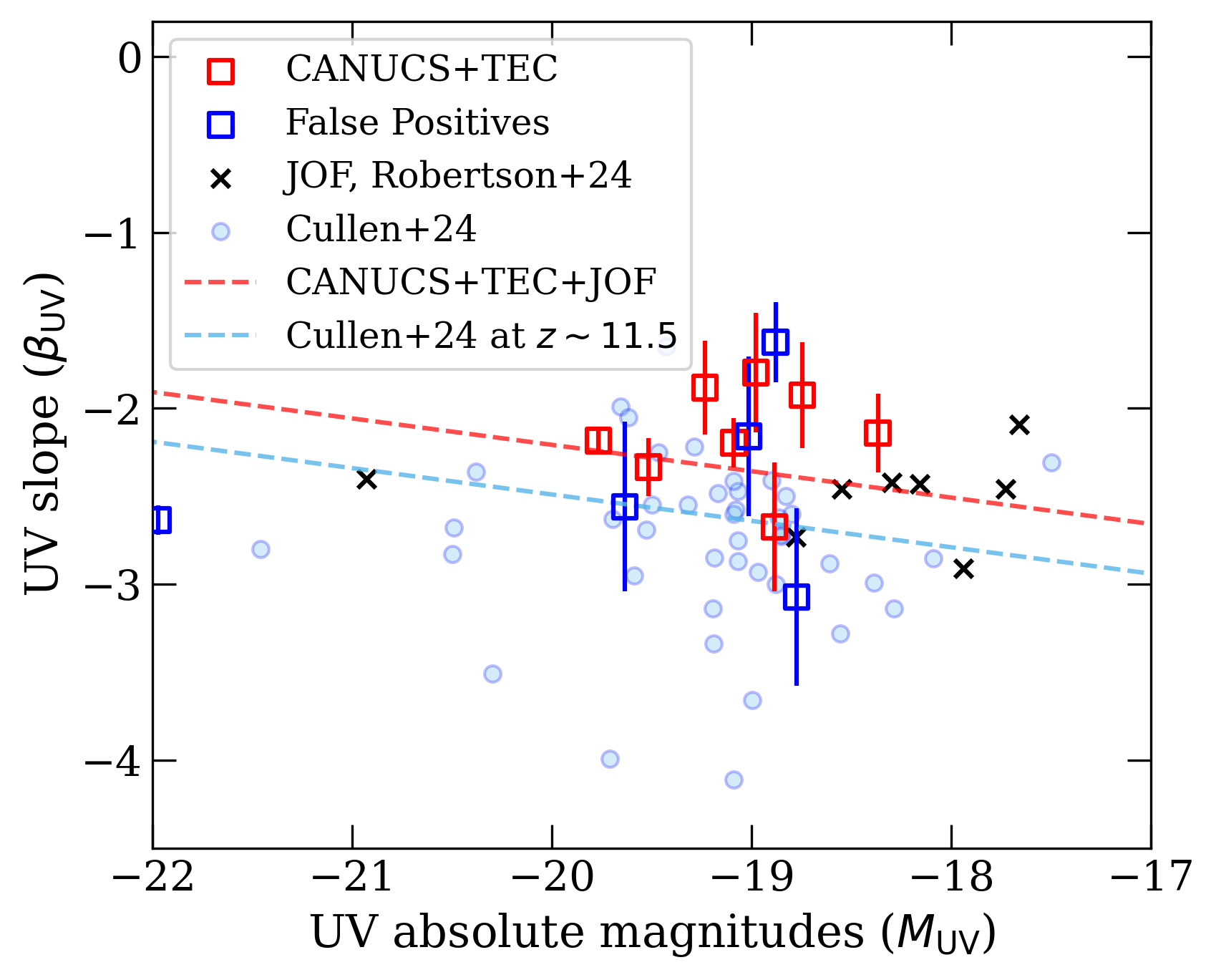}
\caption{The $\beta_{\rm UV}$-$M_{\rm UV}$ relation. 
High-$z$ galaxy candidates based on MB+BB observations (red squares and black crosses) are less blue than those of a comparison sample mostly based on BB-only data \citep[cyan circles]{Cullen2024MNRAS}. In particular, there is a population of faint ($M_{\rm UV}\gtrsim-19$), relatively red ($\beta_{\rm UV}\gtrsim-2.5$) galaxies that is largely absent from the \citet{Cullen2024MNRAS} sample. False Positives (blue squares) cover most of the observed range of $\beta_{\rm UV}$.
The linear regression of the $\beta_{\rm UV}$-$M_{\rm UV}$ relation based on the main MB+BB selection sample (red dashed line) is less blue than found previously (cyan dashed line).
}
\label{fig:Muv_beta}
\end{figure}

Figure \ref{fig:Muv_beta} shows the distribution of the robust high-$z$ sample (CANUCS+TEC, red squares; JOF, black crosses) in the $\beta_{\rm UV}$-$M_{\rm UV}$ plane, compared to the $z\sim11.5$ galaxy sample from \citet{Cullen2024MNRAS} (light blue circles), which mostly used BB-only photometry data in the sample selection.
Interestingly, the high-$z$ galaxies selected with MB+BB data distributes in the relatively less blue and fainter locus in the $\beta_{\rm UV}$-$M_{\rm UV}$ plane ($M_{\rm UV}\gtrsim-19$ mag and $\beta_{\rm UV}\gtrsim-2.5$), and this population is absent in the \citet{Cullen2024MNRAS} sample.
This can be because relatively red high-$z$ galaxies close to the detection limit of the observation can be degenerate with the low-$z$ Balmer break galaxy solution, particularly when only BB data is available, and they could be missed from high-$z$ selection with BB-only data.
A representative example is the relatively red $z\sim15.4$ galaxy candidate CANUCS-3214552 (Sec.~\ref{subsec:CANUCS_z15_0}), whose low-$z$ probability is substantial ($P_{\rm BB}(z<7)=0.13$) when only BB data is used, but negligible with the full MB+BB data ($P(z<7)=0.005$).
Therefore, BB-only selection of high-$z$ galaxy candidates can be relatively incomplete for faint and less blue galaxies.

On the contrary, some False Positives can mimic very blue high-$z$ SEDs with $\beta_{\rm UV}<-2.5$.
Blue squares in Figure \ref{fig:Muv_beta} show the locations of False Positives in this diagram, where $\beta_{\rm UV}$ is computed from the BB-only photometry catalog, assuming the photo-$z$ from the BB-only data.
This could suggest that low-$z$ interlopers can bias the sample even bluer when only BB data is available.

Combining these two potential effects, incompleteness for faint and less blue galaxies and contamination of low-$z$ interlopers that could mimic very blue colors, the $\beta_{\rm UV}$-$M_{\rm UV}$ relation can thus be biased bluer with the BB-only data.
When only robust MB+BB selection candidates are used (red square and black crosses in Figure \ref{fig:Muv_beta}), the linear regression of the $\beta_{\rm UV}$-$M_{\rm UV}$ relation becomes
\begin{equation}
    \beta_{\rm UV} = -0.15(M_{\rm UV}+19) - (2.36\pm0.04),
\end{equation}
fixing the slope $d\beta_{\rm UV}/dM_{\rm UV}=-0.15$ (red dashed line in Figure \ref{fig:Muv_beta}).
The intercept of $\beta_{\rm UV}(M_{\rm UV}=-19) = - 2.36\pm0.04$ at $\langle z \rangle=11.5$ is redder by $\Delta \beta\sim0.3$ than \citet[cyan dashed line]{Cullen2024MNRAS} at this redshift.


\subsection{Future galaxy surveys at $z\gtrsim10$}\label{subsec:future}

By exploiting the robust high-$z$ candidates selected with rich MB+BB data, we demonstrate the importance of MBs in studies of $z\gtrsim10$ galaxies. However, the number of known galaxies is still small and enlarging the sample size is necessary.
The bright-end at $z\gtrsim14$ needs particular caveats, but bright $z\gtrsim14$ galaxy candidates are not always interlopers (e.g., GS-z14-0, \citealt{Carniani2024,Hainline2024}; MoM-z14, \citealt{Naidu2025}).
Securing a high-$z$ bright galaxy sample is crucial in various aspects (e.g., bright-end of LFs, scaling relations such as size-$M_{\rm UV}$ relation), and the effect of CV can be considerable in such a rare population search (Sec.~\ref{subsec:CV}).

\begin{figure}[t]
\centering
\includegraphics[width=0.95\linewidth]{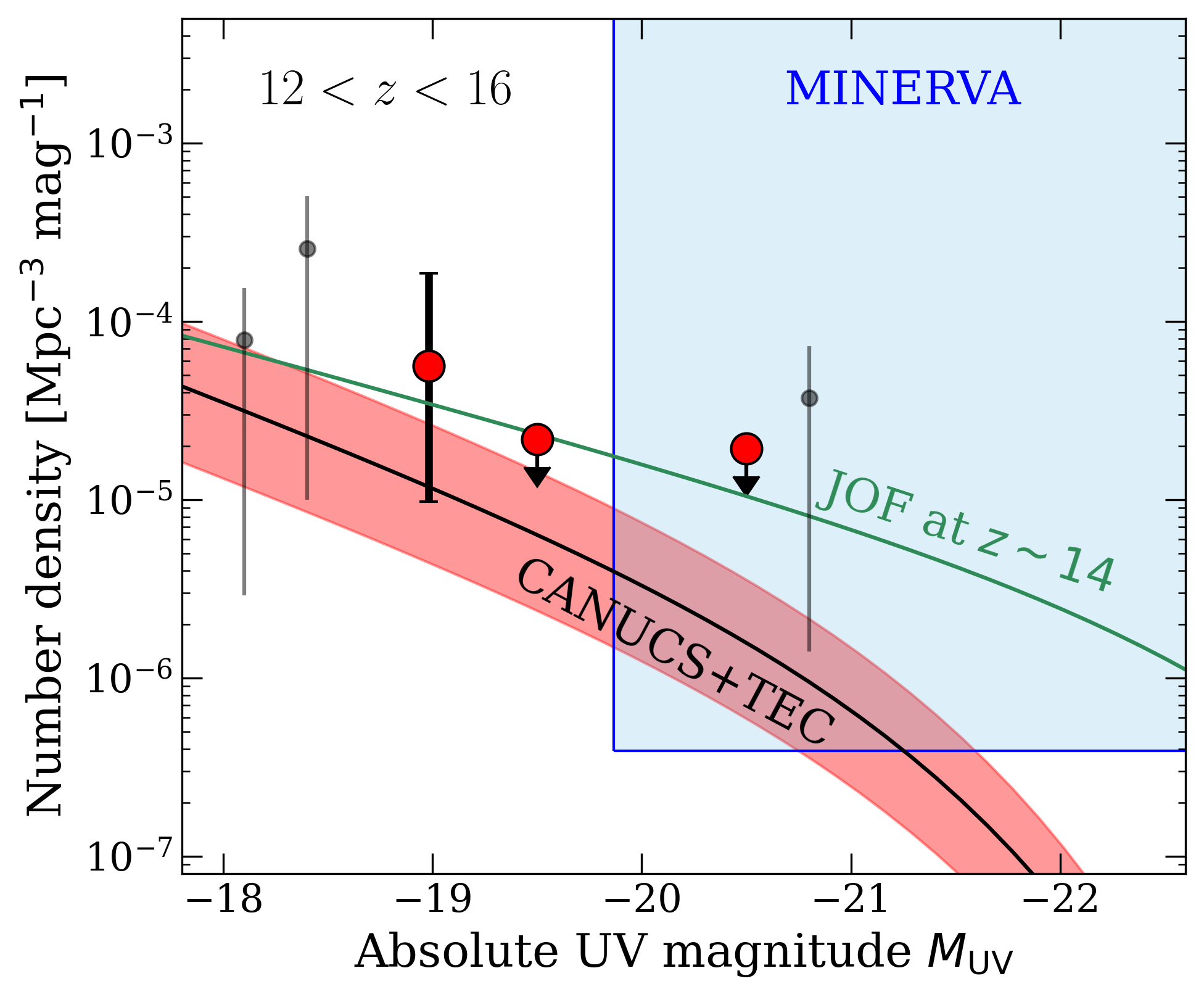}
\caption{UV LFs at $z\sim14$ from the two surveys with extensive MB+BB filter observations (CANUCS+TEC, this work; JOF, \citealt{Robertson2024ApJ}), compared with the MINERVA survey detectability of $z\sim14$ galaxies.
The two UV LF measurements agree at the faint end ($M_{\rm UV}>-19$) but show a large discrepancy at the bright end. The JWST Cycle 4 MINERVA survey will provide $\sim500$ arcmin$^{2}$ with almost all MB+BB filters across four independent line of sights, and will give a tight constraint at the bright end of the highest-$z$ UV LF by building a very robust bright galaxy sample at $z\gtrsim14$ (blue square area).
}
\label{fig:UVLF_w_MINERVA}
\end{figure}

Figure \ref{fig:UVLF_w_MINERVA} compares the UV LF measurements at $z\sim14$ from this work (black) and from JOF \citep[green;][]{Robertson2024ApJ}, both are based on MB+BB observations.
Although they agree with each other at the faint-end, the bright-end of UV LFs is poorly constrained due to their paucity and the two measurements are largely discrepant at $M_{\rm UV}\lesssim-20$ mag.
The large discrepant at the bright-end is mainly due to the presence of one bright galaxy, GS-z14-0, in the small footprint of JOF. The galaxy is thus far the brightest galaxy at this redshift range even across the all JADES field-of-views that covers $\sim\times15$ area of the JOF, and it can significantly affect the number density of bright galaxies in JOF.
It is thus required to search for these bright high-$z$ galaxies with MB+BBs over wide areas in independent lines of sight, mitigating the effect of field-to-field variance.

For example, the MINERVA survey (GO-7814, PI: Muzzin) is scheduled to observe with most MB filters over $\sim500$ arcmin$^2$ across four independent fields previously observed with the BB set of filters.
These observations will give extensive MB+BB filter set observations with $\sim28$ mag depth over the wide field, and will be able to probe the bright-end of highest-$z$ UV LFs (blue square in Figure \ref{fig:UVLF_w_MINERVA}).
The observations are expected to find $N\sim4$ (in a pessimistic case; the CANUCS+TEC LF) to $N\sim10$ (optimistic case; the JOF LF) galaxies at this redshift brighter than $M_{\rm UV}\lesssim-20$ mag, and will give an invaluable sample of bright, high-$z$ galaxies selected with MB+BB observations, like GS-z14-0.

\section{Summary}\label{sec:summary}
We use the NIRCam Medium-band + Broad-band observations in three independent line-of-sight fields from the combination of the CANUCS program in Cycle 1 and the Technicolor program (TEC) in Cycle 2, to build a robust sample of galaxies at $z>9.5$ and study the early evolution of the UV LF at $10\lesssim z\lesssim16$.
The combination of CANUCS+TEC NIRCam observations provides the full set of MB+BB filter images over $\sim23\ {\rm arcmin}^2$ total unmasked area reaching $\sim29$ mag detection limit. 
The extensive NIRCam data with all MB+BB filters is useful both in removing low-$z$ interlopers and in selecting faint, relatively red, high-$z$ galaxies. Having MB+BB observations is thus in particular crucial to obtain a robust sample of galaxy candidates at $z>9.5$, where all strong rest-optical emission lines are redshifted out of the NIRCam wavelength coverage and the Lyman-$\alpha$ continuum break is the only accessible SED feature.
We also discuss the potential bias in several aspects of galaxy statistics at $z\gtrsim10$ due to low-$z$ interlopers when building the sample from BB-only data, by generating a degraded photometry catalog with all MB filters removed from our data and comparing to our main high-$z$ galaxy sample obtained based on the full MB+BB observations.

Our main findings are as follows:
\begin{enumerate}
    \item We obtain a sample of eight galaxies at $9.5<z<16$ with the selection 50~\% complete down to $M_{\rm UV}\sim-18.7$ mag over the full redshift range (Figure \ref{fig:Completeness}). 
    The UV LF from CANUCS+TEC data shows a moderate redshift evolution at $z>10$. The lack of relatively bright galaxies (brighter than $M_{\rm UV}=-20$ mag) and the paucity of $z>12$ galaxies over $\sim23\ {\rm arcmin}^2$ places tight constraints on the redshift evolution, and the redshift-dependent UV LF normalization parameter is found to be $\log\phi_\star(z) = -4.89 - 0.21(z-10)$ (Table \ref{tab:LF_params} and Figure \ref{fig:UV_LFs}). 
    Our UV LF at $10<z<16$ is $\sim0.6$ dex lower than a previous work in the JADES Origins Field (JOF), where similarly rich NIRCam MB+BB filters are available, suggesting that the cosmic variance can affect galaxy abundance measurement by at least 0.6 dex in the very high-$z$ universe (Sec.~\ref{subsec:LFparams}, \ref{subsec:binLF}).
    
    \item The highest redshift candidate in our sample (CANUCS-3214552) is found at $z_{\rm phot}\sim15.4$ with a sharp dropout between F210M and F200W (Figure \ref{fig:3214552}). The galaxy CANUCS-3214552 is faint ($M_{\rm UV}=-19$ mag) and relatively red ($\beta=-1.8$), and, due to the relatively red color, it could not be identified as a reliable high-$z$ candidate only with the BB filter data.
    Also it shows no variability between Cycle 1 and Cycle 2 NIRCam observations, thus CANUCS-3214552 is one of the most reliable candidates of $z>15$ galaxies, which can be secured only with multi-epoch, full NIRCam MB+BB observations (Sec.~\ref{sec:selection}). It is absolutely necessary to follow-up this galaxy with NIRSpec Prism to confirm its redshift and probe the detailed physical properties.
    
    \item The UV luminosity density $\rho_{\rm UV}$ measured with CANUCS+TEC observations is at the lower envelope of recent JWST studies and suggests significant modifications of dust, IMF, or star formation efficiencies are not necessarily needed in model assumptions in cosmological simulations (Sec.~\ref{subsec:UV_lum_dens}).
    \item The log-linear rate of the $\rho_{\rm UV}$ redshift evolution at $z\sim10-16$ from this work is $-0.21$. This is somewhat shallower than at lower redshift of $z\sim8-10$ and is consistent with the previous work at the similar redshift in the JOF. Comparing with a prediction of $\rho_{\rm UV}$ evolution assuming the constant star formation efficiency (SFE), it could suggest the SFE starts to deviate from the simple constant extrapolation at $z\gtrsim11$ (Figure \ref{fig:rho_UVs} and Sec.~\ref{subsec:UV_lum_dens}).
    \item By comparing high-$z$ galaxy candidates from the BB-only degraded catalog and from the full MB+BB catalog, the low-$z$ interloper contamination is found to be non-negligible when selecting $z>9.5$ galaxies based only on the BB data.
    The major population of interlopers is very dusty EELGs that starts to populate at intermediate redshift of $z>2$, and roughly a half of $z>9.5$ galaxy candidates from BB-only photometry could be low-$z$ interlopers (Sec.~\ref{sec:discussion}).

    \item The most clean galaxy samples at $z\gtrsim10$ identified from the rich MB+BB observations suggest that highest-$z$ galaxies could be rarer, more compact, and less blue than previously claimed:
    the UV LF normalization could be overestimated by $\sim0.6$ dex when BB-only data is used (Figure \ref{fig:phistar_comp});
    the low-$z$ interlopers appears systematically larger than real high-$z$ sources, leading to an overestimation of the size-$M_{\rm UV}$ relation by $\sim0.6$ dex (Figure \ref{fig:size_mag});
    the main high-$z$ galaxy sample from the MB+BB selection is less blue while low-$z$ interlopers can masquerade very blue color in the BB-only SEDs, resulting in a systematically bluer estimation of the $\beta_{\rm UV}-M_{\rm UV}$ relation by $\Delta\beta\sim-0.3$ (Figure \ref{fig:Muv_beta}).
    \item Noteworthy, adding MB observations is effective not only in removing low-$z$ interlopers but also in identifying real high-$z$ candidates that are missed with BB-only selections. Selecting $z\gtrsim10$ galaxies can be relatively incomplete for faint and less blue galaxies, because their SEDs can be more easily degenerate with low-$z$ Balmer break solutions, particularly when only BB data is available. With MB+BB data, the sharp continuum dropout of the Ly-$\alpha$ break can be captured and the presence of emission lines is strictly ruled out, making low-$z$ solutions prohibited and securing them as reliable high-$z$ candidates (Sec.~\ref{subsubsec:UV_slope}).
\end{enumerate}

Although we show the significance of the MB observations in obtaining a clean high-$z$ galaxy sample, the number statistics is still small particularly at the bright end of the LF.
Enlarging the survey area with NIRCam MB+BB observations is essential to improve our understanding of the earliest galaxy evolution.
Future JWST programs scheduled in Cycle 4 such as the MINERVA project will give an invaluable insight (Figure \ref{fig:UVLF_w_MINERVA}).

\begin{acknowledgments}
We thank the anonymous referees for their thoughtful comments and suggestions to improve the quality of the paper.
This work is based on observations made with the NASA/ESA/CSA JWST. The data were obtained from the Mikulski Archive for Space Telescopes at the Space Telescope Science Institute, which is operated by the Association of Universities for Research in Astronomy, Inc., under NASA contract NAS5-03127 for JWST. This research was supported by grants 18JWST-GTO1, 23JWGO2A13, and 23JWGO2B15 from the Canadian Space Agency (CSA), and funding from the Natural Sciences and Engineering Research Council of Canada (NSERC). Support for program JWST-GO-03362 was provided through a grant from the STScI under NASA contract NAS5-03127. This research used the Canadian Advanced Network For Astronomy Research (CANFAR) operated in partnership by the Canadian Astronomy Data Centre and The Digital Research Alliance of Canada with support from the National Research Council of Canada, the Canadian Space Agency, CANARIE and the Canadian Foundation for Innovation.
The Dunlap Institute is funded through an endowment established by the David Dunlap family and the University of Toronto.
YA is supported by JSPS KAKENHI Grant Number 23H00131. MB, NM, GF, JJ acknowledge support from the ERC Grant FIRSTLIGHT, Slovenian national research agency ARIS through grants N1-0238 and P1-0188, and the program HST-GO-16667, provided through a grant from the STScI under NASA contract NAS5-26555.
D.M. and A.L.R. acknowledge support from program JWST-GO-03362, provided through a grant from the STScI under NASA contract NAS5-03127.
\end{acknowledgments}





%
\facilities{HST/ACS, JWST/NIRCam}

\software{astropy \citep{2013A&A...558A..33A,2018AJ....156..123A,2022ApJ...935..167A},
          EAzY \citep{Brammer2008ApJ},
          photutils \citep{larry_bradley_2023_7946442}, 
          Source Extractor \citep{1996A&AS..117..393B},
          emcee \citep{emcee}
          }

\section*{Data Availability}

The data products used in this paper were obtained from the Mikulski Archive for Space Telescopes (MAST) at the Space Telescope Science Institute, via \dataset[doi: 10.17909/ph4n-6n76]{https://doi.org/10.17909/ph4n-6n76} and \dataset[doi: 10.17909/cyh7-mm53]{https://doi.org/10.17909/cyh7-mm53}. 
The CANUCS data release HLSP DOI is \dataset[doi:10.17909/18nv-np70]{https://doi.org/10.17909/18nv-np70}. 
The DR1 data release products can also be found at \dataset[the CANUCS website]{https://niriss.github.io/data_release1.html}.


\appendix

\section{False Positives in the BB-only catalog selection}\label{apx:low-z}
This appendix lists the "False Positives" that are selected as $z>9.5$ in the BB-only catalog but not in the full MB+BB catalog.

\subsection{CANUCS-2217931}

\begin{figure*}[t]
\plotone{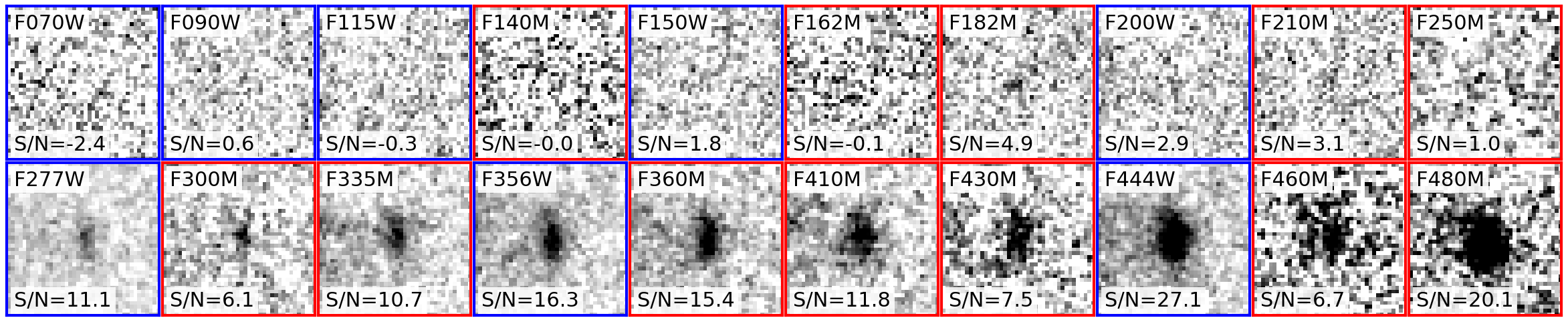}
\plotone{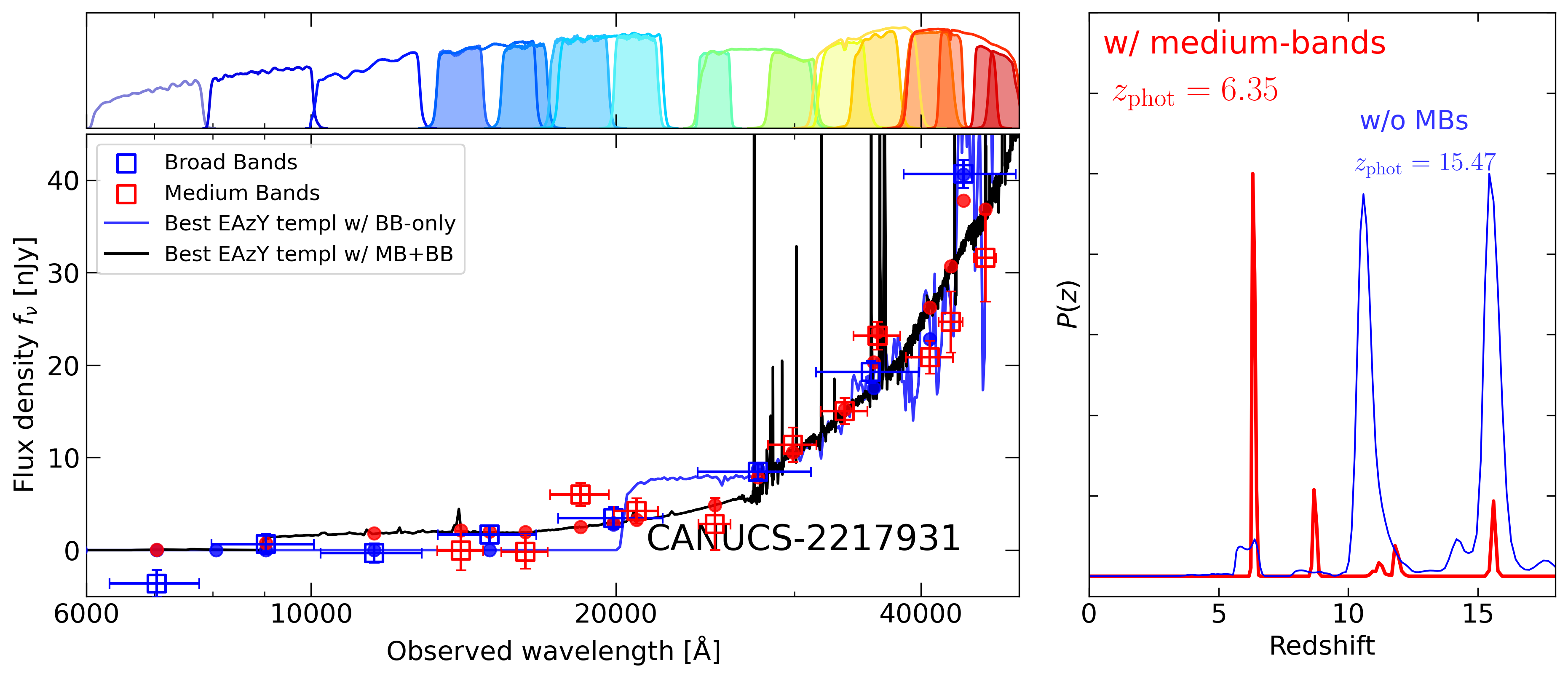}
\caption{A False Positive CANUCS-2217931. \textit{Top:} cutouts of the source in all NIRCam images. BB images are displayed with blue frames, while MB images are with red frames. Cutouts are $1.\!\!^{\prime\prime}6$ on the side.
\textit{Bottom left:} SED of the source. Photometry in BB and MB filters are shown by blue and red open squares, respectively. The best-fit model spectrum when only BB data are used is shown by blue slid curve, while the best model for all MB+BB data is shown by black.
\textit{Bottom right:} $P(z)$ of the source. Blue curve shows the $P(z)$ when only BB data is used, and red shows that for the all MB+BB data SED.
}
\label{fig:2217931}
\end{figure*}

CANUCS-2217931 is a $z\sim15$ candidate found in the BB-only catalog with small low-$z$ solution probability (0.0456).
The SED shows a break in the F200W-F277W color and is not detected in F070W, F090W, F115W, or F150W filters, which favors $z>10$ solutions when only BB data is used (blue solid curves in Figure \ref{fig:2217931}).
However, with the MB observations, this source shows clear excess in F360M and F480M photometry, and is most likely a very dusty EELG at $z\sim6$ (black curve in the bottom left).

\subsection{CANUCS-2225353}

\begin{figure*}[t]
\plotone{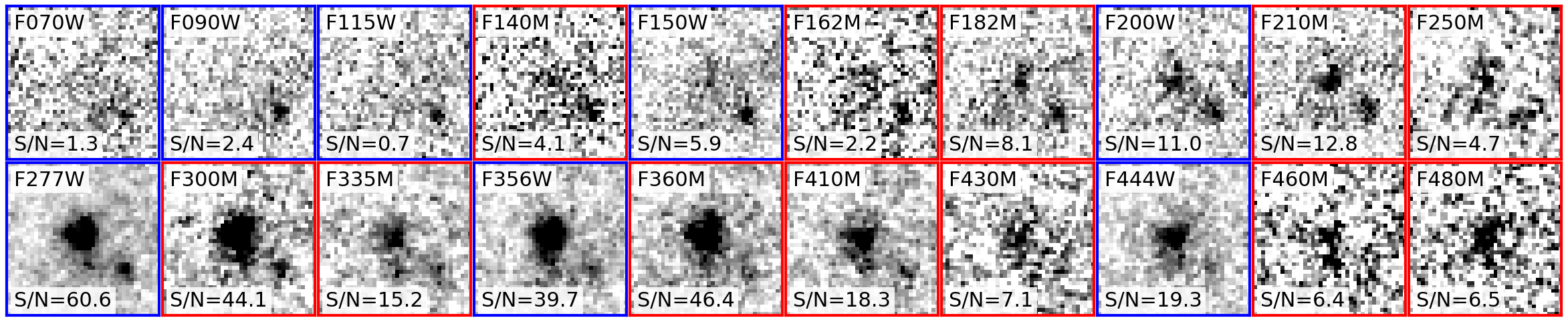}
\plotone{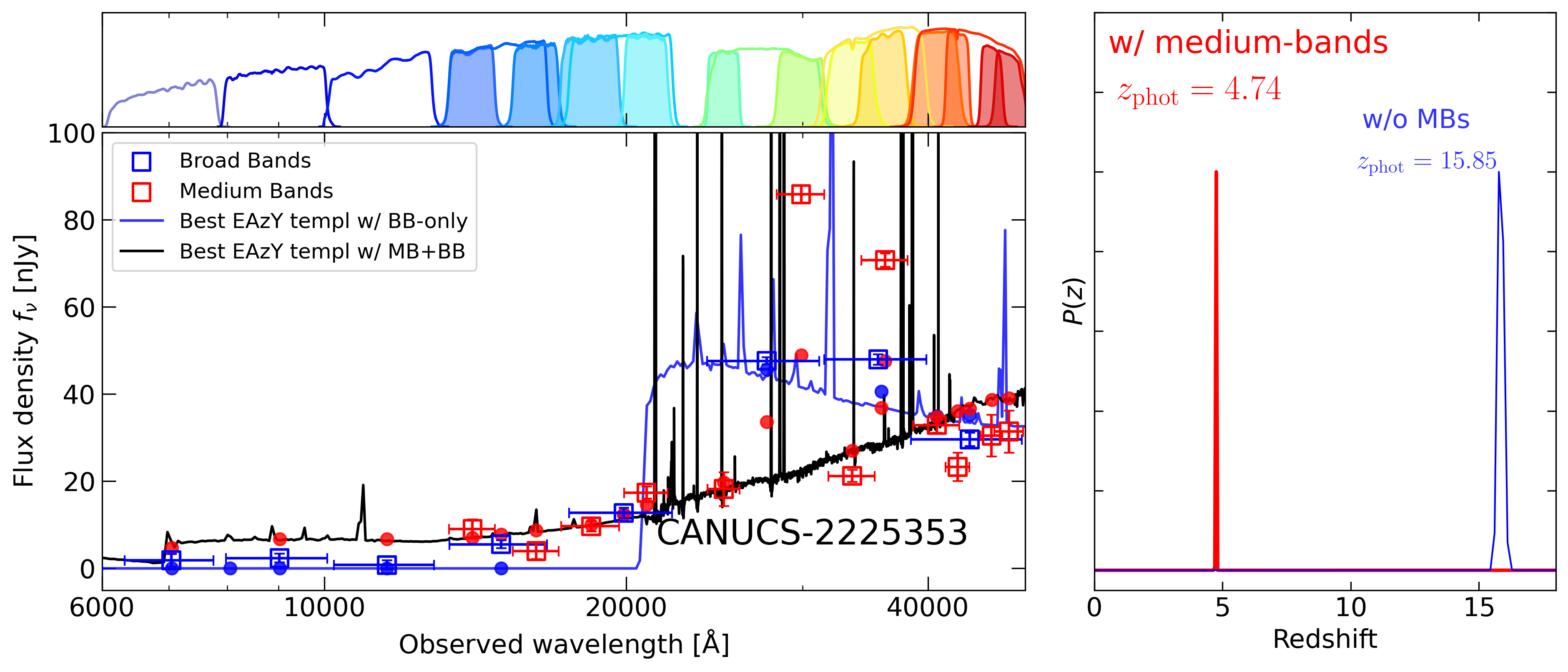}
\caption{Same as Figure \ref{fig:2217931} but for CANUCS-2225353.
}
\label{fig:2225353}
\end{figure*}

CANUCS-2225353 is a $z\sim16$ galaxy candidate found in the BB-only catalog with almost zero low-$z$ solution probability ($8.1\times10^{-5}$).
The SED shows a clear break in the F200W-F277W color and apparently blue SED from F277W to F444W, which favors $z\sim16$ solutions when only BB data is used (blue solid curves in Figure \ref{fig:2225353}).
With the MB observations, though, the SED shows extreme excess in F300M and F360M photometry, which clearly indicates the source is a very dusty EELG at $z\sim5$.
The source is detected in F150W with S/N $=6$, thus CANUCS-2225353 is removed from the \textit{high-$z$ sample} of the BB-only catalog in Section \ref{subsubsec:BB_only_LFs}.

\subsection{CANUCS-5217969}

\begin{figure*}[t]
\plotone{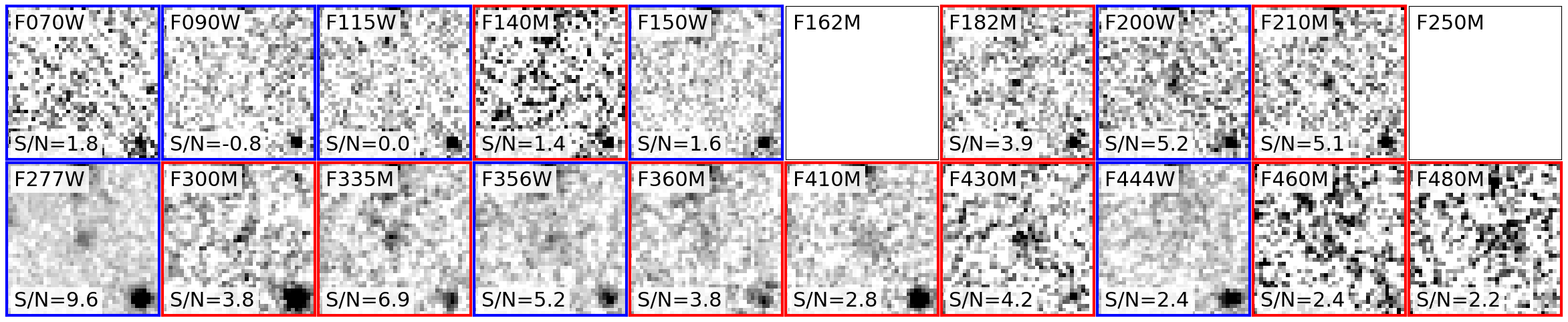}
\plotone{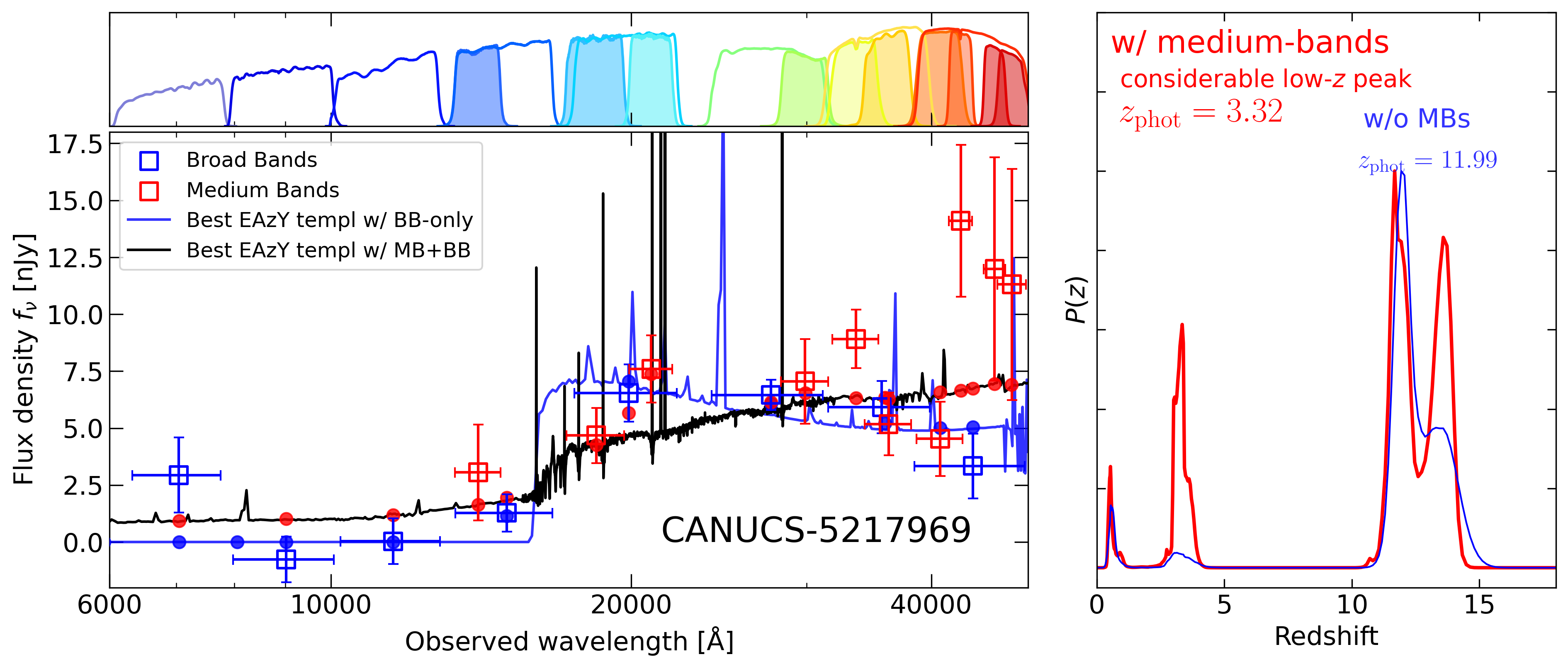}
\caption{Same as Figure \ref{fig:2217931} but for  CANUCS-5217969.
}
\label{fig:5217969}
\end{figure*}

CANUCS-5217969 is a $z\sim12$ candidate found in the BB-only catalog with small low-$z$ solution probability (0.0546).
The SED shows a break in the F150W-F200W color and not detected in F070W, F090W, or F115W images.
Together with the relatively blue SED from F200W to F444W, the galaxy is identified as a $z\sim12$ candidate in the BB-only catalog (Figure \ref{fig:5217969}).
However, the F182M - F210M color indicates the presence of emission lines at this wavelength, and the MB+BB SED can also be reproduced well with a Balmer break galaxy having emission lines at $z\sim3$.
The MB photometry leads to considerably large low-$z$ probability (0.189).

\subsection{CANUCS-5219487}

\begin{figure*}[t]
\plotone{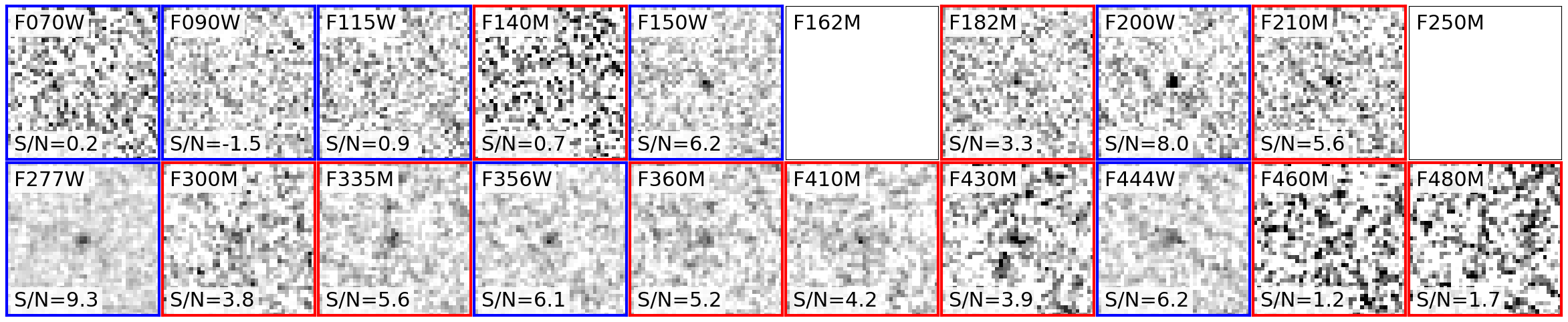}
\plotone{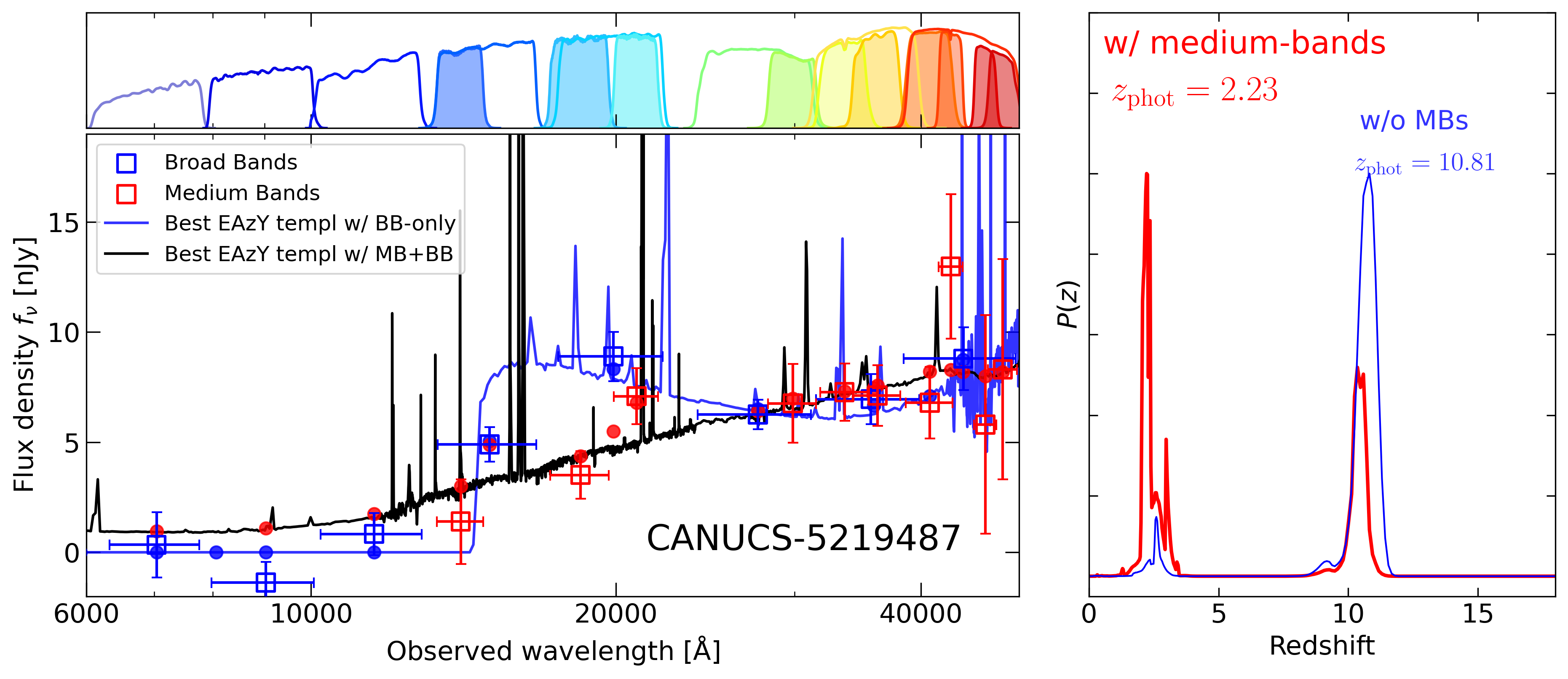}
\caption{Same as Figure \ref{fig:2217931} but for  CANUCS-5219487.
}
\label{fig:5219487}
\end{figure*}

CANUCS-5219487 is a $z\sim11$ candidate from the BB-only catalog with small low-$z$ solution probability (0.0536).
The galaxy is not detected in F070W, F090W, or F115W images, and shows a half drop-out in F150W image, which favors the $z\sim11$ solution  (Figure \ref{fig:5219487}).
However with MB observations, the F182M photometry is not bright as the $z\sim11$ solution predicts, and rather indicates the F200W photometry is boosted by a strong H$\alpha$ emission line at $z\sim2$.

\subsection{CANUCS-3211626 and CANUCS-3211653}

CANUCS-3211626 and CANUCS-3211653 are not catastrophic failures in the BB-only photo-$z$ estimation, but their photo-$z$ is slightly lower $\Delta z\sim0.5-1$ with the MB+BB data than BB-only data.
The photo-$z$ of CANUCS-3211626 with MB+BB photometry (BB-only photometry) is 9.36 (9.79), and that of CANUCS-3211653 is 9.41 (10.37).
They thus are not selected in the MB+BB selection, but are in the BB-only selection.
Conversely, there are two galaxies that are selected in the MB+BB selection but not in the BB-only selection, and thus the effect of missing/recovering candidates due to the random errors on photo-$z$ should be negligible in the comparison of the MB+BB selection vs the BB-only selection candidates.


\bibliography{sample7}{}

@ARTICLE{Naidu2022ApJ,
       author = {{Naidu}, Rohan P. and {Oesch}, Pascal A. and {van Dokkum}, Pieter and {Nelson}, Erica J. and {Suess}, Katherine A. and {Brammer}, Gabriel and {Whitaker}, Katherine E. and {Illingworth}, Garth and {Bouwens}, Rychard and {Tacchella}, Sandro and {Matthee}, Jorryt and {Allen}, Natalie and {Bezanson}, Rachel and {Conroy}, Charlie and {Labbe}, Ivo and {Leja}, Joel and {Leonova}, Ecaterina and {Magee}, Dan and {Price}, Sedona H. and {Setton}, David J. and {Strait}, Victoria and {Stefanon}, Mauro and {Toft}, Sune and {Weaver}, John R. and {Weibel}, Andrea},
        title = "{Two Remarkably Luminous Galaxy Candidates at z {\ensuremath{\approx}} 10-12 Revealed by JWST}",
      journal = {\apjl},
         year = 2022,
        month = nov,
       volume = {940},
       number = {1},
          eid = {L14},
        pages = {L14},
          doi = {10.3847/2041-8213/ac9b22},
archivePrefix = {arXiv},
       eprint = {2207.09434},
 primaryClass = {astro-ph.GA},
       adsurl = {https://ui.adsabs.harvard.edu/abs/2022ApJ...940L..14N}
}

@ARTICLE{Bouwens2023MNRAS,
       author = {{Bouwens}, Rychard J. and {Stefanon}, Mauro and {Brammer}, Gabriel and {Oesch}, Pascal A. and {Herard-Demanche}, Thomas and {Illingworth}, Garth D. and {Matthee}, Jorryt and {Naidu}, Rohan P. and {van Dokkum}, Pieter G. and {van Leeuwen}, Ivana F.},
        title = "{Evolution of the UV LF from z   15 to z   8 using new JWST NIRCam medium-band observations over the HUDF/XDF}",
      journal = {\mnras},
         year = 2023,
        month = jul,
       volume = {523},
       number = {1},
        pages = {1036-1055},
          doi = {10.1093/mnras/stad1145},
archivePrefix = {arXiv},
       eprint = {2211.02607},
 primaryClass = {astro-ph.GA},
       adsurl = {https://ui.adsabs.harvard.edu/abs/2023MNRAS.523.1036B}
}

@ARTICLE{Donnan2023MNRAS,
       author = {{Donnan}, C.~T. and {McLeod}, D.~J. and {McLure}, R.~J. and {Dunlop}, J.~S. and {Carnall}, A.~C. and {Cullen}, F. and {Magee}, D.},
        title = "{The abundance of z {\ensuremath{\gtrsim}} 10 galaxy candidates in the HUDF using deep JWST NIRCam medium-band imaging}",
      journal = {\mnras},
         year = 2023,
        month = apr,
       volume = {520},
       number = {3},
        pages = {4554-4561},
          doi = {10.1093/mnras/stad471},
archivePrefix = {arXiv},
       eprint = {2212.10126},
 primaryClass = {astro-ph.GA},
       adsurl = {https://ui.adsabs.harvard.edu/abs/2023MNRAS.520.4554D}
}

@ARTICLE{Harikane2022ApJS,
       author = {{Harikane}, Yuichi and {Ono}, Yoshiaki and {Ouchi}, Masami and {Liu}, Chengze and {Sawicki}, Marcin and {Shibuya}, Takatoshi and {Behroozi}, Peter S. and {He}, Wanqiu and {Shimasaku}, Kazuhiro and {Arnouts}, Stephane and {Coupon}, Jean and {Fujimoto}, Seiji and {Gwyn}, Stephen and {Huang}, Jiasheng and {Inoue}, Akio K. and {Kashikawa}, Nobunari and {Komiyama}, Yutaka and {Matsuoka}, Yoshiki and {Willott}, Chris J.},
        title = "{GOLDRUSH. IV. Luminosity Functions and Clustering Revealed with  4,000,000 Galaxies at z   2-7: Galaxy-AGN Transition, Star Formation Efficiency, and Implication for Evolution at z > 10}",
      journal = {\apjs},
         year = 2022,
        month = mar,
       volume = {259},
       number = {1},
          eid = {20},
        pages = {20},
          doi = {10.3847/1538-4365/ac3dfc},
archivePrefix = {arXiv},
       eprint = {2108.01090},
 primaryClass = {astro-ph.GA},
       adsurl = {https://ui.adsabs.harvard.edu/abs/2022ApJS..259...20H}
}

@ARTICLE{Drlica-Wagner2018ApJS,
       author = {{Drlica-Wagner}, A. and {Sevilla-Noarbe}, I. and {Rykoff}, E.~S. and {Gruendl}, R.~A. and {Yanny}, B. and {Tucker}, D.~L. and {Hoyle}, B. and {Carnero Rosell}, A. and {Bernstein}, G.~M. and {Bechtol}, K. and {Becker}, M.~R. and {Benoit-L{\'e}vy}, A. and {Bertin}, E. and {Carrasco Kind}, M. and {Davis}, C. and {de Vicente}, J. and {Diehl}, H.~T. and {Gruen}, D. and {Hartley}, W.~G. and {Leistedt}, B. and {Li}, T.~S. and {Marshall}, J.~L. and {Neilsen}, E. and {Rau}, M.~M. and {Sheldon}, E. and {Smith}, J. and {Troxel}, M.~A. and {Wyatt}, S. and {Zhang}, Y. and {Abbott}, T.~M.~C. and {Abdalla}, F.~B. and {Allam}, S. and {Banerji}, M. and {Brooks}, D. and {Buckley-Geer}, E. and {Burke}, D.~L. and {Capozzi}, D. and {Carretero}, J. and {Cunha}, C.~E. and {D'Andrea}, C.~B. and {da Costa}, L.~N. and {DePoy}, D.~L. and {Desai}, S. and {Dietrich}, J.~P. and {Doel}, P. and {Evrard}, A.~E. and {Fausti Neto}, A. and {Flaugher}, B. and {Fosalba}, P. and {Frieman}, J. and {Garc{\'\i}a-Bellido}, J. and {Gerdes}, D.~W. and {Giannantonio}, T. and {Gschwend}, J. and {Gutierrez}, G. and {Honscheid}, K. and {James}, D.~J. and {Jeltema}, T. and {Kuehn}, K. and {Kuhlmann}, S. and {Kuropatkin}, N. and {Lahav}, O. and {Lima}, M. and {Lin}, H. and {Maia}, M.~A.~G. and {Martini}, P. and {McMahon}, R.~G. and {Melchior}, P. and {Menanteau}, F. and {Miquel}, R. and {Nichol}, R.~C. and {Ogando}, R.~L.~C. and {Plazas}, A.~A. and {Romer}, A.~K. and {Roodman}, A. and {Sanchez}, E. and {Scarpine}, V. and {Schindler}, R. and {Schubnell}, M. and {Smith}, M. and {Smith}, R.~C. and {Soares-Santos}, M. and {Sobreira}, F. and {Suchyta}, E. and {Tarle}, G. and {Vikram}, V. and {Walker}, A.~R. and {Wechsler}, R.~H. and {Zuntz}, J. and {DES Collaboration}},
        title = "{Dark Energy Survey Year 1 Results: The Photometric Data Set for Cosmology}",
      journal = {\apjs},
         year = 2018,
        month = apr,
       volume = {235},
       number = {2},
          eid = {33},
        pages = {33},
          doi = {10.3847/1538-4365/aab4f5},
archivePrefix = {arXiv},
       eprint = {1708.01531},
 primaryClass = {astro-ph.CO},
       adsurl = {https://ui.adsabs.harvard.edu/abs/2018ApJS..235...33D}
}

@ARTICLE{Bisigello2025AA,
       author = {{Bisigello}, L. and {Gandolfi}, G. and {Feltre}, A. and {Arrabal Haro}, P. and {Calabr{\`o}}, A. and {Cleri}, N.~J. and {Costantin}, L. and {Girardi}, G. and {Giulietti}, M. and {Grazian}, A. and {Gruppioni}, C. and {Hathi}, N.~P. and {Holwerda}, B.~W. and {Llerena}, M. and {Lucas}, R.~A. and {Pacucci}, F. and {Prandoni}, I. and {Rodighiero}, G. and {Seill{\'e}}, L. -M. and {Wilkins}, S.~M. and {Bagley}, M. and {Dickinson}, M. and {Finkelstein}, S.~L. and {Kartaltepe}, J. and {Koekemoer}, A.~M. and {Papovich}, C. and {Pirzkal}, N.},
        title = "{Spectroscopic confirmation of a dust-obscured, possibly metal-rich dwarf galaxy at z {\ensuremath{\sim}} 5}",
      journal = {\aap},
         year = 2025,
        month = jan,
       volume = {693},
          eid = {L18},
        pages = {L18},
          doi = {10.1051/0004-6361/202452604},
archivePrefix = {arXiv},
       eprint = {2410.10954},
 primaryClass = {astro-ph.GA},
       adsurl = {https://ui.adsabs.harvard.edu/abs/2025A&A...693L..18B}
}

@ARTICLE{emcee,
   author = {{Foreman-Mackey}, D. and {Hogg}, D.~W. and {Lang}, D. and {Goodman}, J.},
    title = {emcee: The MCMC Hammer},
  journal = {PASP},
     year = 2013,
   volume = 125,
    pages = {306-312},
   eprint = {1202.3665},
      doi = {10.1086/670067}
}

@ARTICLE{Finkelstein2024ApJ,
       author = {{Finkelstein}, Steven L. and {Leung}, Gene C.~K. and {Bagley}, Micaela B. and {Dickinson}, Mark and {Ferguson}, Henry C. and {Papovich}, Casey and {Akins}, Hollis B. and {Arrabal Haro}, Pablo and {Dav{\'e}}, Romeel and {Dekel}, Avishai and {Kartaltepe}, Jeyhan S. and {Kocevski}, Dale D. and {Koekemoer}, Anton M. and {Pirzkal}, Nor and {Somerville}, Rachel S. and {Yung}, L.~Y. Aaron and {Amor{\'\i}n}, Ricardo O. and {Backhaus}, Bren E. and {Behroozi}, Peter and {Bisigello}, Laura and {Bromm}, Volker and {Casey}, Caitlin M. and {Ch{\'a}vez Ortiz}, {\'O}scar A. and {Cheng}, Yingjie and {Chworowsky}, Katherine and {Cleri}, Nikko J. and {Cooper}, M.~C. and {Davis}, Kelcey and {de la Vega}, Alexander and {Elbaz}, David and {Franco}, Maximilien and {Fontana}, Adriano and {Fujimoto}, Seiji and {Giavalisco}, Mauro and {Grogin}, Norman A. and {Holwerda}, Benne W. and {Huertas-Company}, Marc and {Hirschmann}, Michaela and {Iyer}, Kartheik G. and {Jogee}, Shardha and {Jung}, Intae and {Larson}, Rebecca L. and {Lucas}, Ray A. and {Mobasher}, Bahram and {Morales}, Alexa M. and {Morley}, Caroline V. and {Mukherjee}, Sagnick and {P{\'e}rez-Gonz{\'a}lez}, Pablo G. and {Ravindranath}, Swara and {Rodighiero}, Giulia and {Rowland}, Melanie J. and {Tacchella}, Sandro and {Taylor}, Anthony J. and {Trump}, Jonathan R. and {Wilkins}, Stephen M.},
        title = "{The Complete CEERS Early Universe Galaxy Sample: A Surprisingly Slow Evolution of the Space Density of Bright Galaxies at z {\ensuremath{\sim}} 8.5{\textendash}14.5}",
      journal = {\apjl},
         year = 2024,
        month = jul,
       volume = {969},
       number = {1},
          eid = {L2},
        pages = {L2},
          doi = {10.3847/2041-8213/ad4495},
archivePrefix = {arXiv},
       eprint = {2311.04279},
 primaryClass = {astro-ph.GA},
       adsurl = {https://ui.adsabs.harvard.edu/abs/2024ApJ...969L...2F}
}

@ARTICLE{Sarrouh2025arXiv,
       author = {{Sarrouh}, Ghassan T.~E. and {Asada}, Yoshihisa and {Martis}, Nicholas S. and {Willott}, Chris J. and {Iyer}, Kartheik G. and {Noirot}, Ga{\"e}l and {Muzzin}, Adam and {Sawicki}, Marcin and {Brammer}, Gabriel and {Desprez}, Guillaume and {Rihtar{\v{s}}i{\v{c}}}, Gregor and {Zabl}, Johannes and {Abraham}, Roberto and {Brada{\v{c}}}, Maru{\v{s}}a and {Doyon}, Ren{\'e} and {Antwi-Danso}, Jacqueline and {Berek}, Samantha and {Brown}, Westley and {Estrada-Carpenter}, Vince and {Favaro}, Jeremy and {Felicioni}, Giordano and {Forrest}, Ben and {Gaspar}, Gaia and {Gould}, Katriona M.~L. and {Gledhill}, Rachel and {Harshan}, Anishya and {Jahan}, Nusrath and {Jagga}, Naadiyah and {Jude{\v{z}}}, Jon and {Marchesini}, Danilo and {Markov}, Vladan and {Matharu}, Jasleen and {MacFarland}, Shannon and {Merchant}, Maya and {M{\'e}rida}, Rosa M. and {Mowla}, Lamiya and {Myers}, Katherine and {Omori}, Kiyoaki C. and {Pacifici}, Camilla and {Ravindranath}, Swara and {Robbins}, Luke and {Strait}, Victoria and {Sok}, Visal and {Tan}, Vivian Yun Yan and {Tripodi}, Roberta and {Wilson}, Gillian and {Withers}, Sunna},
        title = "{CANUCS/Technicolor Data Release 1: Imaging, Photometry, Slit Spectroscopy, and Stellar Population Parameters}",
      journal = {arXiv e-prints},
         year = 2025,
        month = jun,
          eid = {arXiv:2506.21685},
        pages = {arXiv:2506.21685},
archivePrefix = {arXiv},
       eprint = {2506.21685},
 primaryClass = {astro-ph.GA},
       adsurl = {https://ui.adsabs.harvard.edu/abs/2025arXiv250621685S}
}

@ARTICLE{Peng2010AJ,
       author = {{Peng}, Chien Y. and {Ho}, Luis C. and {Impey}, Chris D. and {Rix}, Hans-Walter},
        title = "{Detailed Decomposition of Galaxy Images. II. Beyond Axisymmetric Models}",
      journal = {\aj},
         year = 2010,
        month = jun,
       volume = {139},
       number = {6},
        pages = {2097-2129},
          doi = {10.1088/0004-6256/139/6/2097},
archivePrefix = {arXiv},
       eprint = {0912.0731},
 primaryClass = {astro-ph.CO},
       adsurl = {https://ui.adsabs.harvard.edu/abs/2010AJ....139.2097P}
}

@ARTICLE{Sarrouh2024ApJ,
       author = {{Sarrouh}, Ghassan T.~E. and {Muzzin}, Adam and {Iyer}, Kartheik G. and {Mowla}, Lamiya and {Withers}, Sunna and {Martis}, Nicholas S. and {Abraham}, Roberto and {Asada}, Yoshihisa and {Brada{\v{c}}}, Maru{\v{s}}a and {Brammer}, Gabriel B. and {Desprez}, Guillaume and {Estrada-Carpenter}, Vince and {Matharu}, Jasleen and {Noirot}, Ga{\"e}l and {Sawicki}, Marcin and {Strait}, Victoria and {Willott}, Chris J. and {Zabl}, Johannes},
        title = "{Exposing Line Emission: The Systematic Differences of Measuring Galaxy Stellar Masses with JWST NIRCam Medium versus Wide Band Photometry}",
      journal = {\apjl},
         year = 2024,
        month = may,
       volume = {967},
       number = {1},
          eid = {L17},
        pages = {L17},
          doi = {10.3847/2041-8213/ad43e8},
archivePrefix = {arXiv},
       eprint = {2401.08781},
 primaryClass = {astro-ph.GA},
       adsurl = {https://ui.adsabs.harvard.edu/abs/2024ApJ...967L..17S}
}

@ARTICLE{Suess2024ApJ,
       author = {{Suess}, Katherine A. and {Weaver}, John R. and {Price}, Sedona H. and {Pan}, Richard and {Wang}, Bingjie and {Bezanson}, Rachel and {Brammer}, Gabriel and {Cutler}, Sam E. and {Labb{\'e}}, Ivo and {Leja}, Joel and {Williams}, Christina C. and {Whitaker}, Katherine E. and {Atek}, Hakim and {Dayal}, Pratika and {de Graaff}, Anna and {Feldmann}, Robert and {Franx}, Marijn and {Fudamoto}, Yoshinobu and {Fujimoto}, Seiji and {Furtak}, Lukas J. and {Goulding}, Andy D. and {Greene}, Jenny E. and {Khullar}, Gourav and {Kokorev}, Vasily and {Kriek}, Mariska and {Lorenz}, Brian and {Marchesini}, Danilo and {Maseda}, Michael V. and {Matthee}, Jorryt and {Miller}, Tim B. and {Mitsuhashi}, Ikki and {Mowla}, Lamiya A. and {Muzzin}, Adam and {Naidu}, Rohan P. and {Nanayakkara}, Themiya and {Nelson}, Erica J. and {Oesch}, Pascal A. and {Setton}, David J. and {Shipley}, Heath and {Smit}, Renske and {Spilker}, Justin S. and {van Dokkum}, Pieter and {Zitrin}, Adi},
        title = "{Medium Bands, Mega Science: A JWST/NIRCam Medium-band Imaging Survey of A2744}",
      journal = {\apj},
         year = 2024,
        month = nov,
       volume = {976},
       number = {1},
          eid = {101},
        pages = {101},
          doi = {10.3847/1538-4357/ad75fe},
archivePrefix = {arXiv},
       eprint = {2404.13132},
 primaryClass = {astro-ph.GA},
       adsurl = {https://ui.adsabs.harvard.edu/abs/2024ApJ...976..101S}
}

@ARTICLE{Martis2025arXiv,
       author = {{Martis}, Nicholas and {Withers}, Sunna and {Felicioni}, Giordano and {Muzzin}, Adam and {Brada{\v{c}}}, Maru{\v{s}}a and {Abraham}, Roberto and {Asada}, Yoshihisa and {Desprez}, Guillaume and {Iyer}, Kartheik and {Noirot}, Gael and {Sarrouh}, Ghassan T.~E. and {Sawicki}, Marcin and {Strait}, Victoria and {Willot}, Chris and {Jagga}, Naadiyah and {Jude{\v{z}}}, Jon and {Harshan}, Anishya and {Marchesini}, Danilo and {Markov}, Vladan and {M{\'e}rida}, Rosa M. and {Rihtar{\v{s}}i{\v{c}}}, Gregor and {Tripodi}, Roberta},
        title = "{CANUCS/Technicolor: JWST Medium Band Photometry Finds Half of the Star Formation at $z>7.5$ is Obscured}",
      journal = {arXiv e-prints},
         year = 2025,
        month = mar,
          eid = {arXiv:2503.01579},
        pages = {arXiv:2503.01579},
          doi = {10.48550/arXiv.2503.01579},
archivePrefix = {arXiv},
       eprint = {2503.01579},
 primaryClass = {astro-ph.GA},
       adsurl = {https://ui.adsabs.harvard.edu/abs/2025arXiv250301579M}
}

@ARTICLE{Whitler2025arXiv,
       author = {{Whitler}, Lily and {Stark}, Daniel P. and {Topping}, Michael W. and {Robertson}, Brant and {Rieke}, Marcia and {Hainline}, Kevin N. and {Endsley}, Ryan and {Chen}, Zuyi and {Baker}, William M. and {Bhatawdekar}, Rachana and {Bunker}, Andrew J. and {Carniani}, Stefano and {Charlot}, St{\'e}phane and {Chevallard}, Jacopo and {Curtis-Lake}, Emma and {Egami}, Eiichi and {Eisenstein}, Daniel J. and {Helton}, Jakob M. and {Ji}, Zhiyuan and {Johnson}, Benjamin D. and {P{\'e}rez-Gonz{\'a}lez}, Pablo G. and {Rinaldi}, Pierluigi and {Tacchella}, Sandro and {Williams}, Christina C. and {Willmer}, Christopher N.~A. and {Willott}, Chris and {Witstok}, Joris},
        title = "{The $z rsim 9$ galaxy UV luminosity function from the JWST Advanced Deep Extragalactic Survey: insights into early galaxy evolution and reionization}",
      journal = {arXiv e-prints},
         year = 2025,
        month = jan,
          eid = {arXiv:2501.00984},
        pages = {arXiv:2501.00984},
          doi = {10.48550/arXiv.2501.00984},
archivePrefix = {arXiv},
       eprint = {2501.00984},
 primaryClass = {astro-ph.GA},
       adsurl = {https://ui.adsabs.harvard.edu/abs/2025arXiv250100984W}
}

@ARTICLE{Eisenstein2023arXiv,
       author = {{Eisenstein}, Daniel J. and {Johnson}, Benjamin D. and {Robertson}, Brant and {Tacchella}, Sandro and {Hainline}, Kevin and {Jakobsen}, Peter and {Maiolino}, Roberto and {Bonaventura}, Nina and {Bunker}, Andrew J. and {Cameron}, Alex J. and {Cargile}, Phillip A. and {Curtis-Lake}, Emma and {Hausen}, Ryan and {Pusk{\'a}s}, D{\'a}vid and {Rieke}, Marcia and {Sun}, Fengwu and {Willmer}, Christopher N.~A. and {Willott}, Chris and {Alberts}, Stacey and {Arribas}, Santiago and {Baker}, William M. and {Baum}, Stefi and {Bhatawdekar}, Rachana and {Carniani}, Stefano and {Charlot}, Stephane and {Chen}, Zuyi and {Chevallard}, Jacopo and {Curti}, Mirko and {DeCoursey}, Christa and {D'Eugenio}, Francesco and {de Graaff}, Anna and {Egami}, Eiichi and {Helton}, Jakob M. and {Ji}, Zhiyuan and {Jones}, Gareth C. and {Kumari}, Nimisha and {L{\"u}tzgendorf}, Nora and {Laseter}, Isaac and {Looser}, Tobias J. and {Lyu}, Jianwei and {Maseda}, Michael V. and {Nelson}, Erica and {Parlanti}, Eleonora and {Rauscher}, Bernard J. and {Rawle}, Tim and {Rieke}, George and {Rix}, Hans-Walter and {Rujopakarn}, Wiphu and {Sandles}, Lester and {Saxena}, Aayush and {Scholtz}, Jan and {Sharpe}, Katherine and {Shivaei}, Irene and {Simmonds}, Charlotte and {Smit}, Renske and {Topping}, Michael W. and {{\"U}bler}, Hannah and {Venturi}, Giacomo and {Williams}, Christina C. and {Witstok}, Joris and {Woodrum}, Charity},
        title = "{The JADES Origins Field: A New JWST Deep Field in the JADES Second NIRCam Data Release}",
      journal = {arXiv e-prints},
         year = 2023,
        month = oct,
          eid = {arXiv:2310.12340},
        pages = {arXiv:2310.12340},
          doi = {10.48550/arXiv.2310.12340},
archivePrefix = {arXiv},
       eprint = {2310.12340},
 primaryClass = {astro-ph.GA},
       adsurl = {https://ui.adsabs.harvard.edu/abs/2023arXiv231012340E}
}

@ARTICLE{Jespersen2025ApJ,
       author = {{Jespersen}, Christian Kragh and {Steinhardt}, Charles L. and {Somerville}, Rachel S. and {Lovell}, Christopher C.},
        title = "{On the Significance of Rare Objects at High Redshift: The Impact of Cosmic Variance}",
      journal = {\apj},
         year = 2025,
        month = mar,
       volume = {982},
       number = {1},
          eid = {23},
        pages = {23},
          doi = {10.3847/1538-4357/adb422},
archivePrefix = {arXiv},
       eprint = {2403.00050},
 primaryClass = {astro-ph.GA},
       adsurl = {https://ui.adsabs.harvard.edu/abs/2025ApJ...982...23J}
}

@software{larry_bradley_2023_7946442,
author       = {Larry Bradley and
                Brigitta Sip{\H o}cz and
                Thomas Robitaille and
                Erik Tollerud and
                Z\`e Vin{\'{\i}}cius and
                Christoph Deil and
                Kyle Barbary and
                Tom J Wilson and
                Ivo Busko and
                Axel Donath and
                Hans Moritz G{\"u}nther and
                Mihai Cara and
                P. L. Lim and
                Sebastian Me{\ss}linger and
                Simon Conseil and
                Azalee Bostroem and
                Michael Droettboom and
                E. M. Bray and
                Lars Andersen Bratholm and
                Geert Barentsen and
                Matt Craig and
                Shivangee Rathi and
                Sergio Pascual and
                Gabriel Perren and
                Iskren Y. Georgiev and
                Miguel de Val-Borro and
                Wolfgang Kerzendorf and
                Yoonsoo P. Bach and
                Bruno Quint and
                Harrison Souchereau},
title        = {astropy/photutils: 1.8.0},
month        = may,
year         = 2023,
publisher    = {Zenodo},
version      = {1.8.0},
doi          = {10.5281/zenodo.7946442},
url          = {https://doi.org/10.5281/zenodo.7946442}
}

@ARTICLE{Oke1983ApJ,
       author = {{Oke}, J.~B. and {Gunn}, J.~E.},
        title = "{Secondary standard stars for absolute spectrophotometry.}",
      journal = {\apj},
         year = 1983,
        month = mar,
       volume = {266},
        pages = {713-717},
          doi = {10.1086/160817},
       adsurl = {https://ui.adsabs.harvard.edu/abs/1983ApJ...266..713O}
}

@ARTICLE{Willott2022PASP,
       author = {{Willott}, Chris J. and {Doyon}, Ren{\'e} and {Albert}, Loic and {Brammer}, Gabriel B. and {Dixon}, William V. and {Muzic}, Koraljka and {Ravindranath}, Swara and {Scholz}, Aleks and {Abraham}, Roberto and {Artigau}, {\'E}tienne and {Brada{\v{c}}}, Maru{\v{s}}a and {Goudfrooij}, Paul and {Hutchings}, John B. and {Iyer}, Kartheik G. and {Jayawardhana}, Ray and {LaMassa}, Stephanie and {Martis}, Nicholas and {Meyer}, Michael R. and {Morishita}, Takahiro and {Mowla}, Lamiya and {Muzzin}, Adam and {Noirot}, Ga{\"e}l and {Pacifici}, Camilla and {Rowlands}, Neil and {Sarrouh}, Ghassan and {Sawicki}, Marcin and {Taylor}, Joanna M. and {Volk}, Kevin and {Zabl}, Johannes},
        title = "{The Near-infrared Imager and Slitless Spectrograph for the James Webb Space Telescope. II. Wide Field Slitless Spectroscopy}",
      journal = {\pasp},
         year = 2022,
        month = feb,
       volume = {134},
       number = {1032},
          eid = {025002},
        pages = {025002},
          doi = {10.1088/1538-3873/ac5158},
archivePrefix = {arXiv},
       eprint = {2202.01714},
 primaryClass = {astro-ph.IM},
       adsurl = {https://ui.adsabs.harvard.edu/abs/2022PASP..134b5002W}
}

@ARTICLE{Williams2024ApJ,
       author = {{Williams}, Christina C. and {Alberts}, Stacey and {Ji}, Zhiyuan and {Hainline}, Kevin N. and {Lyu}, Jianwei and {Rieke}, George and {Endsley}, Ryan and {Suess}, Katherine A. and {Sun}, Fengwu and {Johnson}, Benjamin D. and {Florian}, Michael and {Shivaei}, Irene and {Rujopakarn}, Wiphu and {Baker}, William M. and {Bhatawdekar}, Rachana and {Boyett}, Kristan and {Bunker}, Andrew J. and {Cameron}, Alex J. and {Carniani}, Stefano and {Charlot}, Stephane and {Curtis-Lake}, Emma and {DeCoursey}, Christa and {de Graaff}, Anna and {Egami}, Eiichi and {Eisenstein}, Daniel J. and {Gibson}, Justus L. and {Hausen}, Ryan and {Helton}, Jakob M. and {Maiolino}, Roberto and {Maseda}, Michael V. and {Nelson}, Erica J. and {P{\'e}rez-Gonz{\'a}lez}, Pablo G. and {Rieke}, Marcia J. and {Robertson}, Brant E. and {Saxena}, Aayush and {Tacchella}, Sandro and {Willmer}, Christopher N.~A. and {Willott}, Chris J.},
        title = "{The Galaxies Missed by Hubble and ALMA: The Contribution of Extremely Red Galaxies to the Cosmic Census at 3 < z < 8}",
      journal = {\apj},
         year = 2024,
        month = jun,
       volume = {968},
       number = {1},
          eid = {34},
        pages = {34},
          doi = {10.3847/1538-4357/ad3f17},
archivePrefix = {arXiv},
       eprint = {2311.07483},
 primaryClass = {astro-ph.GA},
       adsurl = {https://ui.adsabs.harvard.edu/abs/2024ApJ...968...34W}
}

@ARTICLE{Carniani2025AAS,
       author = {{Carniani}, Stefano and {D'Eugenio}, Francesco and {Ji}, Xihan and {Parlanti}, Eleonora and {Scholtz}, Jan and {Sun}, Fengwu and {Venturi}, Giacomo and {Bakx}, Tom J.~L.~C. and {Curti}, Mirko and {Maiolino}, Roberto and {Tacchella}, Sandro and {Zavala}, Jorge A. and {Hainline}, Kevin and {Witstok}, Joris and {Johnson}, Benjamin D. and {Alberts}, Stacey and {Bunker}, Andrew J. and {Charlot}, St{\'e}phane and {Eisenstein}, Daniel J. and {Helton}, Jakob M. and {Jakobsen}, Peter and {Kumari}, Nimisha and {Robertson}, Brant and {Saxena}, Aayush and {{\"U}bler}, Hannah and {Williams}, Christina C. and {Willmer}, Christopher N.~A. and {Willott}, Chris},
        title = "{The eventful life of a luminous galaxy at z = 14: metal enrichment, feedback, and low gas fraction?}",
      journal = {\aap},
         year = 2025,
        month = apr,
       volume = {696},
          eid = {A87},
        pages = {A87},
          doi = {10.1051/0004-6361/202452451},
archivePrefix = {arXiv},
       eprint = {2409.20533},
 primaryClass = {astro-ph.GA},
       adsurl = {https://ui.adsabs.harvard.edu/abs/2025A&A...696A..87C}
}

@ARTICLE{Bouwens2021,
       author = {{Bouwens}, R.~J. and {Oesch}, P.~A. and {Stefanon}, M. and {Illingworth}, G. and {Labb{\'e}}, I. and {Reddy}, N. and {Atek}, H. and {Montes}, M. and {Naidu}, R. and {Nanayakkara}, T. and {Nelson}, E. and {Wilkins}, S.},
        title = "{New Determinations of the UV Luminosity Functions from z   9 to 2 Show a Remarkable Consistency with Halo Growth and a Constant Star Formation Efficiency}",
      journal = {\aj},
         year = 2021,
        month = aug,
       volume = {162},
       number = {2},
          eid = {47},
        pages = {47},
          doi = {10.3847/1538-3881/abf83e},
archivePrefix = {arXiv},
       eprint = {2102.07775},
 primaryClass = {astro-ph.GA},
       adsurl = {https://ui.adsabs.harvard.edu/abs/2021AJ....162...47B}
}

@ARTICLE{Naidu2025,
       author = {{Naidu}, Rohan P. and {Oesch}, Pascal A. and {Brammer}, Gabriel and {Weibel}, Andrea and {Li}, Yijia and {Matthee}, Jorryt and {Chisholm}, John and {Pollock}, Clara L. and {Heintz}, Kasper E. and {Johnson}, Benjamin D. and {Shen}, Xuejian and {Hviding}, Raphael E. and {Leja}, Joel and {Tacchella}, Sandro and {Ganguly}, Arpita and {Witten}, Callum and {Atek}, Hakim and {Belli}, Sirio and {Bose}, Sownak and {Bouwens}, Rychard and {Dayal}, Pratika and {Decarli}, Roberto and {de Graaff}, Anna and {Fudamoto}, Yoshinobu and {Giovinazzo}, Emma and {Greene}, Jenny E. and {Illingworth}, Garth and {Inoue}, Akio K. and {Kane}, Sarah G. and {Labbe}, Ivo and {Leonova}, Ecaterina and {Marques-Chaves}, Rui and {Meyer}, Romain A. and {Nelson}, Erica J. and {Roberts-Borsani}, Guido and {Schaerer}, Daniel and {Simcoe}, Robert A. and {Stefanon}, Mauro and {Sugahara}, Yuma and {Toft}, Sune and {van der Wel}, Arjen and {van Dokkum}, Pieter and {Walter}, Fabian and {Watson}, Darach and {Weaver}, John R. and {Whitaker}, Katherine E.},
        title = "{A Cosmic Miracle: A Remarkably Luminous Galaxy at $z_{\rm{spec}}=14.44$ Confirmed with JWST}",
      journal = {arXiv e-prints},
         year = 2025,
        month = may,
          eid = {arXiv:2505.11263},
        pages = {arXiv:2505.11263},
          doi = {10.48550/arXiv.2505.11263},
archivePrefix = {arXiv},
       eprint = {2505.11263},
 primaryClass = {astro-ph.GA},
       adsurl = {https://ui.adsabs.harvard.edu/abs/2025arXiv250511263N}
}

@ARTICLE{Sun2023ApJ,
       author = {{Sun}, Guochao and {Faucher-Gigu{\`e}re}, Claude-Andr{\'e} and {Hayward}, Christopher C. and {Shen}, Xuejian and {Wetzel}, Andrew and {Cochrane}, Rachel K.},
        title = "{Bursty Star Formation Naturally Explains the Abundance of Bright Galaxies at Cosmic Dawn}",
      journal = {\apjl},
         year = 2023,
        month = oct,
       volume = {955},
       number = {2},
          eid = {L35},
        pages = {L35},
          doi = {10.3847/2041-8213/acf85a},
archivePrefix = {arXiv},
       eprint = {2307.15305},
 primaryClass = {astro-ph.GA},
       adsurl = {https://ui.adsabs.harvard.edu/abs/2023ApJ...955L..35S}
}

@ARTICLE{Withers2023,
       author = {{Withers}, Sunna and {Muzzin}, Adam and {Ravindranath}, Swara and {Sarrouh}, Ghassan T.~E. and {Abraham}, Roberto and {Asada}, Yoshihisa and {Brada{\v{c}}}, Maru{\v{s}}a and {Brammer}, Gabriel and {Desprez}, Guillaume and {Iyer}, Kartheik and {Martis}, Nicholas and {Mowla}, Lamiya and {Noirot}, Ga{\"e}l and {Sawicki}, Marcin and {Strait}, Victoria and {Willott}, Chris J.},
        title = "{Spectroscopy from Photometry: A Population of Extreme Emission Line Galaxies at 1.7 {\ensuremath{\lesssim}} z {\ensuremath{\lesssim}} 6.7 Selected with JWST Medium Band Filters}",
      journal = {\apjl},
         year = 2023,
        month = nov,
       volume = {958},
       number = {1},
          eid = {L14},
        pages = {L14},
          doi = {10.3847/2041-8213/ad01c0},
archivePrefix = {arXiv},
       eprint = {2304.11181},
 primaryClass = {astro-ph.GA},
       adsurl = {https://ui.adsabs.harvard.edu/abs/2023ApJ...958L..14W}
}

@ARTICLE{Arrabal_Haro2023,
       author = {{Arrabal Haro}, Pablo and {Dickinson}, Mark and {Finkelstein}, Steven L. and {Kartaltepe}, Jeyhan S. and {Donnan}, Callum T. and {Burgarella}, Denis and {Carnall}, Adam C. and {Cullen}, Fergus and {Dunlop}, James S. and {Fern{\'a}ndez}, Vital and {Fujimoto}, Seiji and {Jung}, Intae and {Krips}, Melanie and {Larson}, Rebecca L. and {Papovich}, Casey and {P{\'e}rez-Gonz{\'a}lez}, Pablo G. and {Amor{\'\i}n}, Ricardo O. and {Bagley}, Micaela B. and {Buat}, V{\'e}ronique and {Casey}, Caitlin M. and {Chworowsky}, Katherine and {Cohen}, Seth H. and {Ferguson}, Henry C. and {Giavalisco}, Mauro and {Huertas-Company}, Marc and {Hutchison}, Taylor A. and {Kocevski}, Dale D. and {Koekemoer}, Anton M. and {Lucas}, Ray A. and {McLeod}, Derek J. and {McLure}, Ross J. and {Pirzkal}, Norbert and {Seill{\'e}}, Lise-Marie and {Trump}, Jonathan R. and {Weiner}, Benjamin J. and {Wilkins}, Stephen M. and {Zavala}, Jorge A.},
        title = "{Confirmation and refutation of very luminous galaxies in the early Universe}",
      journal = {\nat},
         year = 2023,
        month = oct,
       volume = {622},
       number = {7984},
        pages = {707-711},
          doi = {10.1038/s41586-023-06521-7},
archivePrefix = {arXiv},
       eprint = {2303.15431},
 primaryClass = {astro-ph.GA},
       adsurl = {https://ui.adsabs.harvard.edu/abs/2023Natur.622..707A}
}

@ARTICLE{Harikane2023,
       author = {{Harikane}, Yuichi and {Ouchi}, Masami and {Oguri}, Masamune and {Ono}, Yoshiaki and {Nakajima}, Kimihiko and {Isobe}, Yuki and {Umeda}, Hiroya and {Mawatari}, Ken and {Zhang}, Yechi},
        title = "{A Comprehensive Study of Galaxies at z   9-16 Found in the Early JWST Data: Ultraviolet Luminosity Functions and Cosmic Star Formation History at the Pre-reionization Epoch}",
      journal = {\apjs},
         year = 2023,
        month = mar,
       volume = {265},
       number = {1},
          eid = {5},
        pages = {5},
          doi = {10.3847/1538-4365/acaaa9},
archivePrefix = {arXiv},
       eprint = {2208.01612},
 primaryClass = {astro-ph.GA},
       adsurl = {https://ui.adsabs.harvard.edu/abs/2023ApJS..265....5H}
}

@ARTICLE{Donnan2023,
       author = {{Donnan}, C.~T. and {McLeod}, D.~J. and {Dunlop}, J.~S. and {McLure}, R.~J. and {Carnall}, A.~C. and {Begley}, R. and {Cullen}, F. and {Hamadouche}, M.~L. and {Bowler}, R.~A.~A. and {Magee}, D. and {McCracken}, H.~J. and {Milvang-Jensen}, B. and {Moneti}, A. and {Targett}, T.},
        title = "{The evolution of the galaxy UV luminosity function at redshifts z ≃ 8 - 15 from deep JWST and ground-based near-infrared imaging}",
      journal = {\mnras},
         year = 2023,
        month = feb,
       volume = {518},
       number = {4},
        pages = {6011-6040},
          doi = {10.1093/mnras/stac3472},
archivePrefix = {arXiv},
       eprint = {2207.12356},
 primaryClass = {astro-ph.GA},
       adsurl = {https://ui.adsabs.harvard.edu/abs/2023MNRAS.518.6011D}
}

@ARTICLE{Steinhardt2021,
       author = {{Steinhardt}, Charles L. and {Jespersen}, Christian Kragh and {Linzer}, Nora B.},
        title = "{Finding High-redshift Galaxies with JWST}",
      journal = {\apj},
         year = 2021,
        month = dec,
       volume = {923},
       number = {1},
          eid = {8},
        pages = {8},
          doi = {10.3847/1538-4357/ac2a2f},
archivePrefix = {arXiv},
       eprint = {2111.14865},
 primaryClass = {astro-ph.GA},
       adsurl = {https://ui.adsabs.harvard.edu/abs/2021ApJ...923....8S}
}

@ARTICLE{Carniani2024,
       author = {{Carniani}, Stefano and {Hainline}, Kevin and {D'Eugenio}, Francesco and {Eisenstein}, Daniel J. and {Jakobsen}, Peter and {Witstok}, Joris and {Johnson}, Benjamin D. and {Chevallard}, Jacopo and {Maiolino}, Roberto and {Helton}, Jakob M. and {Willott}, Chris and {Robertson}, Brant and {Alberts}, Stacey and {Arribas}, Santiago and {Baker}, William M. and {Bhatawdekar}, Rachana and {Boyett}, Kristan and {Bunker}, Andrew J. and {Cameron}, Alex J. and {Cargile}, Phillip A. and {Charlot}, St{\'e}phane and {Curti}, Mirko and {Curtis-Lake}, Emma and {Egami}, Eiichi and {Giardino}, Giovanna and {Isaak}, Kate and {Ji}, Zhiyuan and {Jones}, Gareth C. and {Kumari}, Nimisha and {Maseda}, Michael V. and {Parlanti}, Eleonora and {P{\'e}rez-Gonz{\'a}lez}, Pablo G. and {Rawle}, Tim and {Rieke}, George and {Rieke}, Marcia and {Del Pino}, Bruno Rodr{\'\i}guez and {Saxena}, Aayush and {Scholtz}, Jan and {Smit}, Renske and {Sun}, Fengwu and {Tacchella}, Sandro and {{\"U}bler}, Hannah and {Venturi}, Giacomo and {Williams}, Christina C. and {Willmer}, Christopher N.~A.},
        title = "{Spectroscopic confirmation of two luminous galaxies at a redshift of 14}",
      journal = {\nat},
         year = 2024,
        month = sep,
       volume = {633},
       number = {8029},
        pages = {318-322},
          doi = {10.1038/s41586-024-07860-9},
archivePrefix = {arXiv},
       eprint = {2405.18485},
 primaryClass = {astro-ph.GA},
       adsurl = {https://ui.adsabs.harvard.edu/abs/2024Natur.633..318C}
}

@ARTICLE{Hainline2024,
       author = {{Hainline}, Kevin N. and {Johnson}, Benjamin D. and {Robertson}, Brant and {Tacchella}, Sandro and {Helton}, Jakob M. and {Sun}, Fengwu and {Eisenstein}, Daniel J. and {Simmonds}, Charlotte and {Topping}, Michael W. and {Whitler}, Lily and {Willmer}, Christopher N.~A. and {Rieke}, Marcia and {Suess}, Katherine A. and {Hviding}, Raphael E. and {Cameron}, Alex J. and {Alberts}, Stacey and {Baker}, William M. and {Baum}, Stefi and {Bhatawdekar}, Rachana and {Bonaventura}, Nina and {Boyett}, Kristan and {Bunker}, Andrew J. and {Carniani}, Stefano and {Charlot}, Stephane and {Chevallard}, Jacopo and {Chen}, Zuyi and {Curti}, Mirko and {Curtis-Lake}, Emma and {D'Eugenio}, Francesco and {Egami}, Eiichi and {Endsley}, Ryan and {Hausen}, Ryan and {Ji}, Zhiyuan and {Looser}, Tobias J. and {Lyu}, Jianwei and {Maiolino}, Roberto and {Nelson}, Erica and {Pusk{\'a}s}, D{\'a}vid and {Rawle}, Tim and {Sandles}, Lester and {Saxena}, Aayush and {Smit}, Renske and {Stark}, Daniel P. and {Williams}, Christina C. and {Willott}, Chris and {Witstok}, Joris},
        title = "{The Cosmos in Its Infancy: JADES Galaxy Candidates at z > 8 in GOODS-S and GOODS-N}",
      journal = {\apj},
         year = 2024,
        month = mar,
       volume = {964},
       number = {1},
          eid = {71},
        pages = {71},
          doi = {10.3847/1538-4357/ad1ee4},
archivePrefix = {arXiv},
       eprint = {2306.02468},
 primaryClass = {astro-ph.GA},
       adsurl = {https://ui.adsabs.harvard.edu/abs/2024ApJ...964...71H}
}

@ARTICLE{Desprez2024,
       author = {{Desprez}, Guillaume and {Martis}, Nicholas S. and {Asada}, Yoshihisa and {Sawicki}, Marcin and {Willott}, Chris J. and {Muzzin}, Adam and {Abraham}, Roberto G. and {Brada{\v{c}}}, Maru{\v{s}}a and {Brammer}, Gabe and {Estrada-Carpenter}, Vicente and {Iyer}, Kartheik G. and {Matharu}, Jasleen and {Mowla}, Lamiya and {Noirot}, Ga{\"e}l and {Sarrouh}, Ghassan T.~E. and {Strait}, Victoria and {Gledhill}, Rachel and {Rihtar{\v{s}}i{\v{c}}}, Gregor},
        title = "{{\ensuremath{\Lambda}}CDM not dead yet: massive high-z Balmer break galaxies are less common than previously reported}",
      journal = {\mnras},
         year = 2024,
        month = may,
       volume = {530},
       number = {3},
        pages = {2935-2952},
          doi = {10.1093/mnras/stae1084},
archivePrefix = {arXiv},
       eprint = {2310.03063},
 primaryClass = {astro-ph.GA},
       adsurl = {https://ui.adsabs.harvard.edu/abs/2024MNRAS.530.2935D}
}

@ARTICLE{Trenti2008,
       author = {{Trenti}, M. and {Stiavelli}, M.},
        title = "{Cosmic Variance and Its Effect on the Luminosity Function Determination in Deep High-z Surveys}",
      journal = {\apj},
         year = 2008,
        month = apr,
       volume = {676},
       number = {2},
        pages = {767-780},
          doi = {10.1086/528674},
archivePrefix = {arXiv},
       eprint = {0712.0398},
 primaryClass = {astro-ph},
       adsurl = {https://ui.adsabs.harvard.edu/abs/2008ApJ...676..767T}
}

@ARTICLE{Topping2024,
       author = {{Topping}, Michael W. and {Stark}, Daniel P. and {Endsley}, Ryan and {Whitler}, Lily and {Hainline}, Kevin and {Johnson}, Benjamin D. and {Robertson}, Brant and {Tacchella}, Sandro and {Chen}, Zuyi and {Alberts}, Stacey and {Baker}, William M. and {Bunker}, Andrew J. and {Carniani}, Stefano and {Charlot}, Stephane and {Chevallard}, Jacopo and {Curtis-Lake}, Emma and {DeCoursey}, Christa and {Egami}, Eiichi and {Eisenstein}, Daniel J. and {Ji}, Zhiyuan and {Maiolino}, Roberto and {Williams}, Christina C. and {Willmer}, Christopher N.~A. and {Willott}, Chris and {Witstok}, Joris},
        title = "{The UV continuum slopes of early star-forming galaxies in JADES}",
      journal = {\mnras},
         year = 2024,
        month = apr,
       volume = {529},
       number = {4},
        pages = {4087-4103},
          doi = {10.1093/mnras/stae800},
archivePrefix = {arXiv},
       eprint = {2307.08835},
 primaryClass = {astro-ph.GA},
       adsurl = {https://ui.adsabs.harvard.edu/abs/2024MNRAS.529.4087T}
}

@ARTICLE{Adams2025,
       author = {{Adams}, N.~J. and {Austin}, D. and {Harvey}, T. and {Conselice}, C.~J. and {Trussler}, J.~A.~A. and {Li}, Q. and {Westcott}, L. and {Ferreira}, L. and {Rusakov}, V. and {Goolsby}, C.~M.},
        title = "{The impact of medium-width bands on the selection, and subsequent luminosity function measurements, of high-z galaxies}",
      journal = {arXiv e-prints},
         year = 2025,
        month = feb,
          eid = {arXiv:2502.10282},
        pages = {arXiv:2502.10282},
          doi = {10.48550/arXiv.2502.10282},
archivePrefix = {arXiv},
       eprint = {2502.10282},
 primaryClass = {astro-ph.GA},
       adsurl = {https://ui.adsabs.harvard.edu/abs/2025arXiv250210282A}
}

@ARTICLE{Yung2024,
       author = {{Yung}, L.~Y. Aaron and {Somerville}, Rachel S. and {Finkelstein}, Steven L. and {Wilkins}, Stephen M. and {Gardner}, Jonathan P.},
        title = "{Are the ultra-high-redshift galaxies at z > 10 surprising in the context of standard galaxy formation models?}",
      journal = {\mnras},
         year = 2024,
        month = jan,
       volume = {527},
       number = {3},
        pages = {5929-5948},
          doi = {10.1093/mnras/stad3484},
archivePrefix = {arXiv},
       eprint = {2304.04348},
 primaryClass = {astro-ph.GA},
       adsurl = {https://ui.adsabs.harvard.edu/abs/2024MNRAS.527.5929Y}
}

@ARTICLE{Kannan2025,
       author = {{Kannan}, Rahul and {Puchwein}, Ewald and {Smith}, Aaron and {Borrow}, Josh and {Garaldi}, Enrico and {Keating}, Laura and {Vogelsberger}, Mark and {Zier}, Oliver and {McClymont}, William and {Shen}, Xuejian and {Popovic}, Filip and {Tacchella}, Sandro and {Hernquist}, Lars and {Springel}, Volker},
        title = "{Introducing the THESAN-ZOOM project: radiation-hydrodynamic simulations of high-redshift galaxies with a multi-phase interstellar medium}",
      journal = {arXiv e-prints},
         year = 2025,
        month = feb,
          eid = {arXiv:2502.20437},
        pages = {arXiv:2502.20437},
          doi = {10.48550/arXiv.2502.20437},
archivePrefix = {arXiv},
       eprint = {2502.20437},
 primaryClass = {astro-ph.GA},
       adsurl = {https://ui.adsabs.harvard.edu/abs/2025arXiv250220437K}
}

@ARTICLE{Behroozi2020,
       author = {{Behroozi}, Peter and {Conroy}, Charlie and {Wechsler}, Risa H. and {Hearin}, Andrew and {Williams}, Christina C. and {Moster}, Benjamin P. and {Yung}, L.~Y. Aaron and {Somerville}, Rachel S. and {Gottl{\"o}ber}, Stefan and {Yepes}, Gustavo and {Endsley}, Ryan},
        title = "{The Universe at z > 10: predictions for JWST from the UNIVERSEMACHINE DR1}",
      journal = {\mnras},
         year = 2020,
        month = dec,
       volume = {499},
       number = {4},
        pages = {5702-5718},
          doi = {10.1093/mnras/staa3164},
archivePrefix = {arXiv},
       eprint = {2007.04988},
 primaryClass = {astro-ph.GA},
       adsurl = {https://ui.adsabs.harvard.edu/abs/2020MNRAS.499.5702B}
}

@ARTICLE{Wilkins2023,
       author = {{Wilkins}, Stephen M. and {Vijayan}, Aswin P. and {Lovell}, Christopher C. and {Roper}, William J. and {Irodotou}, Dimitrios and {Caruana}, Joseph and {Seeyave}, Louise T.~C. and {Kuusisto}, Jussi K. and {Thomas}, Peter A. and {Parris}, Shedeur A.~K.},
        title = "{First light and reionization epoch simulations (FLARES) V: the redshift frontier}",
      journal = {\mnras},
         year = 2023,
        month = feb,
       volume = {519},
       number = {2},
        pages = {3118-3128},
          doi = {10.1093/mnras/stac3280},
archivePrefix = {arXiv},
       eprint = {2204.09431},
 primaryClass = {astro-ph.GA},
       adsurl = {https://ui.adsabs.harvard.edu/abs/2023MNRAS.519.3118W}
}

@ARTICLE{Vijayan2021,
       author = {{Vijayan}, Aswin P. and {Lovell}, Christopher C. and {Wilkins}, Stephen M. and {Thomas}, Peter A. and {Barnes}, David J. and {Irodotou}, Dimitrios and {Kuusisto}, Jussi and {Roper}, William J.},
        title = "{First Light And Reionization Epoch Simulations (FLARES) - II: The photometric properties of high-redshift galaxies}",
      journal = {\mnras},
         year = 2021,
        month = mar,
       volume = {501},
       number = {3},
        pages = {3289-3308},
          doi = {10.1093/mnras/staa3715},
archivePrefix = {arXiv},
       eprint = {2008.06057},
 primaryClass = {astro-ph.GA},
       adsurl = {https://ui.adsabs.harvard.edu/abs/2021MNRAS.501.3289V}
}

@ARTICLE{Harikane2022,
       author = {{Harikane}, Yuichi and {Ono}, Yoshiaki and {Ouchi}, Masami and {Liu}, Chengze and {Sawicki}, Marcin and {Shibuya}, Takatoshi and {Behroozi}, Peter S. and {He}, Wanqiu and {Shimasaku}, Kazuhiro and {Arnouts}, Stephane and {Coupon}, Jean and {Fujimoto}, Seiji and {Gwyn}, Stephen and {Huang}, Jiasheng and {Inoue}, Akio K. and {Kashikawa}, Nobunari and {Komiyama}, Yutaka and {Matsuoka}, Yoshiki and {Willott}, Chris J.},
        title = "{GOLDRUSH. IV. Luminosity Functions and Clustering Revealed with  4,000,000 Galaxies at z   2-7: Galaxy-AGN Transition, Star Formation Efficiency, and Implication for Evolution at z > 10}",
      journal = {\apjs},
         year = 2022,
        month = mar,
       volume = {259},
       number = {1},
          eid = {20},
        pages = {20},
          doi = {10.3847/1538-4365/ac3dfc},
archivePrefix = {arXiv},
       eprint = {2108.01090},
 primaryClass = {astro-ph.GA},
       adsurl = {https://ui.adsabs.harvard.edu/abs/2022ApJS..259...20H}
}

@ARTICLE{Gehrels1986,
       author = {{Gehrels}, N.},
        title = "{Confidence Limits for Small Numbers of Events in Astrophysical Data}",
      journal = {\apj},
         year = 1986,
        month = apr,
       volume = {303},
        pages = {336},
          doi = {10.1086/164079},
       adsurl = {https://ui.adsabs.harvard.edu/abs/1986ApJ...303..336G}
}

@ARTICLE{Avni1980,
       author = {{Avni}, Y. and {Bahcall}, J.~N.},
        title = "{On the simultaneous analysis of several complete samples. The V/Vmax and Ve/Va variables, with applications to quasars.}",
      journal = {\apj},
         year = 1980,
        month = feb,
       volume = {235},
        pages = {694-716},
          doi = {10.1086/157673},
       adsurl = {https://ui.adsabs.harvard.edu/abs/1980ApJ...235..694A}
}

@ARTICLE{Marshall1983,
       author = {{Marshall}, H.~L. and {Tananbaum}, H. and {Avni}, Y. and {Zamorani}, G.},
        title = "{Analysis of complete quasar samples to obtain parameters of luminosity and evolution functions}",
      journal = {\apj},
         year = 1983,
        month = jun,
       volume = {269},
        pages = {35-41},
          doi = {10.1086/161016},
       adsurl = {https://ui.adsabs.harvard.edu/abs/1983ApJ...269...35M}
}

@ARTICLE{Adams2024ApJ,
       author = {{Adams}, Nathan J. and {Conselice}, Christopher J. and {Austin}, Duncan and {Harvey}, Thomas and {Ferreira}, Leonardo and {Trussler}, James and {Juod{\v{z}}balis}, Ignas and {Li}, Qiong and {Windhorst}, Rogier and {Cohen}, Seth H. and {Jansen}, Rolf A. and {Summers}, Jake and {Tompkins}, Scott and {Driver}, Simon P. and {Robotham}, Aaron and {D'Silva}, Jordan C.~J. and {Yan}, Haojing and {Coe}, Dan and {Frye}, Brenda and {Grogin}, Norman A. and {Koekemoer}, Anton M. and {Marshall}, Madeline A. and {Pirzkal}, Nor and {Ryan}, Russell E. and {Maksym}, W. Peter and {Rutkowski}, Michael J. and {Willmer}, Christopher N.~A. and {Hammel}, Heidi B. and {Nonino}, Mario and {Bhatawdekar}, Rachana and {Wilkins}, Stephen M. and {Bradley}, Larry D. and {Broadhurst}, Tom and {Cheng}, Cheng and {Dole}, Herv{\'e} and {Hathi}, Nimish P. and {Zitrin}, Adi},
        title = "{EPOCHS. II. The Ultraviolet Luminosity Function from 7.5 < z < 13.5 Using 180 arcmin$^{2}$ of Deep, Blank Fields from the PEARLS Survey and Public JWST Data}",
      journal = {\apj},
         year = 2024,
        month = apr,
       volume = {965},
       number = {2},
          eid = {169},
        pages = {169},
          doi = {10.3847/1538-4357/ad2a7b},
archivePrefix = {arXiv},
       eprint = {2304.13721},
 primaryClass = {astro-ph.GA},
       adsurl = {https://ui.adsabs.harvard.edu/abs/2024ApJ...965..169A}
}

@ARTICLE{Donnan2024,
       author = {{Donnan}, C.~T. and {McLure}, R.~J. and {Dunlop}, J.~S. and {McLeod}, D.~J. and {Magee}, D. and {Arellano-C{\'o}rdova}, K.~Z. and {Barrufet}, L. and {Begley}, R. and {Bowler}, R.~A.~A. and {Carnall}, A.~C. and {Cullen}, F. and {Ellis}, R.~S. and {Fontana}, A. and {Illingworth}, G.~D. and {Grogin}, N.~A. and {Hamadouche}, M.~L. and {Koekemoer}, A.~M. and {Liu}, F. -Y. and {Mason}, C. and {Santini}, P. and {Stanton}, T.~M.},
        title = "{JWST PRIMER: a new multifield determination of the evolving galaxy UV luminosity function at redshifts z ≃ 9 - 15}",
      journal = {\mnras},
         year = 2024,
        month = sep,
       volume = {533},
       number = {3},
        pages = {3222-3237},
          doi = {10.1093/mnras/stae2037},
archivePrefix = {arXiv},
       eprint = {2403.03171},
 primaryClass = {astro-ph.GA},
       adsurl = {https://ui.adsabs.harvard.edu/abs/2024MNRAS.533.3222D}
}

@ARTICLE{Perez_Gonzalez2023,
       author = {{P{\'e}rez-Gonz{\'a}lez}, Pablo G. and {Costantin}, Luca and {Langeroodi}, Danial and {Rinaldi}, Pierluigi and {Annunziatella}, Marianna and {Ilbert}, Olivier and {Colina}, Luis and {N{\o}rgaard-Nielsen}, Hans Ulrik and {Greve}, Thomas R. and {{\"O}stlin}, G{\"o}ran and {Wright}, Gillian and {Alonso-Herrero}, Almudena and {{\'A}lvarez-M{\'a}rquez}, Javier and {Caputi}, Karina I. and {Eckart}, Andreas and {Le F{\`e}vre}, Olivier and {Labiano}, {\'A}lvaro and {Garc{\'\i}a-Mar{\'\i}n}, Macarena and {Hjorth}, Jens and {Kendrew}, Sarah and {Pye}, John P. and {Tikkanen}, Tuomo and {van der Werf}, Paul and {Walter}, Fabian and {Ward}, Martin and {Bik}, Arjan and {Boogaard}, Leindert and {Bosman}, Sarah E.~I. and {G{\'o}mez}, Alejandro Crespo and {Gillman}, Steven and {Iani}, Edoardo and {Jermann}, Iris and {Melinder}, Jens and {Meyer}, Romain A. and {Moutard}, Thibaud and {van Dishoek}, Ewine and {Henning}, Thomas and {Lagage}, Pierre-Olivier and {Guedel}, Manuel and {Peissker}, Florian and {Ray}, Tom and {Vandenbussche}, Bart and {Garc{\'\i}a-Argum{\'a}nez}, {\'A}ngela and {Mar{\'\i}a M{\'e}rida}, Rosa},
        title = "{Life beyond 30: Probing the -20 < M $_{UV}$ < -17 Luminosity Function at 8 < z < 13 with the NIRCam Parallel Field of the MIRI Deep Survey}",
      journal = {\apjl},
         year = 2023,
        month = jul,
       volume = {951},
       number = {1},
          eid = {L1},
        pages = {L1},
          doi = {10.3847/2041-8213/acd9d0},
archivePrefix = {arXiv},
       eprint = {2302.02429},
 primaryClass = {astro-ph.GA},
       adsurl = {https://ui.adsabs.harvard.edu/abs/2023ApJ...951L...1P}
}

@ARTICLE{Harikane2025ApJ,
       author = {{Harikane}, Yuichi and {Inoue}, Akio K. and {Ellis}, Richard S. and {Ouchi}, Masami and {Nakazato}, Yurina and {Yoshida}, Naoki and {Ono}, Yoshiaki and {Sun}, Fengwu and {Sato}, Riku A. and {Ferrami}, Giovanni and {Fujimoto}, Seiji and {Kashikawa}, Nobunari and {McLeod}, Derek J. and {P{\'e}rez-Gonz{\'a}lez}, Pablo G. and {Sawicki}, Marcin and {Sugahara}, Yuma and {Xu}, Yi and {Yamanaka}, Satoshi and {Carnall}, Adam C. and {Cullen}, Fergus and {Dunlop}, James S. and {Egami}, Eiichi and {Grogin}, Norman and {Isobe}, Yuki and {Koekemoer}, Anton M. and {Laporte}, Nicolas and {Lee}, Chien-Hsiu and {Magee}, Dan and {Matsuo}, Hiroshi and {Matsuoka}, Yoshiki and {Mawatari}, Ken and {Nakajima}, Kimihiko and {Nakane}, Minami and {Tamura}, Yoichi and {Umeda}, Hiroya and {Yanagisawa}, Hiroto},
        title = "{JWST, ALMA, and Keck Spectroscopic Constraints on the UV Luminosity Functions at z {\ensuremath{\sim}} 7{\textendash}14: Clumpiness and Compactness of the Brightest Galaxies in the Early Universe}",
      journal = {\apj},
         year = 2025,
        month = feb,
       volume = {980},
       number = {1},
          eid = {138},
        pages = {138},
          doi = {10.3847/1538-4357/ad9b2c},
archivePrefix = {arXiv},
       eprint = {2406.18352},
 primaryClass = {astro-ph.GA},
       adsurl = {https://ui.adsabs.harvard.edu/abs/2025ApJ...980..138H}
}

@ARTICLE{Perez_Gonzalez2025,
       author = {{P{\'e}rez-Gonz{\'a}lez}, Pablo G. and {{\"O}stlin}, G{\"o}ran and {Costantin}, Luca and {Melinder}, Jens and {Finkelstein}, Steven L. and {Somerville}, Rachel S. and {Annunziatella}, Marianna and {{\'A}lvarez-M{\'a}rquez}, Javier and {Colina}, Luis and {Dekel}, Avishai and {Ferguson}, Henry C. and {Li}, Zhaozhou and {Yung}, L.~Y. Aaron and {Bagley}, Mic B. and {Boogard}, Leindert A. and {Burgarella}, Denis and {Calabr{\`o}}, Antonello and {Caputi}, Karina I. and {Cheng}, Yingjie and {Eckart}, Andreas and {Giavalisco}, Mauro and {Gillman}, Steven and {Greve}, Thomas R. and {Hathi}, Nimish P. and {Hjorth}, Jens and {Huertas-Company}, Marc and {Kartaltepe}, Jeyhan and {Koekemoer}, Anton M. and {Kokorev}, Vasily and {Labiano}, {\'A}lvaro and {Langeroodi}, Danial and {Leung}, Gene C.~K. and {Natarajan}, Priyamvada and {Papovich}, Casey and {Peissker}, Florian and {Pentericci}, Laura and {Pirzkal}, Nor and {Rinaldi}, Pierluigi and {van der Werf}, Paul and {Walter}, Fabian},
        title = "{The rise of the galactic empire: luminosity functions at $z\sim17$ and $z\sim25$ estimated with the MIDIS$+$NGDEEP ultra-deep JWST/NIRCam dataset}",
      journal = {arXiv e-prints},
         year = 2025,
        month = mar,
          eid = {arXiv:2503.15594},
        pages = {arXiv:2503.15594},
          doi = {10.48550/arXiv.2503.15594},
archivePrefix = {arXiv},
       eprint = {2503.15594},
 primaryClass = {astro-ph.GA},
       adsurl = {https://ui.adsabs.harvard.edu/abs/2025arXiv250315594P}
}

@ARTICLE{Castellano2025arXiv,
       author = {{Castellano}, M. and {Fontana}, A. and {Merlin}, E. and {Santini}, P. and {Napolitano}, L. and {Menci}, N. and {Calabr{\`o}}, A. and {Paris}, D. and {Pentericci}, L. and {Zavala}, J. and {Dickinson}, M. and {Finkelstein}, S.~L. and {Treu}, T. and {Amorin}, R.~O. and {Arrabal Haro}, P. and {Bergamini}, P. and {Bisigello}, L. and {Daddi}, E. and {Dayal}, P. and {Dekel}, A. and {Ferrara}, A. and {Fortuni}, F. and {Gandolfi}, G. and {Giavalisco}, M. and {Grillo}, C. and {Guida}, S.~T. and {Hathi}, N.~P. and {Holwerda}, B.~W. and {Koekemoer}, A.~M. and {Kokorev}, V. and {Li}, Z. and {Llerena}, M. and {Lucas}, R.~A. and {Mascia}, S. and {Metha}, B. and {Morishita}, T. and {Nanayakkara}, T. and {Pacucci}, F. and {P{\'e}rez-Gonz{\'a}lez}, P.~G. and {Roberts-Borsani}, G. and {Rodighiero}, G. and {Rosati}, P. and {Salazar}, V. and {Schneider}, R. and {Somerville}, R.~S. and {Taylor}, A. and {Trenti}, M. and {Trinca}, A. and {Wang}, X. and {Watson}, P.~J. and {Yang}, L. and {Yung}, L.~Y.~A.},
        title = "{Pushing JWST to the extremes: search and scrutiny of bright galaxy candidates at z$\simeq$15-30}",
      journal = {arXiv e-prints},
         year = 2025,
        month = apr,
          eid = {arXiv:2504.05893},
        pages = {arXiv:2504.05893},
          doi = {10.48550/arXiv.2504.05893},
archivePrefix = {arXiv},
       eprint = {2504.05893},
 primaryClass = {astro-ph.GA},
       adsurl = {https://ui.adsabs.harvard.edu/abs/2025arXiv250405893C}
}

@ARTICLE{Robertson2024ApJ,
       author = {{Robertson}, Brant and {Johnson}, Benjamin D. and {Tacchella}, Sandro and {Eisenstein}, Daniel J. and {Hainline}, Kevin and {Arribas}, Santiago and {Baker}, William M. and {Bunker}, Andrew J. and {Carniani}, Stefano and {Cargile}, Phillip A. and {Carreira}, Courtney and {Charlot}, Stephane and {Chevallard}, Jacopo and {Curti}, Mirko and {Curtis-Lake}, Emma and {D'Eugenio}, Francesco and {Egami}, Eiichi and {Hausen}, Ryan and {Helton}, Jakob M. and {Jakobsen}, Peter and {Ji}, Zhiyuan and {Jones}, Gareth C. and {Maiolino}, Roberto and {Maseda}, Michael V. and {Nelson}, Erica and {P{\'e}rez-Gonz{\'a}lez}, Pablo G. and {Pusk{\'a}s}, D{\'a}vid and {Rieke}, Marcia and {Smit}, Renske and {Sun}, Fengwu and {{\"U}bler}, Hannah and {Whitler}, Lily and {Williams}, Christina C. and {Willmer}, Christopher N.~A. and {Willott}, Chris and {Witstok}, Joris},
        title = "{Earliest Galaxies in the JADES Origins Field: Luminosity Function and Cosmic Star Formation Rate Density 300 Myr after the Big Bang}",
      journal = {\apj},
         year = 2024,
        month = jul,
       volume = {970},
       number = {1},
          eid = {31},
        pages = {31},
          doi = {10.3847/1538-4357/ad463d},
archivePrefix = {arXiv},
       eprint = {2312.10033},
 primaryClass = {astro-ph.GA},
       adsurl = {https://ui.adsabs.harvard.edu/abs/2024ApJ...970...31R}
}

@ARTICLE{Schechter1976ApJ,
       author = {{Schechter}, P.},
        title = "{An analytic expression for the luminosity function for galaxies.}",
      journal = {\apj},
         year = 1976,
        month = jan,
       volume = {203},
        pages = {297-306},
          doi = {10.1086/154079},
       adsurl = {https://ui.adsabs.harvard.edu/abs/1976ApJ...203..297S}
}

@ARTICLE{Morishita2024,
       author = {{Morishita}, Takahiro and {Stiavelli}, Massimo and {Chary}, Ranga-Ram and {Trenti}, Michele and {Bergamini}, Pietro and {Chiaberge}, Marco and {Leethochawalit}, Nicha and {Roberts-Borsani}, Guido and {Shen}, Xuejian and {Treu}, Tommaso},
        title = "{Enhanced Subkiloparsec-scale Star Formation: Results from a JWST Size Analysis of 341 Galaxies at 5 < z < 14}",
      journal = {\apj},
         year = 2024,
        month = mar,
       volume = {963},
       number = {1},
          eid = {9},
        pages = {9},
          doi = {10.3847/1538-4357/ad1404},
archivePrefix = {arXiv},
       eprint = {2308.05018},
 primaryClass = {astro-ph.GA},
       adsurl = {https://ui.adsabs.harvard.edu/abs/2024ApJ...963....9M}
}

@ARTICLE{Cullen2024MNRAS,
       author = {{Cullen}, F. and {McLeod}, D.~J. and {McLure}, R.~J. and {Dunlop}, J.~S. and {Donnan}, C.~T. and {Carnall}, A.~C. and {Keating}, L.~C. and {Magee}, D. and {Arellano-Cordova}, K.~Z. and {Bowler}, R.~A.~A. and {Begley}, R. and {Flury}, S.~R. and {Hamadouche}, M.~L. and {Stanton}, T.~M.},
        title = "{The ultraviolet continuum slopes of high-redshift galaxies: evidence for the emergence of dust-free stellar populations at z > 10}",
      journal = {\mnras},
         year = 2024,
        month = jun,
       volume = {531},
       number = {1},
        pages = {997-1020},
          doi = {10.1093/mnras/stae1211},
archivePrefix = {arXiv},
       eprint = {2311.06209},
 primaryClass = {astro-ph.GA},
       adsurl = {https://ui.adsabs.harvard.edu/abs/2024MNRAS.531..997C}
}

@ARTICLE{DeCoursey2025ApJ,
       author = {{DeCoursey}, Christa and {Egami}, Eiichi and {Pierel}, Justin D.~R. and {Sun}, Fengwu and {Rest}, Armin and {Coulter}, David A. and {Engesser}, Michael and {Siebert}, Matthew R. and {Hainline}, Kevin N. and {Johnson}, Benjamin D. and {Bunker}, Andrew J. and {Cargile}, Phillip A. and {Charlot}, Stephane and {Chen}, Wenlei and {Curti}, Mirko and {DeFour-Remy}, Shea and {Eisenstein}, Daniel J. and {Fox}, Ori D. and {Gezari}, Suvi and {Gomez}, Sebastian and {Jencson}, Jacob and {Joshi}, Bhavin A. and {Khairnar}, Sanvi and {Lyu}, Jianwei and {Maiolino}, Roberto and {Moriya}, Takashi J. and {Quimby}, Robert M. and {Rieke}, George H. and {Rieke}, Marcia J. and {Robertson}, Brant and {Shahbandeh}, Melissa and {Strolger}, Louis-Gregory and {Tacchella}, Sandro and {Wang}, Qinan and {Williams}, Christina C. and {Willmer}, Christopher N.~A. and {Willott}, Chris and {Zenati}, Yossef},
        title = "{The JADES Transient Survey: Discovery and Classification of Supernovae in the JADES Deep Field}",
      journal = {\apj},
         year = 2025,
        month = feb,
       volume = {979},
       number = {2},
          eid = {250},
        pages = {250},
          doi = {10.3847/1538-4357/ad8fab},
archivePrefix = {arXiv},
       eprint = {2406.05060},
 primaryClass = {astro-ph.HE},
       adsurl = {https://ui.adsabs.harvard.edu/abs/2025ApJ...979..250D}
}

@ARTICLE{Yan2023ApJS,
       author = {{Yan}, Haojing and {Ma}, Zhiyuan and {Sun}, Bangzheng and {Wang}, Lifan and {Kelly}, Patrick and {Diego}, Jos{\'e} M. and {Cohen}, Seth H. and {Windhorst}, Rogier A. and {Jansen}, Rolf A. and {Grogin}, Norman A. and {Beacom}, John F. and {Conselice}, Christopher J. and {Driver}, Simon P. and {Frye}, Brenda and {Coe}, Dan and {Marshall}, Madeline A. and {Koekemoer}, Anton and {Willmer}, Christopher N.~A. and {Robotham}, Aaron and {D'Silva}, Jordan C.~J. and {Summers}, Jake and {Nonino}, Mario and {Pirzkal}, Nor and {Ryan}, Russell E. and {Ortiz}, Rafael and {Tompkins}, Scott and {Bhatawdekar}, Rachana A. and {Cheng}, Cheng and {Zitrin}, Adi and {Willner}, S.~P.},
        title = "{JWST's PEARLS: Transients in the MACS J0416.1-2403 Field}",
      journal = {\apjs},
         year = 2023,
        month = dec,
       volume = {269},
       number = {2},
          eid = {43},
        pages = {43},
          doi = {10.3847/1538-4365/ad0298},
archivePrefix = {arXiv},
       eprint = {2307.07579},
 primaryClass = {astro-ph.GA},
       adsurl = {https://ui.adsabs.harvard.edu/abs/2023ApJS..269...43Y}
}

@ARTICLE{Inoue2014MNRAS,
       author = {{Inoue}, Akio K. and {Shimizu}, Ikkoh and {Iwata}, Ikuru and {Tanaka}, Masayuki},
        title = "{An updated analytic model for attenuation by the intergalactic medium}",
      journal = {\mnras},
         year = 2014,
        month = aug,
       volume = {442},
       number = {2},
        pages = {1805-1820},
          doi = {10.1093/mnras/stu936},
archivePrefix = {arXiv},
       eprint = {1402.0677},
 primaryClass = {astro-ph.CO},
       adsurl = {https://ui.adsabs.harvard.edu/abs/2014MNRAS.442.1805I}
}

@ARTICLE{Willott2024ApJ,
       author = {{Willott}, Chris J. and {Desprez}, Guillaume and {Asada}, Yoshihisa and {Sarrouh}, Ghassan T.~E. and {Abraham}, Roberto and {Brada{\v{c}}}, Maru{\v{s}}a and {Brammer}, Gabe and {Estrada-Carpenter}, Vince and {Iyer}, Kartheik G. and {Martis}, Nicholas S. and {Matharu}, Jasleen and {Mowla}, Lamiya and {Muzzin}, Adam and {Noirot}, Ga{\"e}l and {Sawicki}, Marcin and {Strait}, Victoria and {Rihtar{\v{s}}i{\v{c}}}, Gregor and {Withers}, Sunna},
        title = "{A Steep Decline in the Galaxy Space Density beyond Redshift 9 in the CANUCS UV Luminosity Function}",
      journal = {\apj},
         year = 2024,
        month = may,
       volume = {966},
       number = {1},
          eid = {74},
        pages = {74},
          doi = {10.3847/1538-4357/ad35bc},
archivePrefix = {arXiv},
       eprint = {2311.12234},
 primaryClass = {astro-ph.GA},
       adsurl = {https://ui.adsabs.harvard.edu/abs/2024ApJ...966...74W}
}

@ARTICLE{Larson2023ApJL,
       author = {{Larson}, Rebecca L. and {Hutchison}, Taylor A. and {Bagley}, Micaela and {Finkelstein}, Steven L. and {Yung}, L.~Y. Aaron and {Somerville}, Rachel S. and {Hirschmann}, Michaela and {Brammer}, Gabriel and {Holwerda}, Benne W. and {Papovich}, Casey and {Morales}, Alexa M. and {Wilkins}, Stephen M.},
        title = "{Spectral Templates Optimal for Selecting Galaxies at z > 8 with the JWST}",
      journal = {\apj},
         year = 2023,
        month = dec,
       volume = {958},
       number = {2},
          eid = {141},
        pages = {141},
          doi = {10.3847/1538-4357/acfed4},
archivePrefix = {arXiv},
       eprint = {2211.10035},
 primaryClass = {astro-ph.GA},
       adsurl = {https://ui.adsabs.harvard.edu/abs/2023ApJ...958..141L}
}

@software{Conroy2010ascl,
       author = {{Conroy}, Charlie and {Gunn}, James E.},
        title = "{FSPS: Flexible Stellar Population Synthesis}",
 howpublished = {Astrophysics Source Code Library, record ascl:1010.043},
         year = 2010,
        month = oct,
          eid = {ascl:1010.043},
       adsurl = {https://ui.adsabs.harvard.edu/abs/2010ascl.soft10043C}
}

@ARTICLE{Brammer2008ApJ,
       author = {{Brammer}, Gabriel B. and {van Dokkum}, Pieter G. and {Coppi}, Paolo},
        title = "{EAZY: A Fast, Public Photometric Redshift Code}",
      journal = {\apj},
         year = 2008,
        month = oct,
       volume = {686},
       number = {2},
        pages = {1503-1513},
          doi = {10.1086/591786},
archivePrefix = {arXiv},
       eprint = {0807.1533},
 primaryClass = {astro-ph},
       adsurl = {https://ui.adsabs.harvard.edu/abs/2008ApJ...686.1503B}
}

@ARTICLE{Asada2025ApJL,
       author = {{Asada}, Yoshihisa and {Desprez}, Guillaume and {Willott}, Chris J. and {Sawicki}, Marcin and {Brada{\v{c}}}, Maru{\v{s}}a and {Brammer}, Gabriel and {Dubath}, Florian and {Iyer}, Kartheik G. and {Martis}, Nicholas S. and {Muzzin}, Adam and {Noirot}, Ga{\"e}l and {Paltani}, St{\'e}phane and {Sarrouh}, Ghassan T.~E. and {Harshan}, Anishya and {Markov}, Vladan},
        title = "{Improving Photometric Redshifts of Epoch of Reionization Galaxies: A New Empirical Transmission Curve with Neutral Hydrogen Damping Wing Ly{\ensuremath{\alpha}} Absorption}",
      journal = {\apjl},
         year = 2025,
        month = apr,
       volume = {983},
       number = {1},
          eid = {L2},
        pages = {L2},
          doi = {10.3847/2041-8213/adc388},
archivePrefix = {arXiv},
       eprint = {2410.21543},
 primaryClass = {astro-ph.GA},
       adsurl = {https://ui.adsabs.harvard.edu/abs/2025ApJ...983L...2A}
}

@ARTICLE{Fitzpatrick1999PASP,
       author = {{Fitzpatrick}, Edward L.},
        title = "{Correcting for the Effects of Interstellar Extinction}",
      journal = {\pasp},
         year = 1999,
        month = jan,
       volume = {111},
       number = {755},
        pages = {63-75},
          doi = {10.1086/316293},
archivePrefix = {arXiv},
       eprint = {astro-ph/9809387},
 primaryClass = {astro-ph},
       adsurl = {https://ui.adsabs.harvard.edu/abs/1999PASP..111...63F}
}

@ARTICLE{Lotz2017ApJ,
       author = {{Lotz}, J.~M. and {Koekemoer}, A. and {Coe}, D. and {Grogin}, N. and {Capak}, P. and {Mack}, J. and {Anderson}, J. and {Avila}, R. and {Barker}, E.~A. and {Borncamp}, D. and {Brammer}, G. and {Durbin}, M. and {Gunning}, H. and {Hilbert}, B. and {Jenkner}, H. and {Khandrika}, H. and {Levay}, Z. and {Lucas}, R.~A. and {MacKenty}, J. and {Ogaz}, S. and {Porterfield}, B. and {Reid}, N. and {Robberto}, M. and {Royle}, P. and {Smith}, L.~J. and {Storrie-Lombardi}, L.~J. and {Sunnquist}, B. and {Surace}, J. and {Taylor}, D.~C. and {Williams}, R. and {Bullock}, J. and {Dickinson}, M. and {Finkelstein}, S. and {Natarajan}, P. and {Richard}, J. and {Robertson}, B. and {Tumlinson}, J. and {Zitrin}, A. and {Flanagan}, K. and {Sembach}, K. and {Soifer}, B.~T. and {Mountain}, M.},
        title = "{The Frontier Fields: Survey Design and Initial Results}",
      journal = {\apj},
         year = 2017,
        month = mar,
       volume = {837},
       number = {1},
          eid = {97},
        pages = {97},
          doi = {10.3847/1538-4357/837/1/97},
archivePrefix = {arXiv},
       eprint = {1605.06567},
 primaryClass = {astro-ph.GA},
       adsurl = {https://ui.adsabs.harvard.edu/abs/2017ApJ...837...97L}
}

@ARTICLE{2022ApJ...935..167A,
       author = {{Astropy Collaboration} and {Price-Whelan}, Adrian M. and {Lim}, Pey Lian and {Earl}, Nicholas and {Starkman}, Nathaniel and {Bradley}, Larry and {Shupe}, David L. and {Patil}, Aarya A. and {Corrales}, Lia and {Brasseur}, C.~E. and {N{\"o}the}, Maximilian and {Donath}, Axel and {Tollerud}, Erik and {Morris}, Brett M. and {Ginsburg}, Adam and {Vaher}, Eero and {Weaver}, Benjamin A. and {Tocknell}, James and {Jamieson}, William and {van Kerkwijk}, Marten H. and {Robitaille}, Thomas P. and {Merry}, Bruce and {Bachetti}, Matteo and {G{\"u}nther}, H. Moritz and {Aldcroft}, Thomas L. and {Alvarado-Montes}, Jaime A. and {Archibald}, Anne M. and {B{\'o}di}, Attila and {Bapat}, Shreyas and {Barentsen}, Geert and {Baz{\'a}n}, Juanjo and {Biswas}, Manish and {Boquien}, M{\'e}d{\'e}ric and {Burke}, D.~J. and {Cara}, Daria and {Cara}, Mihai and {Conroy}, Kyle E. and {Conseil}, Simon and {Craig}, Matthew W. and {Cross}, Robert M. and {Cruz}, Kelle L. and {D'Eugenio}, Francesco and {Dencheva}, Nadia and {Devillepoix}, Hadrien A.~R. and {Dietrich}, J{\"o}rg P. and {Eigenbrot}, Arthur Davis and {Erben}, Thomas and {Ferreira}, Leonardo and {Foreman-Mackey}, Daniel and {Fox}, Ryan and {Freij}, Nabil and {Garg}, Suyog and {Geda}, Robel and {Glattly}, Lauren and {Gondhalekar}, Yash and {Gordon}, Karl D. and {Grant}, David and {Greenfield}, Perry and {Groener}, Austen M. and {Guest}, Steve and {Gurovich}, Sebastian and {Handberg}, Rasmus and {Hart}, Akeem and {Hatfield-Dodds}, Zac and {Homeier}, Derek and {Hosseinzadeh}, Griffin and {Jenness}, Tim and {Jones}, Craig K. and {Joseph}, Prajwel and {Kalmbach}, J. Bryce and {Karamehmetoglu}, Emir and {Ka{\l}uszy{\'n}ski}, Miko{\l}aj and {Kelley}, Michael S.~P. and {Kern}, Nicholas and {Kerzendorf}, Wolfgang E. and {Koch}, Eric W. and {Kulumani}, Shankar and {Lee}, Antony and {Ly}, Chun and {Ma}, Zhiyuan and {MacBride}, Conor and {Maljaars}, Jakob M. and {Muna}, Demitri and {Murphy}, N.~A. and {Norman}, Henrik and {O'Steen}, Richard and {Oman}, Kyle A. and {Pacifici}, Camilla and {Pascual}, Sergio and {Pascual-Granado}, J. and {Patil}, Rohit R. and {Perren}, Gabriel I. and {Pickering}, Timothy E. and {Rastogi}, Tanuj and {Roulston}, Benjamin R. and {Ryan}, Daniel F. and {Rykoff}, Eli S. and {Sabater}, Jose and {Sakurikar}, Parikshit and {Salgado}, Jes{\'u}s and {Sanghi}, Aniket and {Saunders}, Nicholas and {Savchenko}, Volodymyr and {Schwardt}, Ludwig and {Seifert-Eckert}, Michael and {Shih}, Albert Y. and {Jain}, Anany Shrey and {Shukla}, Gyanendra and {Sick}, Jonathan and {Simpson}, Chris and {Singanamalla}, Sudheesh and {Singer}, Leo P. and {Singhal}, Jaladh and {Sinha}, Manodeep and {Sip{\H{o}}cz}, Brigitta M. and {Spitler}, Lee R. and {Stansby}, David and {Streicher}, Ole and {{\v{S}}umak}, Jani and {Swinbank}, John D. and {Taranu}, Dan S. and {Tewary}, Nikita and {Tremblay}, Grant R. and {de Val-Borro}, Miguel and {Van Kooten}, Samuel J. and {Vasovi{\'c}}, Zlatan and {Verma}, Shresth and {de Miranda Cardoso}, Jos{\'e} Vin{\'\i}cius and {Williams}, Peter K.~G. and {Wilson}, Tom J. and {Winkel}, Benjamin and {Wood-Vasey}, W.~M. and {Xue}, Rui and {Yoachim}, Peter and {Zhang}, Chen and {Zonca}, Andrea and {Astropy Project Contributors}},
        title = "{The Astropy Project: Sustaining and Growing a Community-oriented Open-source Project and the Latest Major Release (v5.0) of the Core Package}",
      journal = {\apj},
         year = 2022,
        month = aug,
       volume = {935},
       number = {2},
          eid = {167},
        pages = {167},
          doi = {10.3847/1538-4357/ac7c74},
archivePrefix = {arXiv},
       eprint = {2206.14220},
 primaryClass = {astro-ph.IM},
       adsurl = {https://ui.adsabs.harvard.edu/abs/2022ApJ...935..167A}
}

@ARTICLE{2018AJ....156..123A,
       author = {{Astropy Collaboration} and {Price-Whelan}, A.~M. and {Sip{\H{o}}cz}, B.~M. and {G{\"u}nther}, H.~M. and {Lim}, P.~L. and {Crawford}, S.~M. and {Conseil}, S. and {Shupe}, D.~L. and {Craig}, M.~W. and {Dencheva}, N. and {Ginsburg}, A. and {VanderPlas}, J.~T. and {Bradley}, L.~D. and {P{\'e}rez-Su{\'a}rez}, D. and {de Val-Borro}, M. and {Aldcroft}, T.~L. and {Cruz}, K.~L. and {Robitaille}, T.~P. and {Tollerud}, E.~J. and {Ardelean}, C. and {Babej}, T. and {Bach}, Y.~P. and {Bachetti}, M. and {Bakanov}, A.~V. and {Bamford}, S.~P. and {Barentsen}, G. and {Barmby}, P. and {Baumbach}, A. and {Berry}, K.~L. and {Biscani}, F. and {Boquien}, M. and {Bostroem}, K.~A. and {Bouma}, L.~G. and {Brammer}, G.~B. and {Bray}, E.~M. and {Breytenbach}, H. and {Buddelmeijer}, H. and {Burke}, D.~J. and {Calderone}, G. and {Cano Rodr{\'\i}guez}, J.~L. and {Cara}, M. and {Cardoso}, J.~V.~M. and {Cheedella}, S. and {Copin}, Y. and {Corrales}, L. and {Crichton}, D. and {D'Avella}, D. and {Deil}, C. and {Depagne}, {\'E}. and {Dietrich}, J.~P. and {Donath}, A. and {Droettboom}, M. and {Earl}, N. and {Erben}, T. and {Fabbro}, S. and {Ferreira}, L.~A. and {Finethy}, T. and {Fox}, R.~T. and {Garrison}, L.~H. and {Gibbons}, S.~L.~J. and {Goldstein}, D.~A. and {Gommers}, R. and {Greco}, J.~P. and {Greenfield}, P. and {Groener}, A.~M. and {Grollier}, F. and {Hagen}, A. and {Hirst}, P. and {Homeier}, D. and {Horton}, A.~J. and {Hosseinzadeh}, G. and {Hu}, L. and {Hunkeler}, J.~S. and {Ivezi{\'c}}, {\v{Z}}. and {Jain}, A. and {Jenness}, T. and {Kanarek}, G. and {Kendrew}, S. and {Kern}, N.~S. and {Kerzendorf}, W.~E. and {Khvalko}, A. and {King}, J. and {Kirkby}, D. and {Kulkarni}, A.~M. and {Kumar}, A. and {Lee}, A. and {Lenz}, D. and {Littlefair}, S.~P. and {Ma}, Z. and {Macleod}, D.~M. and {Mastropietro}, M. and {McCully}, C. and {Montagnac}, S. and {Morris}, B.~M. and {Mueller}, M. and {Mumford}, S.~J. and {Muna}, D. and {Murphy}, N.~A. and {Nelson}, S. and {Nguyen}, G.~H. and {Ninan}, J.~P. and {N{\"o}the}, M. and {Ogaz}, S. and {Oh}, S. and {Parejko}, J.~K. and {Parley}, N. and {Pascual}, S. and {Patil}, R. and {Patil}, A.~A. and {Plunkett}, A.~L. and {Prochaska}, J.~X. and {Rastogi}, T. and {Reddy Janga}, V. and {Sabater}, J. and {Sakurikar}, P. and {Seifert}, M. and {Sherbert}, L.~E. and {Sherwood-Taylor}, H. and {Shih}, A.~Y. and {Sick}, J. and {Silbiger}, M.~T. and {Singanamalla}, S. and {Singer}, L.~P. and {Sladen}, P.~H. and {Sooley}, K.~A. and {Sornarajah}, S. and {Streicher}, O. and {Teuben}, P. and {Thomas}, S.~W. and {Tremblay}, G.~R. and {Turner}, J.~E.~H. and {Terr{\'o}n}, V. and {van Kerkwijk}, M.~H. and {de la Vega}, A. and {Watkins}, L.~L. and {Weaver}, B.~A. and {Whitmore}, J.~B. and {Woillez}, J. and {Zabalza}, V. and {Astropy Contributors}},
        title = "{The Astropy Project: Building an Open-science Project and Status of the v2.0 Core Package}",
      journal = {\aj},
         year = 2018,
        month = sep,
       volume = {156},
       number = {3},
          eid = {123},
        pages = {123},
          doi = {10.3847/1538-3881/aabc4f},
archivePrefix = {arXiv},
       eprint = {1801.02634},
 primaryClass = {astro-ph.IM},
       adsurl = {https://ui.adsabs.harvard.edu/abs/2018AJ....156..123A}
}

@ARTICLE{2013A&A...558A..33A,
       author = {{Astropy Collaboration} and {Robitaille}, Thomas P. and
         {Tollerud}, Erik J. and {Greenfield}, Perry and {Droettboom}, Michael and
         {Bray}, Erik and {Aldcroft}, Tom and {Davis}, Matt and
         {Ginsburg}, Adam and {Price-Whelan}, Adrian M. and
         {Kerzendorf}, Wolfgang E. and {Conley}, Alexander and {Crighton}, Neil and
         {Barbary}, Kyle and {Muna}, Demitri and {Ferguson}, Henry and
         {Grollier}, Fr{\'e}d{\'e}ric and {Parikh}, Madhura M. and
         {Nair}, Prasanth H. and {Unther}, Hans M. and {Deil}, Christoph and
         {Woillez}, Julien and {Conseil}, Simon and {Kramer}, Roban and
         {Turner}, James E.~H. and {Singer}, Leo and {Fox}, Ryan and
         {Weaver}, Benjamin A. and {Zabalza}, Victor and {Edwards}, Zachary I. and
         {Azalee Bostroem}, K. and {Burke}, D.~J. and {Casey}, Andrew R. and
         {Crawford}, Steven M. and {Dencheva}, Nadia and {Ely}, Justin and
         {Jenness}, Tim and {Labrie}, Kathleen and {Lim}, Pey Lian and
         {Pierfederici}, Francesco and {Pontzen}, Andrew and {Ptak}, Andy and
         {Refsdal}, Brian and {Servillat}, Mathieu and {Streicher}, Ole},
        title = "{Astropy: A community Python package for astronomy}",
      journal = {\aap},
         year = "2013",
        month = "Oct",
       volume = {558},
          eid = {A33},
        pages = {A33},
          doi = {10.1051/0004-6361/201322068},
archivePrefix = {arXiv},
       eprint = {1307.6212},
 primaryClass = {astro-ph.IM},
       adsurl = {https://ui.adsabs.harvard.edu/abs/2013A&A...558A..33A}
}

@ARTICLE{1996A&AS..117..393B,
       author = {{Bertin}, E. and {Arnouts}, S.},
        title = "{SExtractor: Software for source extraction.}",
      journal = {\aaps},
         year = "1996",
        month = "Jun",
       volume = {117},
        pages = {393-404},
          doi = {10.1051/aas:1996164},
       adsurl = {https://ui.adsabs.harvard.edu/abs/1996A&AS..117..393B}
}
\bibliographystyle{aasjournalv7}



\end{document}